\pdfoutput=1
\documentclass[twocolumn]{svjour3}          
\smartqed  
\usepackage{amsmath}
\usepackage{amsfonts}
\usepackage[english]{babel}
\usepackage{booktabs}
\usepackage{float}
\usepackage[T1]{fontenc}
\usepackage{graphicx}
\usepackage{epstopdf} 
\usepackage{grffile}
\usepackage[utf8]{inputenc}
\usepackage{morefloats}
\usepackage{multirow}
\usepackage{placeins}
\usepackage[novbox]{pdfsync} 
\usepackage{pstool}
\usepackage{natbib}
\usepackage[alsoload=abbr]{siunitx}
\usepackage{stfloats}
\usepackage{subfig}
\usepackage[textsize=tiny]{todonotes} 
\usepackage[font=small]{caption} 

\usepackage{upgreek}

\usepackage{hyphenat} 

\usepackage[breaklinks]{hyperref}
\hypersetup{
    bookmarks=true,
    bookmarksnumbered=true,
    colorlinks=true,
    citecolor=blue
}

\makeatletter
\def\cl@chapter{\@elt {theorem}}
\makeatother

\usepackage[capitalize]{cleveref}
    \creflabelformat{equation}{#2#1#3}
    \crefname{figure}{Fig.}{Fig.}
    \crefname{section}{Sec.}{Sec.}
    \crefname{table}{Tab.}{Tab.}
    \crefname{equation}{Eq.}{Eq.}

\usepackage{breakurl}
\usepackage[titletoc, title]{appendix}

\usepackage{color}

\newcommand{\tb}[1]{\textbf{#1}}

\newcommand{\var}[1]{\mathrm{Var} \left[ #1 \right]}

\newcommand{\mean}[1]{\langle #1 \rangle}
\newcommand{\expect}[1]{\mean{#1}}

\newcommand{\Cm}{C_\mathrm{m}}
\newcommand{\tauref}{\tau_\mathrm{refrac}}
\newcommand{\Vspike}{E^\mathrm{spike}}
\newcommand{\Er}{E^{\mathrm{r}}}              
\newcommand{\Vrest}{E_\mathrm{L}}
\newcommand{\taum}{\tau_\mathrm{m}}
\newcommand{\adexdeltaT}{\Delta_\mathrm{T}}
\newcommand{\tauw}{\tau_w}
\newcommand{\Vthresh}{E_{\mathrm{T}}}

\newcommand{\Erev}{E^{\mathrm{rev}}}
\newcommand{\Ereve}{E^{\mathrm{rev,e}}}
\newcommand{\Erevi}{E^{\mathrm{rev,i}}}

\newcommand{\tausyn}{\tau^\mathrm{syn}}
\newcommand{\tausyne}{\tau^{\mathrm{syn,e}}}
\newcommand{\tausyni}{\tau^\mathrm{syn,i}}

\newcommand{\gleak}{g_\mathrm{L}}

\newcommand{\gsyn}{g^\mathrm{syn}}
\newcommand{\Vmem}{V} 
\newcommand{\Isyn}{I^\mathrm{syn}}

\newcommand{\wsyn}{w^\mathrm{syn}}

\newcommand{\taurec}{\tau_\mathrm{rec}}
\newcommand{\taufacil}{\tau_\mathrm{facil}}
\newcommand{\tsoU}{U}
\newcommand{\tsou}{u} 

\newcommand{\tauon}{\tau_\mathrm{on}}

\newcommand{\Nmc}{N_{\mbox{\tiny{MC}}}}
\newcommand{\Nhc}{N_{\mbox{\tiny{HC}}}}
\newcommand{\Npyr}{N_{\mbox{\tiny{PYR}}}}
\newcommand{\Nbas}{N_{\mbox{\tiny{BAS}}}}
\newcommand{\Nrsnp}{N_{\mbox{\tiny{RSNP}}}}
\newcommand{\KCa}{\mathrm{K}_\mathrm {Ca}}
\newcommand{\gext}{\mathrm{g}_\mathrm{ext}}
\newcommand{\ENa}{\mathrm{E}_\mathrm{Na}}
\newcommand{\ECa}{\mathrm{E}_\mathrm{Ca}}
\newcommand{\EK}{\mathrm{E}_\mathrm{K}}
\newcommand{\ECaNMDA}{\mathrm{E}_\mathrm{Ca(NMDA)}}
\newcommand{\gNa}{\mathrm{g}_\mathrm{Na}}
\newcommand{\gK}{\mathrm{g}_\mathrm{K}}
\newcommand{\gNMDA}{\mathrm{g}_\mathrm{NMDA}}
\newcommand{\CaV}{\mathrm{Ca}_\mathrm{V}}
\newcommand{\CaNMDA}{\mathrm{Ca}_\mathrm{NMDA}}
\newcommand{\tauraise}{\tau_{\mathrm{raise}}}
\newcommand{\taudecay}{\tau_{\mathrm{decay}}}

\newcommand{\Vreset}{V_\mathrm{reset}}
\newcommand{\Eleak}{\mathrm{E}_\mathrm{leak}}
\newcommand{\deltaT}{\Delta T}
\newcommand{\VT}{V_T}

\newcommand{\cvisi}{\text{CV}_{\text{ISI}}}
\newcommand{\cvrate}{\text{CV}_{\text{rate}}}
\newcommand{\Ge}{g_\mathrm{exc}}
\newcommand{\Gi}{g_\mathrm{inh}}

\begin{document}


\clearpage

\title{Characterization and Compensation of Network-Level Anomalies in Mixed-Signal Neuromorphic Modeling Platforms}


\author{Mihai A. Petrovici \and
        Bernhard Vogginger \and
        Paul M\"uller \and
        Oliver Breitwieser \and
        Mikael Lundqvist \and
        Lyle Muller \and
        Matthias Ehrlich \and
        Alain Destexhe \and
        Anders Lansner \and
        Ren\'e Sch\"uffny \and
        Johannes Schemmel \and
        Karlheinz Meier
}


\institute{ M.~A.~Petrovici
            \and P.~M\"uller
            \and O.~Breitwieser
            \and J.~Schemmel
            \and K.~Meier
                \at Kirchhoff Institute for Physics\\Ruprecht-Karls-Universit\"at Heidelberg, Germany\\
                    Tel.: +49 6221 549847\\
                    \email{mpedro@kip.uni-heidelberg.de}
            \and M.~Ehrlich \and R.~Sch\"uffny \and B.~Vogginger
                    \at Institute of Circuits and Systems, Technische Universit\"at Dresden, Germany
            \and M.~Lundqvist
            \and A.~Lansner
                \at Computational Biology, KTH Stockholm, Sweden
            \and A.~Destexhe \and L.~Muller
                \at Unit\'e de Neuroscience, Information et Complexit\'e, CNRS, Gif sur Yvette, France
}

\date{October 10, 2014}

\maketitle

\section*{Abstract}

Advancing the size and complexity of neural network models leads to an ever increasing demand for computational resources for their simulation.
Neuromorphic devices offer a number of advantages over conventional computing architectures, such as high emulation speed or low power consumption, but this usually comes at the price of reduced configurability and precision.
In this article, we investigate the consequences of several such factors that are common to neuromorphic devices, more specifically limited hardware resources, limited parameter configurability and parameter variations due to fixed-pattern noise and trial-to-trial variability.
Our final aim is to provide an array of methods for coping with such inevitable distortion mechanisms.
As a platform for testing our proposed strategies, we use an executable system specification (ESS) of the BrainScaleS neuromorphic system, which has been designed as a universal emulation back-end for neuroscientific modeling.
We address the most essential limitations of this device in detail and study their effects on three prototypical benchmark network models within a well-defined, systematic workflow.
For each network model, we start by defining quantifiable functionality measures by which we then assess the effects of typical hardware-specific distortion mechanisms, both in idealized software simulations and on the ESS\@.
For those effects that cause unacceptable deviations from the original network dynamics, we suggest generic compensation mechanisms and demonstrate their effectiveness.
Both the suggested workflow and the investigated compensation mechanisms are largely back-end independent and do not require additional hardware configurability beyond the one required to emulate the benchmark networks in the first place.
We hereby provide a generic methodological environment for configurable neuromorphic devices that are targeted at emulating large-scale, functional neural networks.


\section{Introduction}
\label{intro}

\subsection{Modeling and computational neuroscience}

The limited availability of detailed biological data has always posed a major challenge to the advance of neuroscientific understanding.
The formulation of theories about information processing in the brain has therefore been predominantly model-driven, with much freedom of choice in model architecture and parameters.
As more powerful mathematical and computational tools became available, increasingly detailed and complex cortical models have been proposed.
However, because of the manifest nonlinearity and sheer complexity of interactions that take place in the nervous system, analytically treatable ensemble-based models can only partly cover the vast range of activity patterns and behavioral phenomena that are characteristic for biological nervous systems \cite{laing2009stochastic}.
The high level of model complexity often required for computational proficiency and biological plausibility has led to a rapid development of the field of computational neuroscience, which focuses on the simulation of network models as a powerful complement to the search for analytic solutions \cite{brette2007simulation}.

The feasibility of the computational approach has been facilitated by the development of the hardware devices used to run neural network simulations.
The brisk pace at which available processing speed has been increasing over the past few decades, as allegorized by Moore's Law,
as well as the advancement of computer architectures in general, closely correlate to the size and complexity of simulated models.
Today, network models with tens of thousands of neurons are routinely simulated on desktop machines, with supercomputers allowing several orders of magnitude more \cite{djurfeldt2008bluegene, helias2012supercomputers}.
However, as many authors have pointed out (see e.g. \cite{morrison05distributed}, \cite{brette2007simulation}), the inherently massively parallel structure of biological neural networks becomes progressively difficult to map to conventional architectures based on digital general-purpose CPUs, as network size and complexity increase.

Conventional simulation becomes especially restrictive when considering long time scales, such as are required for modeling long-term network dynamics or when performing statistics-intensive experiments.
Additionally, power consumption can quickly become prohibitive at these scales \cite{bergman2008exascale, hasler2013}.

\subsection{Neuromorphic Hardware}

The above issues can, however, be eluded by reconsidering the fundamental design principles of conventional computer systems.
The core idea of the so-called neuromorphic approach is to implement features (such as connectivity) or components (neurons, synapses) of neural networks directly \emph{in silico}: instead of calculating the dynamics of neural networks, neuromorphic devices contain physical representations of the networks themselves, behaving, by design, according to the same dynamic laws.
An immediate advantage of this approach is its inherent parallelism (emulated network components evolve in parallel, without needing to wait for clock signals or synchronization), which is particularly advantageous in terms of scalability.
First proposed by Mead in the 1980s \cite{mead88silicon, mead89analog, mead90neuromorphic}, the neuromorphic approach has since delivered a multitude of successful applications \cite{renaud2007neuromimetic, indiveri2009artificial, indiveri2011, ieee2014specialissue}.

By far the largest number of neuromorphic systems developed thus far are highly application-specific, such as visual processing systems \cite{serrano_nips2005, merolla2006dynamic, netter02arobotic, delbrueck04silicon} or robotic motor control devices \cite{lewis00toward}.
Several groups have focused on more biological aspects, such as the neuromorphic implementation of biologically-inspired self-organization and learning \cite{hafliger2007adaptive, mitra09realtime}, detailed replication of Hodgkin-Huxley neurons \cite{zou2006analogneurons} or hybrid systems interfacing analog neural networks with living neural tissue \cite{bontorin2007realtime}.

These devices, however, being rather specialized, can not match the flexibility of traditional software simulations.
Adding configurability comes at a high price in terms of hardware resources, due to various hardware-specific limitations, such as physical size and essentially two-dimensional structure.
So far there have only been few attempts at realizing highly configurable hardware emulators \cite{indiveri_tnn2006, vogelstein2007reconfigurable, rocke2008fpaanetworks, schemmel_iscas2010, furber2012}.
This approach alone, however, does not completely resolve the computational bottleneck of software simulators, as scaling neuromorphic neural networks up in size becomes non-trivial when considering bandwidth limitations between multiple interconnected hardware devices \cite{serrano_tcas2007, berge_iscas07, indiveri2008_vlsi, fieres_ijcnn2008, serrano2009caviar}.

\subsection{The BrainScaleS hardware system}

A very efficient way of interconnecting multiple VLSI (Very Large Scale Integration) modules is offered by so-called wafer-scale integration.
This implies the realization of both the modules in question and their communication infrastructure on the same silicon wafer, the latter being done in a separate, post-processing step.
The BrainScaleS wafer-scale hardware \cite{schemmel_iscas2010} uses this process to achieve a high communication bandwidth between individual neuromorphic cores on a wafer, thereby allowing a highly flexible connection topology of the emulated network.
Together with the large available parameter space for neurons and synapses, this creates a neuromorphic architecture that is comparable in flexibility with standard simulation software.
At the same time, it provides a powerful alternative to software simulators by avoiding the abovementioned computational bottleneck, in particular owing to the fact that the emulation duration does not scale with the size of the emulated network, since individual netowrk components operate, inherently, in parallel.
An additional benefit which is inherent to this specific VLSI implementation is the high acceleration with respect to biological real-time, which is facilitated by the high on-wafer bandwidth.
This allows investigating the evolution of network dynamics over long periods of time which would otherwise be strongly prohibitive for software simulations.

\subsection{Hardware-Induced Distortions: A Systematic Investigation}

Along with the many advantages it offers, the neuromorphic approach also comes with limitations of its own.
These have various causes that lie both in the hardware itself and the control software.
We will later identify these causes, which we henceforth refer to as \textit{distortion mechanisms}.
The neural network emulated by the hardware device can therefore differ significantly from the original model, be it in terms of pulse transmission, connectivity between populations or individual neuron or synapse parameters.
We refer to all the changes in network dynamics (i.e., deviations from the original behavior defined by software simulations) caused by hardware-specific effects as \textit{hardware-induced distortions}.

Due to the complexity of state-of-the-art neuromorphic platforms and their control software, as well as the vast landscape of emulable neural network models, a thorough and systematic approach is essential for providing reliable information about causal mechanisms and functional effects of hardware-induced distortions in model dynamics and for ultimately designing effective compensation methods.
In this article, we design and perform such a systematic analysis and compensation for several hardware-specific distortion mechanisms.

First and foremost, we identify and quantify the most important sources of model distortions.
We then proceed to investigate their effect on network functionality.
In order to cover a wide range of possible network dynamics, we have chosen three very different cortical network models to serve as benchmarks.
In particular, these models implement several prototypical cortical paradigms of computation, relying on winner-take-all structures (attractor networks), precise spike timing correlations (synfire chains) or balanced activity (self-sustained asynchronous irregular states).

For every emulated model, we define a set of functionality criteria, based on specific aspects of the network dynamics.
This set should be complex enough to capture the characteristic network behavior, from a microscopic (e.g., membrane potentials) to a mesoscopic level (e.g., firing rates) and, where suitable, computational performance at a specific task.
Most importantly, these criteria need to be precisely quantified, in order to facilitate an accurate comparison between software simulations and hardware emulations or between different simulation/emulation back-ends in general.
The chosen functionality criteria should also be measured, if applicable, for various relevant realizations (i.e.\ for different network sizes, numbers of functional units etc.) of the considered network.

Because multiple distortion mechanisms occur simultaneously in hardware emulations, it is often difficult, if not impossible, to understand the relationship between the observed effects (i.e., modifications in the network dynamics) and their potential underlying causes.
Therefore, we investigate the effects of individual distortion mechanisms by implementing them, separately, in software simulations.
As before, we perform these analyses over a wide range of network realizations, since - as we will show later - these may strongly influence the effects of the examined mechanisms.

After having established the relationship between structural distortions caused by hardware-specific factors and their consequences for network dynamics, we demonstrate various compensation techniques in order to restore the original network behavior.

In the final stage, for each of the studied models, we simulate an implementation on the hardware back-end by running an appropriately configured executable system specification, which includes the full panoply of hardware-specific distortion mechanisms.
Using the proposed compensation techniques, we then attempt to deal with all these effects simultaneously.
The results from these experiments are then compared to results from software simulations, thus allowing a comprehensive assertion of the effectivity of our proposed compensation techniques, as well as of the capabilities and limitations of the neuromorphic emulation device.

\subsection{Article Structure}

In \Cref{sec:hardware}, we describe our testbench neuromorphic modeling platform with its most relevant components, as well as the essential layers of the operation workflow.
We continue by explaining the causes of various network-level distortions that are expected to be common for similar mixed-signal neuromorphic devices.
In the same section, we also introduce the executable system specification of the hardware, which we later use for experimental investigations.

\Cref{sec:models} contains the description of the three benchmark models.
We start the section on each of the models with a short summary of all the relevant findings.
We then describe its architecture and characteristic aspects of its dynamics which we later use as quality controls.
We continue by discussing the effects of individual hardware-specific distortion mechanisms as observed in software simulations, propose various compensation strategies and investigate their efficacy in restoring the functionality of the network model in question.
Subsequently, we apply these methods to large-scale neuromorphic emulations and examine the results.

Finally, we summarize and discuss our findings in \cref{sec:conclusions}.

\section{Neuromorphic testbench and investigated distortion mechanisms}
\label{sec:hardware}
In this section we introduce the BrainScaleS neuromorphic wafer-scale hardware system and its executable system specification, henceforth called the ESS, as the testbench for our studies.
The system's hardware and software components are only described on an abstract level, while highlighting the mechanisms responsible for distortions of the emulated networks.
Finally, we identify the three most relevant causes of distortion as being synapse loss, synaptic weight noise and non-configurable axonal delays.

\subsection{The BrainScaleS wafer-scale hardware}
\label{sec:neuromorphic_components}

\begin{figure}[t]
    \centering          
    \includegraphics{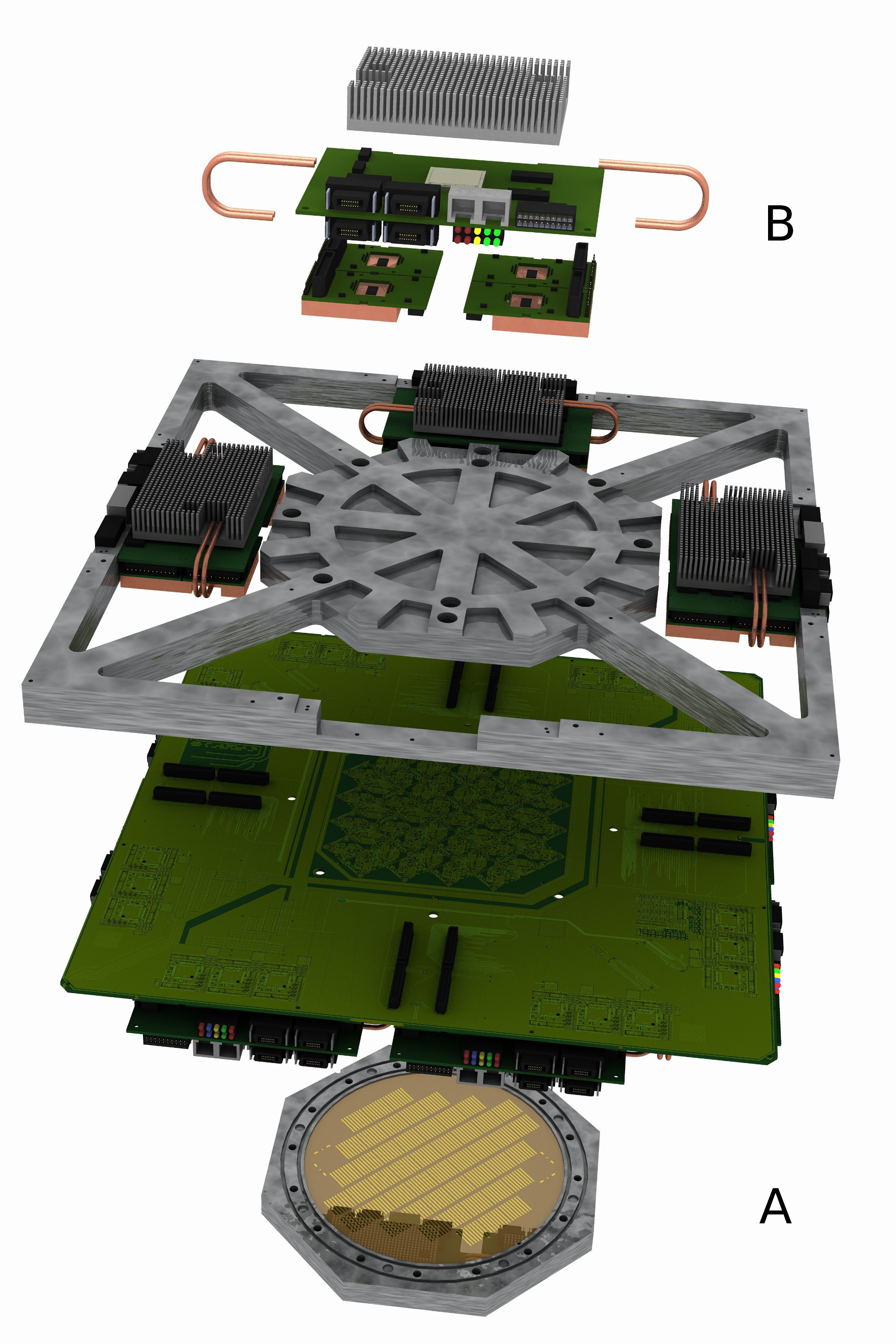}
	\caption{{\bf The BrainScaleS wafer-scale hardware system:} (\textbf{A}) Wafer comprising HICANN building blocks and on-wafer communication infrastructure covered by an aluminium plate, (\textbf{B}) digital inter-wafer and wafer-host communication modules. Also visible: mechanical and electrical support.
    \label{fig:waferscale_system}
	}
\end{figure}

\begin{figure}[t]
    \centering          
    \includegraphics{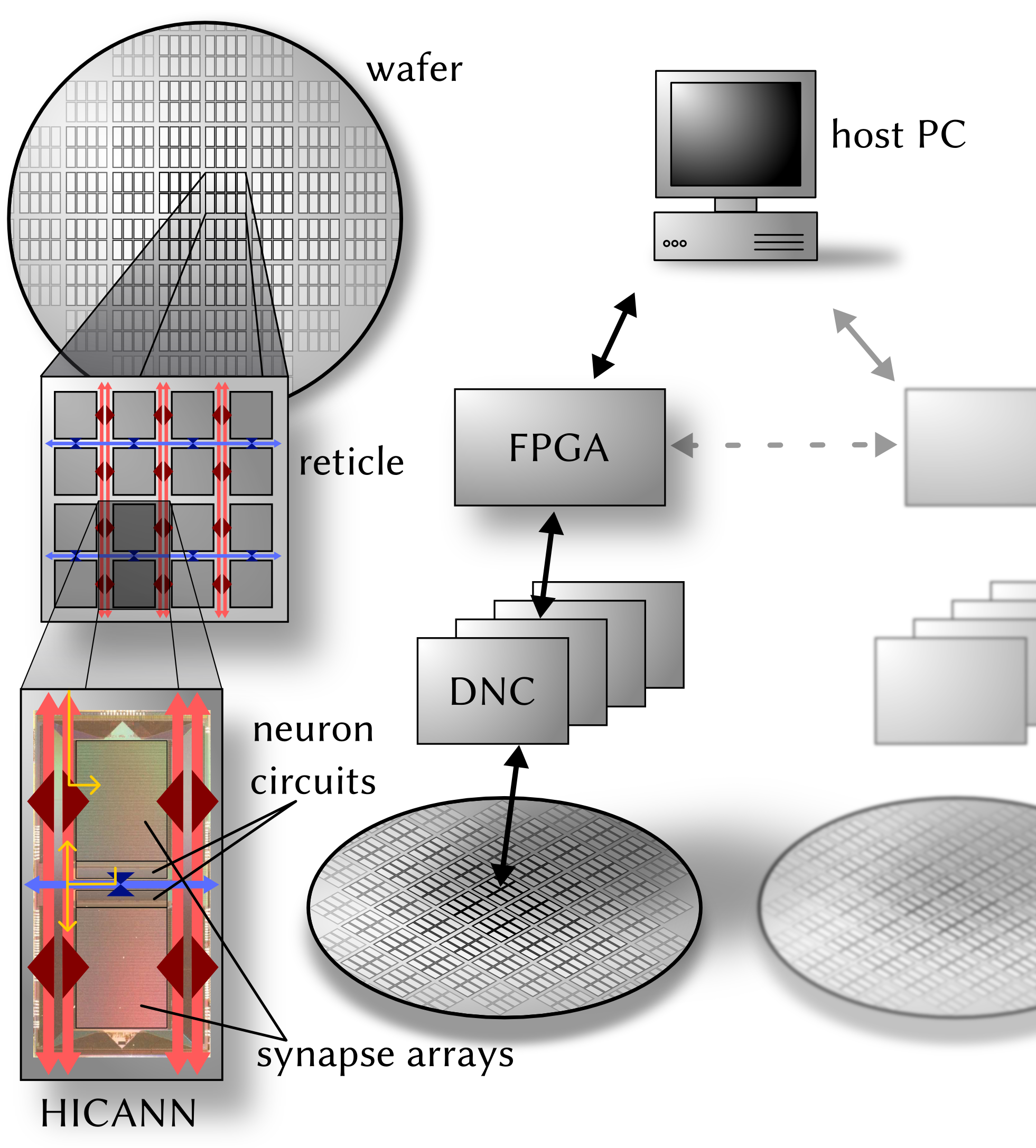}
    \caption{
		{\bf Architecture of the BrainScaleS wafer-scale hardware system.} Left: The HICANN building block has two symmetric halves with synapse arrays and neuron circuits.
        Neural activity is transported horizontally (blue) and vertically (red) via asynchronous buses that span over the entire wafer.
        Exemplary spike paths are shown in yellow on the HICANN: The incoming spike packet is routed to the synapse
        drivers. In the event that a neuron spikes, it emits a spike packet back into the routing network.
        Right: Off-wafer connectivity is established by a hierarchical packed-based network via DNCs and FPGAs.
        It interfaces the on-wafer routing buses on the HICANN building blocks.
        Several wafer modules can be interconnected using routing functionality between the FPGAs.
    \label{fig:comm_infra}
    }
\end{figure}

\Cref{fig:waferscale_system} shows a 3D-rendered image of the BrainScaleS wafer-scale hardware system:
the 8~inch silicon wafer contains \num{196608} neurons and \num{44} million plastic synapses implemented in mixed-signal VLSI circuitry.
Due to the
high integration of the circuits, the capacitances and thus the intrinsic time constants are small, so that
neural dynamics take place approximately \num{10000} faster than biological real time.
The principal building block of the wafer is the so-called HICANN (High Input Count Analog Neural Network) chip \cite{schemmel_iscas2010, schemmel_ijcnn2008}.
During chip fabrication one is limited to a maximum area that can be simultaneously exposed during photolitography, a reticle,
thus usually such a wafer is cut into individual chips after production.
For the BrainScaleS system, however, the wafer is left intact, and additional wiring is applied onto the wafer's surface in a post-processing step.
This process establishes connections betwen all 384 HICANN blocks that allow a very high bandwidth for on-wafer pulse-event communication \cite{schemmel_ijcnn2008}.
The neuromorphic wafer is accompanied by a stack of digital communication modules for the connection of the wafer to the host PC and to other wafers (\Cref{fig:comm_infra} and \cref{sec:communication_infrastructure}).

\subsubsection{HICANN building block}
\label{sec:hicann}
On the HICANN chip (lower left of \Cref{fig:comm_infra}), one can recognize two symmetric blocks which hold the analog core modules.
The upper block is depicted in detail in \Cref{fig:hicann_routing}:
Most of the area is occupied by the synapse array with 224 rows and 256 columns.
All synapses in a column are connected to one of the 256 neuron circuits located at the center of the chip.
For each two adjacent synapse rows, there is one \emph{synapse driver} that forms the input for pre-synaptic pulses to the synapse array.
Synapse drivers are evenly distributed to the left and right side of one synapse array (56 per side).
A grid of horizontal and vertical buses enables the routing of spikes from neuron circuits to synapse drivers.

Up to 64 neuron circuits can be interconnected to form neurons with up to 14336 synapses.
The neurons emulate the dynamics of the Adaptive-Exponential Integrate-and-Fire model (AdEx) \cite{brette_05} in analog circuitry,
defined by equations for the membrane voltage $\Vmem$, the adaption current $w$ and a reset condition that applies when a spike is triggered:

\begin{align}
	\Cm\frac{d\Vmem}{dt} = & -\gleak(\Vmem-\Vrest)+\gleak \adexdeltaT\exp\left(\frac{\Vmem-\Vthresh}{\adexdeltaT}\right) \nonumber \\
                         & - w + \Isyn \label{eqn:adex1} \\
   \tauw\frac{dw}{dt} = &~a(\Vmem-\Vrest) - w \label{eqn:adex2} \\
 \textrm{if} ~ \Vmem \ge &~\Vspike : \quad
    \begin{cases}
        \Vmem \rightarrow \Er  \\
        w \rightarrow w + b 
    \end{cases} \quad ,\label{eqn:adex3} 
\end{align}

where $\Cm$, $\gleak$ and $\Vrest$ denote the membrane capacitance, leak conductance and leak potential, respectively, $\Vthresh$ and $\adexdeltaT$ represent the spike initiation threshold and the threshold slope factor and $\tauw$ and $a$ represent the adaptation time constant and coupling parameter.
When $\Vmem$ reaches a certain threshold value $\Vspike$, a spike is emitted and the membrane potential is reset to $\Er$.
At the same time, the adaptation variable is increased by a fixed amount $b$, thereby allowing for spike-frequency adaptation.
An absolute refractory mechanism is supported by clamping $\Vmem$ to its reset value for the refractory time $\tauref$.
The generated spikes are transmitted digitally to synapse drivers (analog multiplier), synapses (digital multiplier) and finally other neurons, where postsynaptic conductance courses are generated and summed up linearly, resulting in the synaptic current $\Isyn$:

\begin{align}
    & \Isyn = \sum_{\mathrm{synapses} \, i} g_i (\Erev_i - \Vmem) \label{eqn:cond_exp_syn1} \\
    & \tausyn\frac{dg_i}{dt} = -g_i + \wsyn_i \sum_{\mathrm{spikes} \, s} \delta(t-t_s) \quad . \label{eqn:cond_exp_syn2}
\end{align}
Here, $g_i$ represents the synaptic conductance and $\Erev_i$ the synaptic reversal potential of the $i$-th synapse, $\tausyn$ the time constant of the exponential decay and $\wsyn$ the synaptic weight.
In the hardware implementation \cite{millner10}, each neuron features two of such synaptic input circuits, which are typically used for excitatory and inhibitory input.
Nearly all parameters of the neuron model and the synaptic input circuits are individually adjustable by means of analog storage banks based on floating gate technology \cite{lande96}.
In the hardware neuron, both the circuit for the adaption mechanism and the exponential term circuit can be effectively disconnected from the membrane capacitance, such that a simple Leaky Integrate-and-Fire (LIF) model can also be emulated.
The hardware membrane capacitance is fixed to one of two possible values.
As the parameters controlling the temporal dynamics of the neuron such as $\gleak$ and the time constants are configurable within a wide range, the hardware is able to run at a variable speedup factor ($10^3-10^5$) compared to biological real time.
In particular, the translation of the membrane capacitance between the hardware and the biological domain can be chosen freely due to the independent configurability of both membrane and synaptic conductances, thereby effectively allowing the emulation of point neurons of arbitrary size - within the limits imposed by the hardware parameter ranges.

In contrast to neurons, where each parameter is fully configurable within the specified ranges, the \textit{synaptic weights} are adjustable by a combination of analog and digital memories.
The synaptic weight $\wsyn$ is proportional to a row-wise adjustable analog parameter $g_\textrm{max}$ and to a 4-bit digital weight specific to each synapse.
The $g_{\textrm{max}}$ of two adjacent rows can be configured to be a fixed multiple of each other. This way, two synapses of adjacent rows can be combined to offer a weight resolution of 8 bits, at the cost of halving the number of synapses for this synapse driver.

\emph{Long-term learning} is incorporated in every synapse through spike-timing-dependent plasticity (STDP) \cite{poo98stdp}.
The implemented STDP mechanism follows a pairwise update rule with programmable update functions \cite{morrison08_stdp}.
As STDP is not contained in the models investigated in this article (\Cref{sec:models}), we refer to \cite{bruederle_biolcybern2010,schemmel_ijcnn06,schemmel_iscas07} for details on the hardware implementation and to \cite{Pfeil12_90} for an applicability study of these circuits.

In contrast to the long-term learning, the implemented \emph{short-term plasticity} mechanism (STP) decays over several hundreds of milliseconds.
It is motivated by the phenomenological model by \cite{markram98information} and depends only on the pre-synaptic activity, therefore being implemented in the synapse driver.
For every incoming spike, a synapse only has access to a portion $U$ of the recovered partition $R$ of its total synaptic weight $\wsyn_\mathrm{max}$, which then instantly decreases by a factor $1-U$ and recovers slowly along an exponential with the time constant $\taurec$, thus emulating synaptic depression.
Facilitation is implemented by replacing the fixed $U$ with a running variable $\tsou$, which increases with every incoming spike by an amount $U(1-\tsou)$ and then decays exponentially back to U with the time constant $\taufacil$:

\begin{align}
	\wsyn_{n+1} & = \wsyn_\mathrm{max}R_{n+1}\tsou_{n+1} \\
	R_{n+1} & = 1-\left[1-R_n(1-\tsou_n)\right]\exp\left(-\frac{\Delta t}{\taurec}\right) \\
    \tsou_{n+1} & = U+\tsou_n(1-U)\exp\left(-\frac{\Delta t}{\taufacil}\right)
\label{eq:tm_stp}
\end{align}

with $\Delta t$ being the time interval between the $n$th and $(n+1)$st afferent spike.
In contrast to the original Tsodyks-Markram (TSO) mechanism, the hardware implementation does not allow simultaneous depression and facilitation \cite{schemmel_ijcnn2008,bill2010compensating}.
See \cref{sec:methods-hw-stp} for details about the hardware implementation and the translation of the original model to the hardware STP.

All of the neuron and synapse parameters mentioned above are affected by fixed-pattern noise due to transistor-level mismatch in the manufacturing process.
Additionally, the floating gate analog parameter storage reproduces the programmed voltage with a limited precision on each re-write.
This leads to trial-to-trial variation for each experiment (see \cref{sec:methods-hw-measurements} for exemplary measurements).
Limited configurability, such as the discretization of available synaptic weights, is another source for discrepancy between targeted and realized configuration.
The trial-to-trial variability, which cannot be remedied by calibration (\cref{sec:software}), is assumed to be less than \SI{30}{\%} (standard-deviation-to-mean ratio) for synaptic weights.
Other neuron parameters are assumed to have a much smaller variability. $\Vrest$, $\Vthresh$, $\Erev$ have a standard deviation of less than \SI{1}{\milli\volt} in the biological domain (cf. \cref{sec:software,sec:methods-hw-measurements}).
In this publication, we limit all investigations to the variation of synaptic weights, as they are assumed to be the dominant effect.
To accomodate the total effect of trial-to-trial and fixed-pattern variation as well as parameter discretization, we simulate deviations of up to \SI{50}{\%} (cf. \cref{sec:investigated_distortions}).

For technical details about the HICANN chip and its components we refer to \cite{schemmel_iscas2010, schemmel_ijcnn2008}.

\subsubsection{Communication infrastructure}
\label{sec:communication_infrastructure}

\begin{figure}[t]
    \centering
    \includegraphics{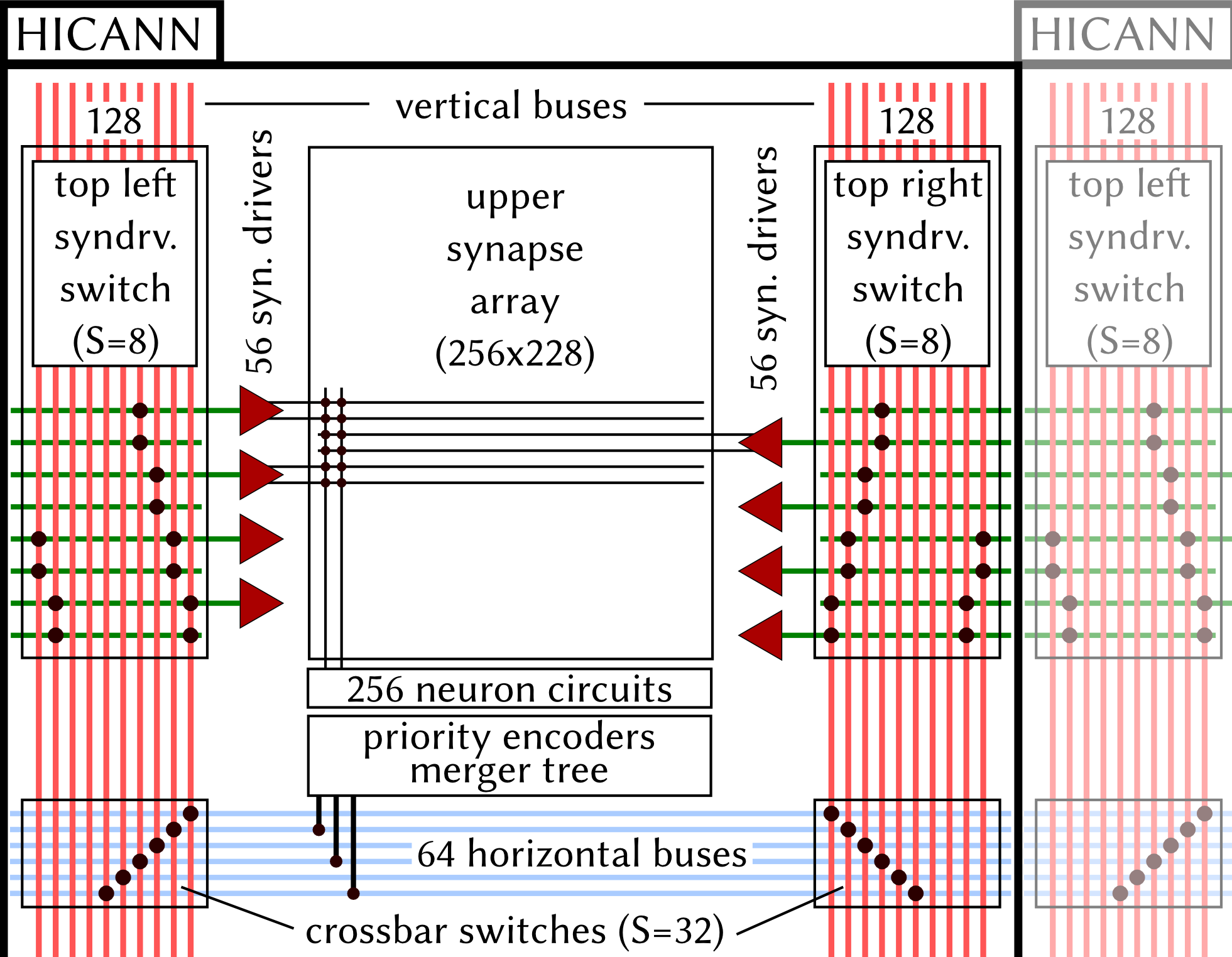}
	\caption{{\bf Components and connectivity of the HICANN building block.}
		The figure shows the upper block of the HICANN chip: most of the area is occupied by the synapse array with 256 columns and 224 rows.
		Each synapse column is connected to one of 256 neuron circuits, from which up to 64 can be interconnected to form larger neurons with up to 14336 input synapses.
		When a neuron fires, a 6-bit address representing this neuron is generated and injected into one of eight accessible horizontal buses after passing a merger stage.
		Via two statically configurable switches (crossbar rsp. synapse driver switch) these pulses are routed to the synapse drivers, which operate two synapse rows each.
		Every synapse is configured to a specific 6-bit address, so that, when a pre-synaptic pulse with matching address arrives, a post-synaptic conductance course is generated at the associated neuron  circuit.
		Both switch matrices are sparse, i.e. configurable switches do not exist at all crossings of horizontal and vertical lines, but e.g. only at every 8th crossing (Sparseness S=8).
		On the wafer, the horizontal and vertical buses, as well as the horizontal lines connected to the synapse drivers do not end at the HICANN borders, but go beyond them.
		\label{fig:hicann_routing}
	}
\end{figure}

The infrastructure for pulse communication in the wafer-scale system is supplied by a two-layer approach:
While the on-wafer network routes pulses between neurons on the same wafer, the off-wafer network connects the wafer to the outside world, i.e. to the host PC or to other wafers.

The backbone of the \emph{on-wafer communication} consists of a grid of horizontal and vertical buses enabling the transport of action potentials by a mixture of time division and space division multiplexing.
Each HICANN building block contains 64 horizontal buses at its center and 128 vertical buses located on each side of the synapse blocks, as can be seen in \cref{fig:hicann_routing}.
A bus can carry the spikes of up to 64 source neurons by transmitting a serial 6-bit signal encoding the currently sending neuron (with an ID from 0 to 63).
When a neuron fires, its pulse is first processed by one of eight priority encoders and finally injected into a horizontal bus after passing a merger stage.
By enabling a static switch of a sparse crossbar between horizontal and vertical buses, the injected serial signal can be made available to a vertical bus next to the synapse array.
Another sparse switch matrix allows to feed the signals from the vertical buses into the synapse array, more precisely into the synapse drivers which represent the data sinks of the routing network.
Synapse drivers can be connected in a chain, forwarding their input to their top or bottom neighbours, thereby allowing to increase the number of synapse rows fed by the same routing bus.
The bus lanes do not end at the HICANN border but run over the whole wafer by edge-connecting the HICANN building blocks (\Cref{fig:comm_infra}).
We note that, due to electrotechnical reasons, the switches could not be implemented as full matrices, thus their sparseness was chosen as a compromise still providing maximum flexibility for implementing various neural network topologies \cite{fieres_ijcnn2008, schemmel_iscas2010}.
Both the sparseness of the switches and the limited number of horizontal and vertical buses represent a possible restriction for the connectivity of network models.
If an emulated network requires a connectivity that exceeds the on-wafer bus capacity, some synapses will be impossible to map to the wafer and will therefore be lost.

Pulse propagation delays in the routing network are small, distance-dependent and not configurable: the time between spike detection and the onset of a post-synaptic potential (PSP) has been measured as 120 \nano\second~ for a recurrent connection on a HICANN.
The additional time needed to transmit a pulse across the entire wafer is typically less than \SI{100}{\nano\second} \cite{schemmel_ijcnn2008}, hence the overall delay sums up to 1.2 - \SI{2.2}{\milli\second} in the biological time domain, assuming a speedup factor of $10^4$.
Also, in case of synchronous bursting of the neurons feeding one bus, some pulses are delayed with respect to others, as they are processed successively:
A priority encoder handles the spikes of 64 hardware neurons with priority fixed by design.
If several neurons have fired, the pulse of the neuron with highest priority is transmitted first to the connected horizontal bus. 
The priority encoder can process one pulse every two clock cycles ($2\times\SI{5}{\nano\second}$), leading to an additional delay for the pulses with lower priority. 
In rare cases some pulses may be completely discarded, e.g., when the total rate of all 64 neurons feeding one bus exceeds \SI{10}{\kilo\hertz} for longer than \SI{6.4}{\milli\second} (in biological real-time).

A hierarchical packet-based network provides the infrastructure for \emph{off- and inter-wafer communication}.
All HICANNs on the wafer are connected to the surrounding system and to other wafers via 12 pulse communication subgroups (PCS).
Each PCS consists of one FPGA (Field Programmable Gate Array) and 4 ASICs (Application Specific Integrated Circuits) that were designed for high-bandwidth pulse-event communication (so-called Digital Network Chips or DNCs).
Being the only communication link to/from the wafer, the off-wafer network also transports the configuration and control information for all the circuits on the wafer.
As depicted in \cref{fig:comm_infra}, the network is hierarchically organized: one FPGA is connected to four DNCs, each of which is connected to 8 HICANNs of a reticle.
Each FPGA is also connected to the host PC and potentially to up to 4 other FPGAs.
When used for pulse-event communication, an FPGA-DNC-HICANN connection supports a throughput of 40 Mevents/s \cite{scholze11b} with a timing precision of \SI{4}{\nano\second}.
In the biological time domain, this corresponds to monitoring the spikes of all 512 neurons on a HICANN firing with a mean rate of 8 Hz each with a resolution of \SI{0.04}{\milli\second}.
The same bandwidth is available simultaneously in the opposite direction, allowing a flexible network stimulation with user-defined spiketrains.
For each FPGA-DNC-HICANN connection there are 512 pulse addresses that have to be subdivided into blocks of 64 used for either stimulation or recording.
For all technical details about the PCS, the FPGA design and the DNC, we refer to \cite{scholze11a,hartmann2010,scholze2010heap}.

Although the off-wafer communication interface allows the interconnection of multiple wafers, we restrict our studies here to the use of  a single wafer.

\subsection{Software framework}
\label{sec:software}

The utilized software stack \cite{bruederle_biolcybern2010} allows the user to define a network
description and maps it to a hardware configuration.

The network definition is accomplished by using PyNN \cite{davison08pynn}, a si\-mu\-la\-tor-independent API (Application Programming Interface) to describe spiking neural network models.
It can interface to several simulation platforms such as NEURON \cite{hines09} or NEST \cite{eppler2008} as well as to neuromorphic hardware platforms \cite{bruederle09establishing,galluppi2010}.

The mapping process \cite{ehrlich2010anniip, bruederle_biolcybern2010} translates the PyNN description of the neural network structure, as well as its neuron and synapse models and parameters, in several steps into a neuromorphic device configuration.
This translation is constrained by the architecture of the device and its available resources.

The first step of the mapping process is to allocate static structural neural network elements to particular neuromorphic components during the so-called \textit{placement}.
Subsequently, a \textit{routing} step is executed for establishing connections in between the placed components.
During the final \textit{parameter transformation} step, all parameters of the network components (neurons and synapses) are translated into hardware parameters.
First, the model parameters are transformed to the voltage and time domain of the hardware, taking into account the acceleration and the voltage range of \SIrange{0}{1.8}{\volt}~\cite{millner10}.
Second, previously obtained \textit{calibration} data is used to reduce mismatches between ideal neuromorphic circuitry behavior and real analogue signal hardware behavior.

The objective of the mapping process is to find a configuration of the hardware that best reproduces the neural network experiment specified in PyNN.
The most relevant constraints are sketched in the following:

Each hardware neuron circuit has a limited number of 224 incoming synapses.
By interconnecting several neuron circuits one can form ``larger'' neurons with more incoming synapses (\Cref{sec:hicann}), with the trade-off that the overall number of neurons is reduced.
Still, each hardware synapse can not be used to implement a connection from an arbitrary neuron but only from a subset of neurons, namely the 64 source neurons whose pulses arrive at the corresponding synapse driver.
For networks larger than \num{10000} neurons it is the \emph{limited number of inputs} to one HICANN that becomes even more restricting, as there are only 224 synapse drivers (cf. \cref{fig:hicann_routing}), yielding a maximum of 14366 different source neurons for all neurons that are placed to the same HICANN.
Hence, one objective of the mapping process is to reduce this number of source neurons per HICANN, thus increasing the number of realized synapses on the hardware.
In general, this criterion is met when neurons with common pre-synaptic partners are placed onto the same HICANN and neurons with common targets inject their pulses into the same on-wafer routing bus.

All of the above, as well as the limited number of on-wafer routing resources (\cref{sec:communication_infrastructure})
make the mapping optimization an NP-hard problem.
The used placement and routing algorithms, which improve upon the ones described in \cite{bruederle_biolcybern2010} and \cite{fieres_ijcnn2008} but are far from being optimal, can minimize the effect of these constraints only to a certain degree.
Thus, depending on the network model size, its connectivity, and the choice of the mapping algorithms, \emph{synapses are lost} during the mapping process;
in other words, some synapses of a network defined in PyNN will be inexistent in the corresponding network emulated on the hardware.
For an estimation of the amount of synapse loss, we first scaled all three benchmark models to sizes between \num{1000} and \num{100000} neurons and mapped them onto the hardware using a simple, not optimized placement strategy.
The results strongly depend on the size and the connectivity structure of the emulated network.
In order to allow a comprehensive discussion within this study, we then used various placement strategies, sometimes optimizing the mapping by hand to minimize the synapse loss, or purposely using a wasteful allocation of resources to generate synapse loss.

\subsection{Executable system specification (ESS)}
\label{sec:ess}

The ESS is a detailed simulation of the hardware platform \cite{ehrlich_ssd07, bruederle_biolcybern2010} that replicates the topology and dynamics of the communication infrastructure as well as the analog synaptic and neuronal components.

The simulation encompasses a numerical solution of the equations that govern the hardware neuron and synapse dynamics (\Cref{eqn:adex1,eqn:adex2,eqn:adex3,eqn:cond_exp_syn1,eqn:cond_exp_syn2,eqn:TSO,eqn:hw_stp1,eqn:hw_stp2}) and a detailed reproduction of the digital communication infrastructure at the level of individual spike transmission in logical hardware modules.
The ESS is a \textit{specification} of the hardware in the sense that its configuration space faithfully maps the possible interconnection topologies, parameter limits, parameter discretization and shared parameters.
Being executable, the ESS also covers dynamic constraints, such as the consecutive processing of spikes which can lead to spike time jitter or spike loss.
Variations in the analog circuits due to production variations are not simulated at transistor level but are rather artificially imposed on ideal hardware parameters.
In this article, only synaptic weight noise is considered, as detailed in \cref{sec:investigated_distortions}.
All of this allows to simultaneously capture the complex dynamic behavior of the hardware and comply with local bandwidth limitations, while allowing relatively quick simulations due to the high level of abstraction.
Simulations on the ESS can be controlled using PyNN (\Cref{sec:software}), similarly to any other PyNN-compatible back-end.
Both for the real hardware and for the ESS, the mapping process translates a PyNN network into a device configuration, which is then used as an input for the respective back-end.
One particular advantage of the ESS is that it allows access to state variables which can otherwise not be read out from the real hardware, such as the logging of lost or jittered time events.

\subsection{Investigated distortion mechanisms}
\label{sec:investigated_distortions}

Reviewing the hardware and software components of the BrainScaleS wafer-scale system (\Cref{sec:neuromorphic_components,sec:software}) leaves us with a number of mechanisms that can affect or impede the emulation of neural network models:
\begin{itemize}
	\item neuron and synapse models are cast into silicon and can not be altered after chip production
	\item limited ranges for neuron and synapse parameters
	\item discretized and shared parameters
	\item limited number of neurons and synapses
	\item restricted connectivity
	\item synapse loss due to non-optimal algorithms for NP-hard mapping
	\item parameter variations due to transistor level mismatch and limited re-write precision
	\item non-configurable pulse delays and jitter
	\item limited bandwidth for stimulation and recording of spikes
\end{itemize}
It is clear that, for all of the above distortion mechanisms, it is possible to find a corner case where network dynamics are influenced strongly.
However, a few of these effects stand out:
on one hand, they are of such fundamental nature to mixed-signal VLSI that they are likely to play some role in any similar neuromorphic device;
on the other hand, they are expected to influence any kind of emulated network to some extent.
We have therefore directed our focus towards these particular effects, which we summarize in the following.
In order to allow general assessments, we investigate various magnitudes of these effects, also beyond the values we expect for our particular hardware implementation.

\paragraph{Neuron and synapse models}
While some network architectures employ relatively simple neuron and synapse models for analytical and/or numerical tractability, others rely on more complex components in order to remain more faithful to their biological archetypes.
Such models may not allow a straightforward translation to those available on the hardware, requiring a certain amount of fitting.
In our particular case, we search for parameters to \Cref{eqn:adex1,eqn:adex2,eqn:adex3,eqn:cond_exp_syn1,eqn:cond_exp_syn2,eqn:TSO,eqn:hw_stp1,eqn:hw_stp2} that best reproduce reproduce low-level dynamics (e.g. membrane potential traces for simple stimulus patterns) and then tweak these as to optimally reproduce high-level network behaviors.
Additionally, further constraints are imposed by the parameter ranges permitted by the hardware (\Cref{table:hardware_parameter_ranges}).

\paragraph{Synapse loss}
\label{dist:synapseloss}
Above a certain network size or density, the mapping process may not be able to find enough hardware resources to realize every single synapse.
We use the term ``synapse loss'' to describe this process, which causes a certain portion of synaptic connections to be lost after mapping.
In a first stage, we model synapse loss as homogeneous, i.e., each synapse is deleted with a fixed probability between 0 and \SI{50}{\%}.
To ease the analysis of distortions, we make an exception for synapses that mediate external input, since, in principle, they can be prioritized in the mapping process such that the probability of losing them practically vanishes.
Ultimately however, the compensation methods designed for homogeneous synapse loss are validated against a concrete mapping scenario.

\paragraph{Non-configurable axonal delays}
Axonal delays on the wafer are not configurable and depend predominantly on the processing speed of digital spikes within one HICANN, but also on the physical distance of the neurons on the wafer.
In all simulations, we assume a constant delay of \SI{1.5}{\milli\second} for all synaptic connections in the network, which represents an average of the expected delays when running the hardware with a speedup of $10^4$ with respect to real time.

\paragraph{Synaptic weight noise}
\label{sec:dist-weights}
As described in \cref{sec:hicann}, the variation of synaptic weights is assumed to be the most significant source of parameter variation within the network.
This is due to the coarser discretization (4-bit weight vs. 10 bit used for writing the analog neuron parameters) as well as the large number of available synapses, which prohibits the storage of calibration data for each individual synapse.
The quality of the calibration only depends on the available time and number of parameter settings, while the trial-to-trial variability and the limited setting resolution remains.
To restrict the parameter space of the following investigations (\cref{sec:models}), only the synaptic weights are assumed to be affected by noise.
In both software and ESS simulations, we model this effect by drawing synaptic weights from a Gaussian centered on the target synaptic weight with a standard-deviation-to-mean-ratio between 0 and \SI{50}{\%}.
Occasionally, this leads to excitatory synapses becoming inhibitory and vice versa, which can not happen on the hardware.
Such weights are clipped to zero.
Note that this effectively leads to an increase of the mean of the distribution, which however can be neglected, e.g., for \SI{50}{\%} noise the mean is increased by \SI{0.425}{\%}.
For ESS simulations we assume a synaptic weight noise of \SI{20}{\%}, as test measurements on the hardware indicate that the noise level can not be reduced to below this number.

It has to be noted that the mechanism of distortion plays a role in the applicability of the compensation mechanisms.
The iterative compensation in \cref{sec:iterative_compensation} is only applicable when the dominant distortion mechanism is fixed-pattern noise.
The other compensation methods, which do not rely on any kind of knowledge of the fixed-pattern distribution, function independently of the distortion mode.

\section{Hardware-induced distortions and compensation strategies}
\label{sec:models}

\begin{figure}
    \centering	
    \includegraphics{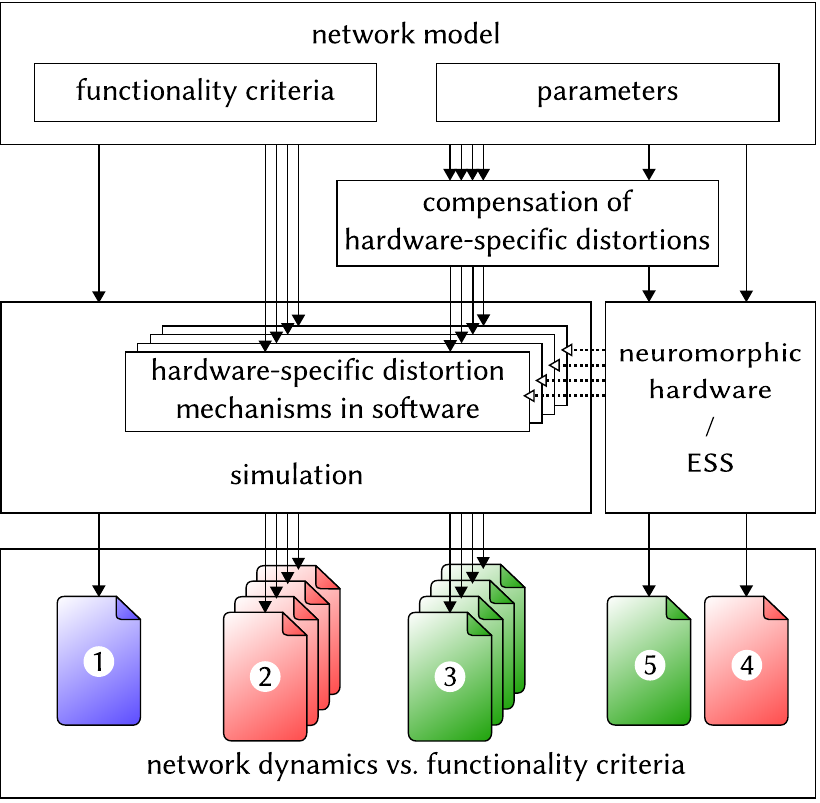}
    \caption{
	{\bf Schematic of the workflow used for studying and compensating hardware-induced distortions of network dynamics.}
    (1) A given network model is defined by providing suitable parameters (for its connectivity and components) and well-defined functionality criteria.
    (2) The distortion mechanisms that are expected to occur natively on the hardware back-end are implemented and studied individually in software simulations.
    (3) Compensation methods are designed and tested, with the aim of recovering the original network dynamics as determined by the functionality criteria.
    (4) The network model is run on the hardware (here: the ESS) without any compensation to evaluate the full effect of the combined distortion mechanisms.
    (5) The compensation methods are combined and applied to the hardware (here: the ESS) simulation in order to restore the original network dynamics.
    \label{fig:workflow_article}
	}
\end{figure}

In the following, we analyze the effects of hardware-specific distortion mechanisms on a set of neuronal network models and propose adequate compensation mechanisms for restoring the original network dynamics.
The aim of these studies is twofold.
On one hand, we propose a generic workflow which can be applied for different neural network models regardless of the neuromorphic substrate, assuming it posesses a certain degree of configurability (\Cref{fig:workflow_article}).
On the other hand, we seek to characterize the universality of the BrainScaleS neuromorphic device by assessing its capability of emulating very different large-scale network models with minimal, if any, impairment to their functionality.

In order to allow a comprehensive overview, the set of benchmark experiments is required to cover a broad range of possible network architectures, parameters and function modi.
To this end, we have chosen three very different network models, each of which highlights crucial aspects of the biology-to-hardware mapping procedure and poses unique challenges for the hardware implementation.
In order to facilitate the comparison between simulations of the original model and their hardware implementation, all experimental setups were implemented in PyNN, running the same set of instructions on either simulation back-end.

For each of our benchmark models we define a set of specific well-quantifiable functionality criteria.
These criteria are measured in software simulations of the ideal, i.e., undistorted network, which is then further referenced as the ``original''.

Assuming that the broad range of hardware-specific distortion mechanisms affects various network parameters, their impact on these measures are investigated in software simulations and various changes to the model structure are proposed in order to recover the original functionality.
The feasibility of these compensation methods is then studied for the BrainScaleS neuromorphic platform with the help of the ESS described in \Cref{sec:ess}.

All software simulations were performed with NEST \cite{diesmann01nest} or Neuron \cite{hines03neuron}.

\subsection{Cortical layer 2/3 attractor memory}
\label{model:l23}

As our first benchmark, we have chosen an attractor network model of the cerebral cortex which exihbits characterisic and well-quantifiable dynamics, both at the single-cell level (membrane voltage UP and DOWN states) and for entire populations (gamma band oscillations, pattern completion, attentional blink).
For this model, the mapping to the hardware was particularly challenging, due to the complex neuron and synapse models required by the original architecture on the one hand, as well as its dense connectivity on the other.
In particular, we observed that the shape of synaptic conductances strongly affects the duration of the attractor states. 
As expected for a model with relatively large populations as functional units, it exhibits a pronounced robustness to synaptic weight noise.
Homogeneous synapse loss, on the other hand, has a direct impact on single-cell dynamics, resulting in significant deviations from the expected high-level functionality, such as the attenuation of attentional blink.
As a compensation for synapse loss, we suggest two methods: increasing the weights of the remaining synapses in order to maintain the total average synaptic conductance and reducing the size of certain populations and thereby decreasing the total number of required synapses.
After mapping to the hardware substrate, synapse loss is not homogeneous, due to the different connectivity patterns of the three neuron types required by the model.
However, we were able to apply a population-wise version of the suggested compensation methods and demonstrate their effectiveness in recovering the previously defined target functionality measures.

\subsubsection{Architecture}
\label{sec:l23architecture}

As described in \cite{lundqvist2006attractor} and \cite{lundqvist2010bistable}, this model (henceforth called L2/3 model) implements a columnar architecture \cite{mountcastle1997columnar, buxhoeveden2002minicolumn}.
The connectivity is compliant with data from cat cerebral cortex connectivity \cite{thomson02synaptic}.
The key aspect of the model is its modularity, which manifests itself on two levels.
On a large scale, the simulated cortical patch is represented by a number $N_{\mbox{\tiny{HC}}}$ of hypercolumns (HCs)
arranged on a hexagonal grid.
On a smaller scale, each HC is further subdivided into a number $N_{\mbox{\tiny{MC}}}$ of minicolumns (MCs)
\cite{mountcastle1997columnar, buxhoeveden2002minicolumn}.
Such MCs should first and foremost be seen as functional units, and could, in biology, also be a group of distributed, but highly interconnected cells \cite{song05nonrandom, kampa04cortical, perin2011synaptic}.
In the model, each MC consists, in turn, of 30 pyramidal (PYR), 2 regular spiking non-pyramidal (RSNP) and 1 basket (BAS) cells \cite{peters1997organization, markram04interneurons}.
Within each MC, PYR neurons are mutually interconnected, with 25\% connectivity, such that they will tend to be co-active and code for similar input.

\begin{figure}
    \centering
    \includegraphics[width=\columnwidth]{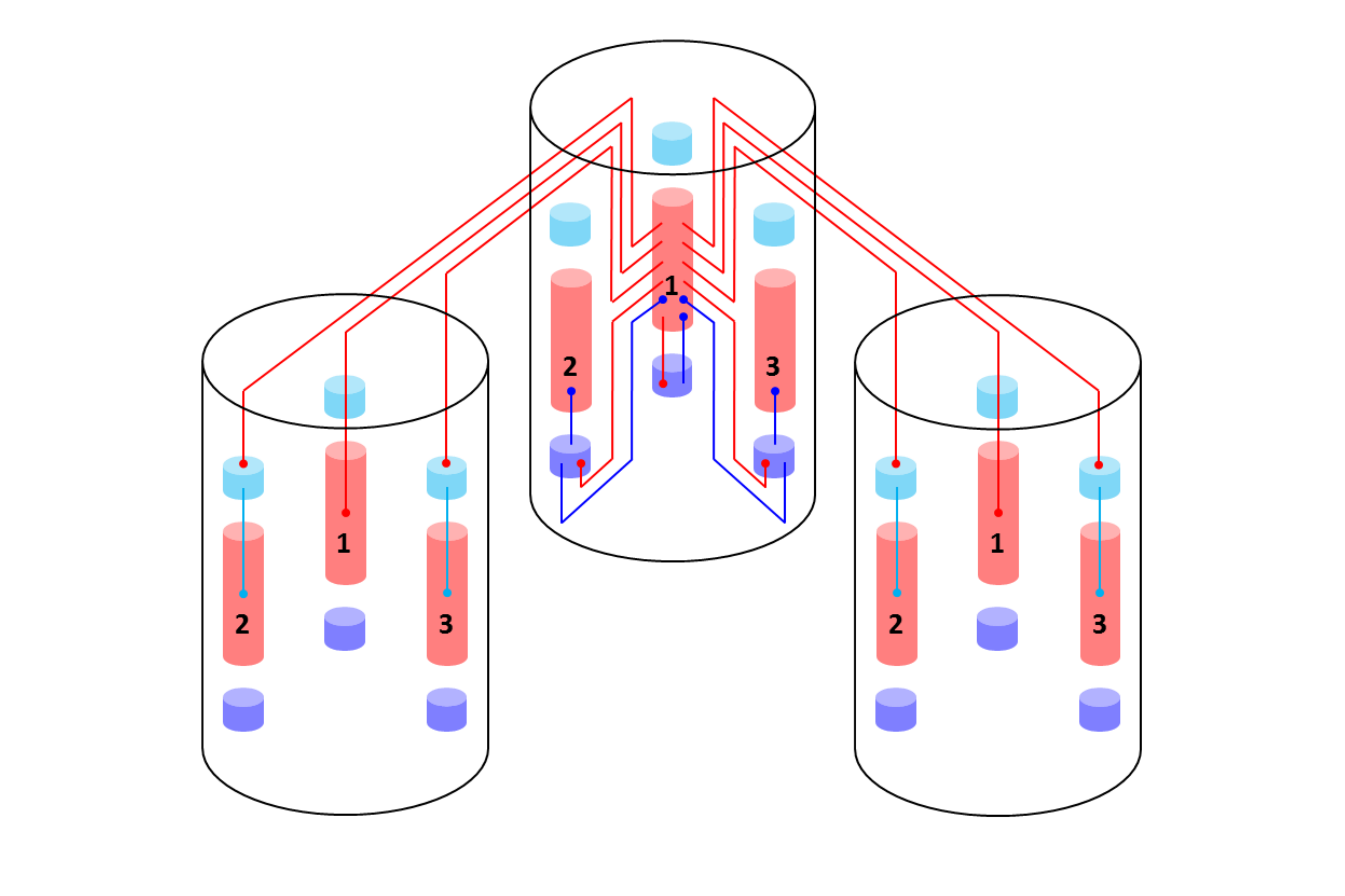}
    \caption{
		{\bf Pseudo-3D schematic of the layer 2/3 model architecture.}
        Excitatory (PYR) cell populations are represented as red cylinders, inhibitory populations as blue ones (BAS: dark, RSNP: light).
        A minicolumn (MC) consists of three vertically aligned populations: one PYR, one BAS and one RSNP.
        Multiple MCs are grouped into hypercolumns (HCs, transparent cylinders).
        MCs with the same ID (one per HC) form a so-called pattern or attractor.
        When active, all PYR cells belonging to an attractor excite each other via short-range (within an MC) and long-range (between MCs) connections.
        The inhibition of PYR cells belonging to other attractors occurs via inhibitory interneurons: locally (within an HC) through BAS cells and globally (between HCs) through RSNP cells.
        Only a subset of connections are shown, namely those which are mainly used during active periods of attractor 1.
    \label{fig:l23schematics}
    }
\end{figure}

The functional units of the network, the MCs, are connected in global, distributed patterns containing a set of MCs in the network (\Cref{fig:l23schematics}).
Here the attractors, or patterns, 
contain exactly one MC from each HC.
We have only considered the case of orthogonal patterns, which implies that no two attractors share any number of MCs.
Due to the mutual excitation within an attractor, the network is able to perform pattern completion, which means that whenever a subset of MCs in an attractor is activated, the activity tends to spread throughout the entire attractor.

Pattern rivalry results from competition between attractors mediated by short and long-range connections via inhibitory interneurons.
Each HC can be viewed as a soft winner-take-all (WTA) module which normalizes activity among its constituent MCs \cite{lundqvist2010bistable}.
This is achieved by the inhibitory BAS cells, which receive input from the PYR cells from the 8 closest MCs and project back onto the PYR cells in all the MCs within the home HC.
Apart from providing long-range connections to PYR cells within the same pattern, the PYR cells within an MC project onto RSNP cells in all the MCs which do not belong to the same pattern and do not lie within the same HC.
The inhibitory RSNP cells, in turn, project onto the PYR cells in their respective MC.
The effect of this connectivity is a disynaptic inhibition between competing patterns.
\Cref{fig:l23schematics} shows a schematic of the default architecture, emphasizing the connectivity pattern described above.
It consists of $N_{\mbox{\tiny{HC}}}=9$ HCs, each containing $N_{\mbox{\tiny{MC}}}=9$ MCs, yielding a total of 2673 neurons.
Due to its modular structure, this default model can easily be scaled up or down in size with preserved dynamics, as described in the Supplement (\Cref{sec:methods-l23-scaling}).

When a pattern receives enough excitation, its PYR cells enter a state reminiscent of a so-called local UP-state \cite{cossart03attractor}, which is characterized by a high average membrane potential, several mV above its rest value, and elevated firing rates.
Pattern rivalry leads to states where only one attractor may be active (with all its PYR cells in an UP-state) at any given time.
Inter-PYR synapses feature an STP mechanism which weakens the mutual activation of PYR cells over time and prevents a single attractor from becoming persistently active.
Additionally, PYR neurons exhibit spike-frequency adaptation, which also suppresses prolonged firing.
These mechanisms impose a finite life-time on the attractors such that after their termination more weakly stimulated or less excitable attractors can become active, in contrast to what happens in classical WTA networks.

The inputs to the layer 2/3 PYR cells arrive from the cortical layer 4, which is represented by 5 cells per MC.
The layer 4 cells project onto the layer 2/3 PYR cells and can be selectively activated by external Poisson spike trains.
Additionally, the network receives unspecific input representing activity in various connected cortical areas outside the modeled patch.
This input is modeled as diffuse noise and generates a background activity of several Hz.

More details on the model architecture, as well as neuron and synapse parameters, can be found in the Supplement (\Cref{sec:l23_original_params_appendix}).

\subsubsection{Functionality criteria}
\label{sec:l23criteria}

\begin{figure*}
    \centering
    \includegraphics[width=\textwidth]{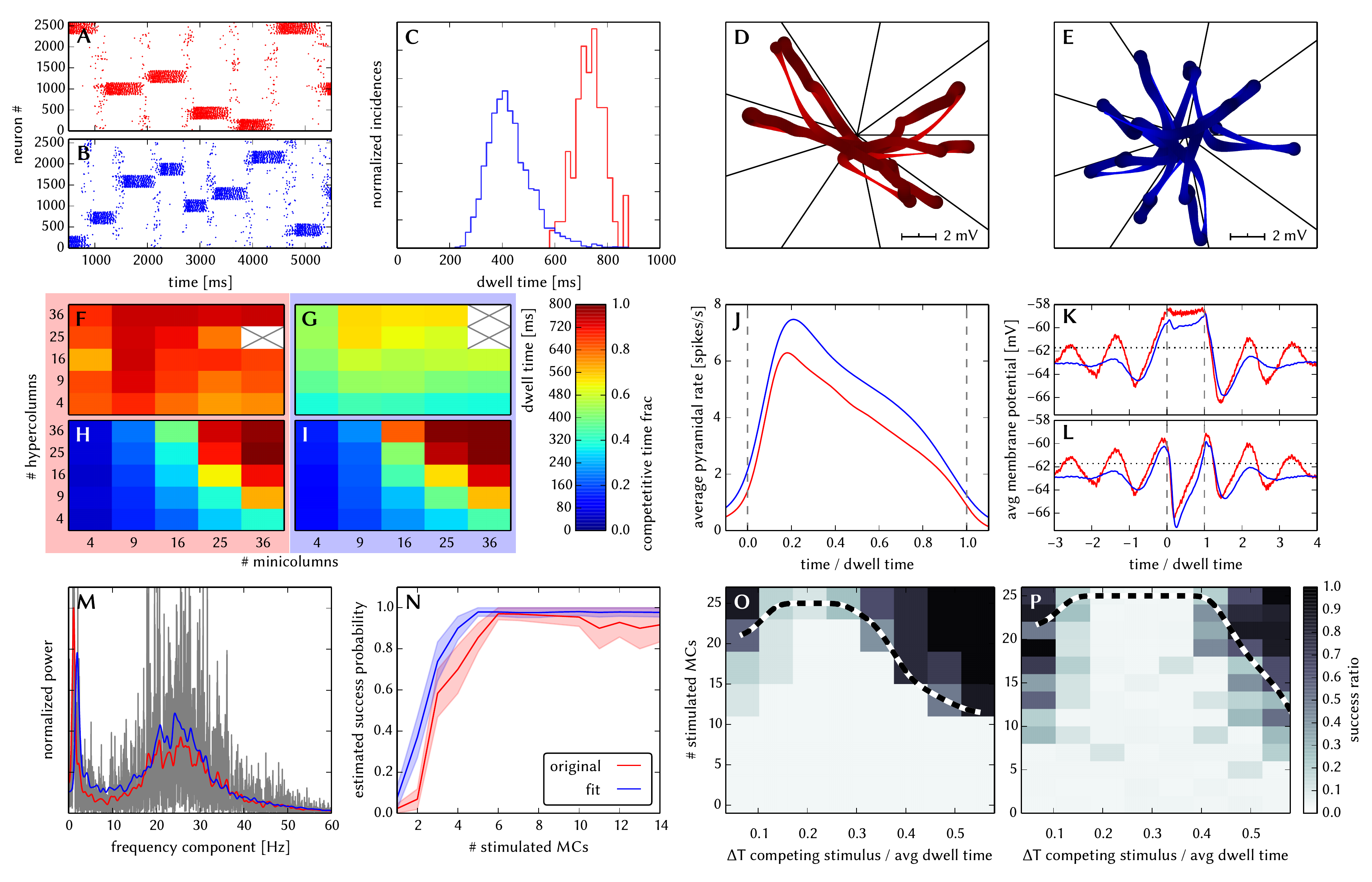}
    \caption{
	{\bf Comparison between original and adapted L2/3 network models.}
    Unless explicitly stated otherwise, the default network model (9HC$\times$9MC) was used.
    Measurements from the original model are depicted (or highlighted) in red, while those from the adapted model are depicted (or highlighted) in blue.
    (\textbf{A}, \textbf{B}): Raster plots of spiking activity.
    Attractors activate spontaneously only due to diffuse background noise.
    Only PYR cells are shown.
    The MCs are ordered such that those belonging to the same attractor (and \textit{not} those within the same HC) are grouped together.
    \textbf{C}: Attractor dwell time distributions.
    The shorter average dwell times in the adapted model are caused by sharper PSPs which miss the long NMDA time constants.
    (\textbf{D}, \textbf{E}): Star plots of average PYR cell voltages from a sample of 5 PYR cells per MC. Details on this representation of multidimensional data can be found in \Cref{sec:l23starplots}.
    (\textbf{F}, \textbf{G}): Average dwell time for various network sizes.
    (\textbf{H}, \textbf{I}): Fraction of time spent in competitive states (i.e. no active attractors) for various network sizes.
    While dwell times remain relatively constant, competition times increase with network size, suppressing spontaneous attractors in large networks.
    (\textbf{J}): Average firing rate of PYR within an active period of their parent attractor.
    (\textbf{K}): Average voltage of PYR cells before, during and after their parent attractor is active (UP state).
    (\textbf{L}): Average voltage of PYR cells before, during and after an attractor they do not belong to is active (DOWN state).
    For subplots \tb{J}, \tb{K} and \tb{L}, the abscissa has been subdivided into multiples of the average attractor dwell time in the respective simulations.
    The oscillations of the average voltages occur due to spike-triggered adaptation: after an active period, PYR cells need to recover before being able to support an active period of their home attractor, during which time they are inhibited by other active attractors.
    The more pronounced attenuation of the oscillations in the adapted model happens due to a higher relative variability of dwell times (compare subplot \tb{C}). In subplots \tb{K} and \tb{L} the dotted line indicates the leak potential $\Vrest$ of the PYR cells.
    (\textbf{M}): Smoothed power spectrum of PYR firing rate averaged over all MCs.
    The grey curve in the background represents the unsmoothed spectral density for the original model.
    Attractor switches ($\approx$ 2 Hz) and gamma oscillations ($\approx$ 25 Hz) can be clearly observed.
    (\textbf{N}): Pattern completion in a 25HC$\times$25MC network.
    Estimated probability of an attractor to fully activate (success ratio) as a function of the number of stimulated constituent MCs, measured over 25 trials per abscissa value.
    (\textbf{O}, \textbf{P}): Attentional blink in a 25HC$\times$25MC network.
    Two attractors are stimulated (the second one only partially, i.e. a certain number of constituent MCs) with a temporal lag of $\Delta T$ in between.
    Activation probability of the second attractor and $p=0.5$ iso-probability contours, measured over 14 trials per ($\Delta T$, \#MCs) pair.
    A detailed description of the data and methods used for all figures concerning the L2/3 model can be found in \Cref{sec:l23_original_params_appendix} to \ref{sec:l23-pattern_rivalry}.
    \label{fig:comparison}
    }
\end{figure*}

\Cref{fig:comparison} shows some characteristic dynamics of the L2/3 model, which have also been chosen as functionality criteria and are described below.

The core functionality of the original model is easily identifiable by its distinctive display of spontaneously activating attractors in e.g. raster plots (\tb{A}) or voltage star plots (\tb{D}, for an explanation of star plots see \Cref{sec:l23starplots}).
However, in particular for large network sizes, spontaneous attractors become increasingly sparse.
Additionally, many further indicators of functionality can be found, such as the average membrane potential or the gamma oscillations observed in UP states.
Finally, when receiving L4 stimulation in addition to the background noise, the original model displays important features such as pattern completion and attentional blink, which need to be reproducible on the hardware as well.
Consequently, we consider several measures of functionality throughout our analyses.

When an attractor becomes active, it remains that way for a characteristic dwell time $\tauon$.
The dwell time depends strongly on the neuron and synapse parameters (as will be discussed in the following sections) and only weakly on the network size (\tb{C}, \tb{F}), since the scaling rules ensure a constant average fan-in for each neuron type.
Conversely, this makes $\tauon$ sensitive to hardware-induced variations in the average synaptic input.
The detection of active attractors is performed automatically using the spike data (for a description of the algorithm, see \Cref{sec:l23-upstate_detection}).

We describe the periods between active attractors as competition phases and the time spent therein as the total competition time.
The competition time varies strongly depending on the network size (\tb{H}).
One can observe that the competition time is a monotonically increasing function of both $\Nhc$ and $\Nmc$.
For an increasing number of HCs, i.e., a larger number of neurons in every pattern, the probability of a spontaneous activation of a sufficiently large number of PYR cells decreases.
For an increasing number of MCs per HC, there is a larger number of competing patterns, leading to a reduced probability of any single pattern becoming dominant.

When an attractor becomes active, the average spike rate of its constituent PYR cells rises sharply and then decays slowly until the attractor becomes inactive again (\tb{J}).
Two independent mechanisms are the cause of this decay: neuron adaptation and synaptic depression.
The characteristic time course of the spike rate depends only weakly on the size of the network.

As described in \Cref{sec:l23architecture}, PYR cells within active attractors enter a so-called local UP state, with an increased average membrane potential and an elevated firing rate (\tb{K}).
While inactive or inhibited by other active attractors, PYR cells are in a DOWN state, with low average membrane potential and almost no spiking at all (\tb{L}).
In addition to these characteristic states, the average PYR membrane potential exhibits oscillations with a period close to $\tauon$.
These occur because the activation probability of individual attractors is an oscillatory function of time as well.
In the immediate temporal vicinity of an active period (i.e., assuming an activation at $t=0$, during $[-\tauon, 0) \cup [\tauon, 2\tauon)$) the same attractor must have been inactive, since PYR populations belonging to an activated attractor need time to recuperate from synaptic depression and spike-triggered adaptation before being able to activate again.

An essential emerging feature of this model are oscillations of the instantaneous PYR spike rate in the gamma band within active attractors (\tb{M}).
The frequency of these oscillations are independent of size and rather depend on excitation levels in the network \cite{lundqvist2010bistable}.
Although the gamma oscillations might suggest periodic spiking, it is important to note that individual PYR cells spike irregularly ($\left\langle\cvisi\right\rangle=1.36 \pm0.36$ within active attractors).

Apart from these statistical measures, two behavioral properties are essential for defining the functionality of the network: the pattern completion and attentional blink mentioned above.
The pattern completion ability of the network can be described as the successful activation probability of individual patterns as a function of the number of stimulated MCs (\tb{N}).
Similarly, the attentional blink phenomenon can also be quantified by the successful activation rate of an attractor as a function of the number of stimulated MCs if it is preceded by the activation of some other attractor with a time lag of
$\deltaT$ (\tb{O}).
For small $\deltaT$, the second attractor is completely ``blinked out'', i.e., it can not be activated regardless of the number of stimulated MCs.
To facilitate the comparison between different realizations of the network with respect to attentional blink, we consider the 50\% iso-line, which represents the locus of the input variable pair which leads to an attractor activation ratio of 50\%.
These functional properties are easiest to observe in large networks, where spontaneous attractors are rare and do not interfere with stimulated ones.

A detailed description of the data and methods used for these figures can be found in the Supplement (\Cref{sec:l23_original_params_appendix} to \Cref{sec:l23-pattern_rivalry}).

\subsubsection{Neuron and synapse model translation}
\label{sec:l23_fit}

\begin{figure}
    \centering
    \includegraphics[width=\columnwidth]{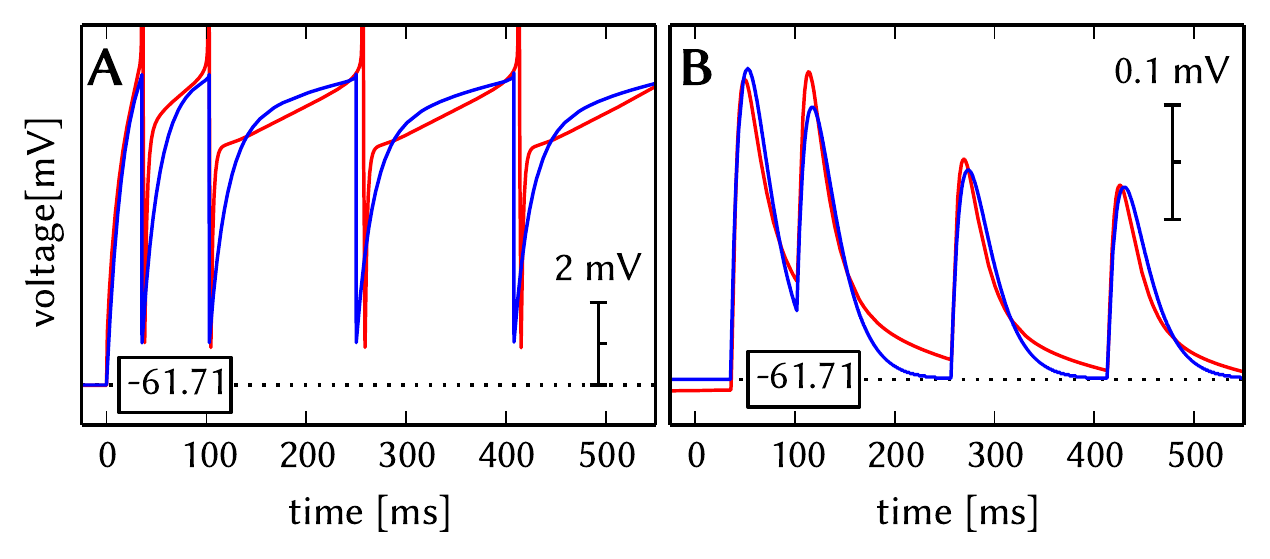}%
    \caption{
		{\bf Comparison of original and fitted neuron and synapse dynamics:}
        Original neuron (multi-compartment HH) and synapse (NMDA+AMPA) dynamics are shown in red, the fitted dynamics of hardware-compatible neuron (point AdEx) and synapse (single decay time constant) models in blue.
        (\textbf{A}) Membrane potential of PYR cells under spike-inducing current stimulation.
        While the precise membrane potential time course of the original neuron model can not be reproduced by a single-compartment AdEx neuron, spike timing and especially firing rates can be recovered.
        (\textbf{B}) PSPs generated by PYR$\rightarrow$PYR synapses between MCs where the spikes from \textbf{A} were used as input.
        As a replacement for the multiple synaptic time constants in the original model, we have chosen an intermediate value for $\tausyn$, which constitutes the main reason for the difference in PSP shapes.
        Additionally, the combination of STP and saturation in the original model had to be replaced by STP alone.
  		\label{fig:l23fit}
        }
\end{figure}

A particular feature of this benchmark model is the complexity of both neuron and synapse models used in its original version.
Therefore, the first required type of compensation concerns the parameter fitting for the models implemented on the hardware.
Some exemplary results of this parameter fit can be seen in \Cref{fig:l23fit}.
More details can be found in the Supplement (\Cref{sec:l23_fit_appendix}).

\paragraph{Neurons}

In general, the typical membrane potential time course during a spike of a Hodgkin-Huxley neuron can be well approximated by the exponential term in the AdEx equation \cite{brette_05}.
However, when fitting for spike timing, we found that spike times were best reproduced when eliminating the exponential term, i.e. setting $\adexdeltaT=0$.

Adaptation is an essential feature of both the PYR and the RSNP cells in the original model, where it is generated by voltage-dependent $\KCa$ channels.
We were able to reproduce the correct equilibrium spike frequency by setting the AdEx adaptation parameters $a$ and $b$ to nonzero values.
One further difference resides in the original neurons being modeled as having several compartments, whereas the hardware only implements point neurons.
The passive neuron properties (membrane capacitances and leak conductances) were therefore determined by fitting the membrane potential time course under stimulation by a step current which was not strong enough to elicit spikes.

\paragraph{Synapses}

We have performed an initial estimation of synaptic weights and time constants by fitting the membrane potential time course of the corresponding neurons in a subthreshold regime.
However, two important differences remain between the synapses in the original model and the ones available on our hardware.

In the original model, PYR-PYR and PYR-RSNP synapses contain two types of neurotransmitters: Kai\-nate/AMPA and NMDA (see \Cref{tab:l23synapseparams}).
Due to the vastly different time constants for neurotransmitter removal at the postsynaptic site (6 ms and 150 ms, respectively), the PSPs have a characteristic shape, with a pronounced peak and a long tail (red curve in \cref{fig:l23fit} \textbf{B}).
While, in priniciple, the HICANN supports several excitatory time constants per neuron (\Cref{sec:hicann}), the PyNN API as well as the mapping process support only one excitatory time constant per neuron.
With this limitation the PSP shape can not be precisely reproduced.

One further difference lies in the saturating nature of the postsynaptic receptor pools after a single presynaptic spike.
In principle, this behavior could be emulated by the TSO plasticity mechanism by setting $U=1$ and $\taurec=\tausyn$.
However, this would conflict with the TSO parameters required for modeling short-term depression of PYR synapses and would also require parameters outside the available hardware ranges.

For these reasons, we have further modified synaptic weights and time constants by performing a behavioral fit, i.e., by optimizing these parameters towards reproducing the correct firing rates of the three neuron types in two scenarios - first without and then subsequently with inhibitory synapses.
Because the original model was characterized by relatively long and stable attractors, we further optimized the excitatory synapse time constants towards this behavior.

\paragraph{Post-fit model behavior}

\Cref{fig:comparison} shows the results of the translation of the original model to hardware-com\-patible dynamics and parameter ranges.
Overall, one can observe a very good qualitative agreeement of characteristic dynamics with the original model.
In the following, we discuss this in more detail and explain the sources of quantitative deviations.

When subject to diffuse background noise only, the default size network clearly exhibits its characteristic spontanous attractors (\tb{B}).
Star plots exhibit the same overall traits, with well-defined attractors, characterized by state space trajectories situated close to the axes and low trajectory velocities within attractors (\tb{E}).
Attractor dwell times remain relatively stable for different network sizes, while the competition times increase along with the network size (\tb{G} and \tb{I}).
The average value of dwell times, however, lies significantly lower than in the original (\tb{C}).
The reason for this lies mainly in the shape of EPSPs: the long EPSP tails enabled by the large NMDA time constants in the original model caused a higher average membrane potential, thereby prolonging the activity of PYR cells. 

Within attractors, active and inactive PYR cells enter well-defined local UP and DOWN states, respectively (\tb{K} and \tb{L}).
Before and after active attractors, the dampened oscillations described in \Cref{sec:l23criteria} can be observed.
In the adapted model, attenuation is stronger due to a higher coefficient of variation of the dwell times ($\frac{\sigma}{\mu}=0.20$ as compared to $0.08$ in the original model).

Average PYR firing rates within active attractors have very similar time courses (\tb{J}), with a small difference in amplitude, which can be attributed to the difference in EPSP shapes discussed earlier.
Both low-fre\-quency switches between attractors (< 3 Hz, equivalent to the incidence rate) and high-frequency gamma oscillations arising from synchronous PYR firing (with a peak around 25 Hz) can be clearly seen in a power spectrum of the PYR firing rate (\tb{M}).

Pattern completion occurs similarly early, with a steep rise and nearly 100\% success rate starting at 25\% of stimulated MCs per attractor (\tb{N}).
Attentional blink follows the same qualitative pattern (\tb{P}, \tb{Q}), although with a slightly more pronounced dominance of the first activated attractor in the case of the adapted network, which happens due to the slightly higher firing rates discussed above.

Having established the quality of the model fit and in order to facilitate a meaningful comparison, all following studies concerning hardware-induced distortions and compensation thereof use data from the adapted model as reference.

\subsubsection{Synapse loss}
\label{sec:l23synloss}

The effects of homogeneous synapse loss and the results of the attempted compensation are depicted in \Cref{fig:synlosscompensation}.
More detailed plots can be found in the Supplement (\Cref{fig:synloss}).

\paragraph{Effects}
\label{sec:l23-loss_effects}

With increasing synapse loss, the functionality of the network gradually deteriorates.
Attractors become shorter or disappear entirely, with longer periods of competition in between (\textbf{D}, \tb{K}, \tb{O}).

While average excitatory conductances are only affected linearly by synaptic loss, inhibitory conductances feel a compound effect of synapse loss, as it affects both afferent and efferent connections of inhibitory interneurons.
Therefore, synapse loss has a stronger effect on inhibition, leading to a net increase in the average PYR membrane potential (\tb{R}, \tb{S}).
Additionally, since all connections become weaker, the variance of the membrane potential becomes smaller, as observed in the corresponding star plots as well (\tb{E}).
The weaker connections also decrease the self-excitation of active attractors while decreasing the inhibition of inactive ones, thereby leading to shorter attractor dwell times (\tb{P}).
Somewhat surprisingly, the maximum average PYR firing rate in active attractors remains almost unchanged when subjected to synapse loss.
However, the temporal evolution of the PYR firing rate changes significantly (\tb{Q}).

The pattern completion ability of the network suffers particularly in the region of weak stimuli, due to weaker internal excitation of individual attractors.
The probability of triggering a partially stimulated pattern can drop by more than 50\% (\tb{T}).
Due to the decreased stability of individual attractors discussed above, rival attractors are easier to excite, thereby significantly suppressing the attentional blink phenomenon (\tb{U}).

\paragraph{Compensation}
\label{sec:l23-loss_comp}

As a first-order approximation, we can consider the population average of the neuron conductance as the determining factor in the model dynamics.
For synapses with exponential conductance courses, the average conductance generated by the $i$th synapse is proportional to both synaptic weight $w_{ij}$ and afferent firing rate $\nu_j$.
Because conductances sum up linearly, the total conductance that a neuron from population $i$ receives from some other population $j$ is, on average (see \Cref{eqn:cond_mean_multi})
\begin{equation}
    \mean{g} =  N_j p_{ij} \mean{w_{ij}} \mean{\nu_j} \tausyn \quad ,
\end{equation}
where $N_j$ represents the size of the presynaptic population and $p_{ij}$ represents the probability of a neuron from the presynaptic population to project onto a neuron from the postsynaptic population.
Since homogeneous synapse loss is equivalent to a decrease in $p_{ij}$, we can compensate for synapse loss that occurs with probability $p_\mathrm{loss}$ by increasing the weights of the remaining synapses by a factor $1/(1-p_\mathrm{loss})$.
\Cref{fig:synlosscompensation} shows the results of this compensation strategy for $p_\mathrm{loss}=0.5$.
In all aspects, a clear improvement can be observed.
The remaining deviations can be mainly attributed to two effects.
First of all, preserving the average conductance by compensating homogeneous synapse loss with increasing synaptic weights leads to an increase in the variance of the membrane potential (\Cref{eqn:cond_var_single}).
Secondly, finite population sizes coupled with random elimination of synapses lead to locally inhomogeneous synapse loss and further increase the variability of neuronal activity.

\begin{figure*}
    \centering
    \includegraphics[width=\textwidth]{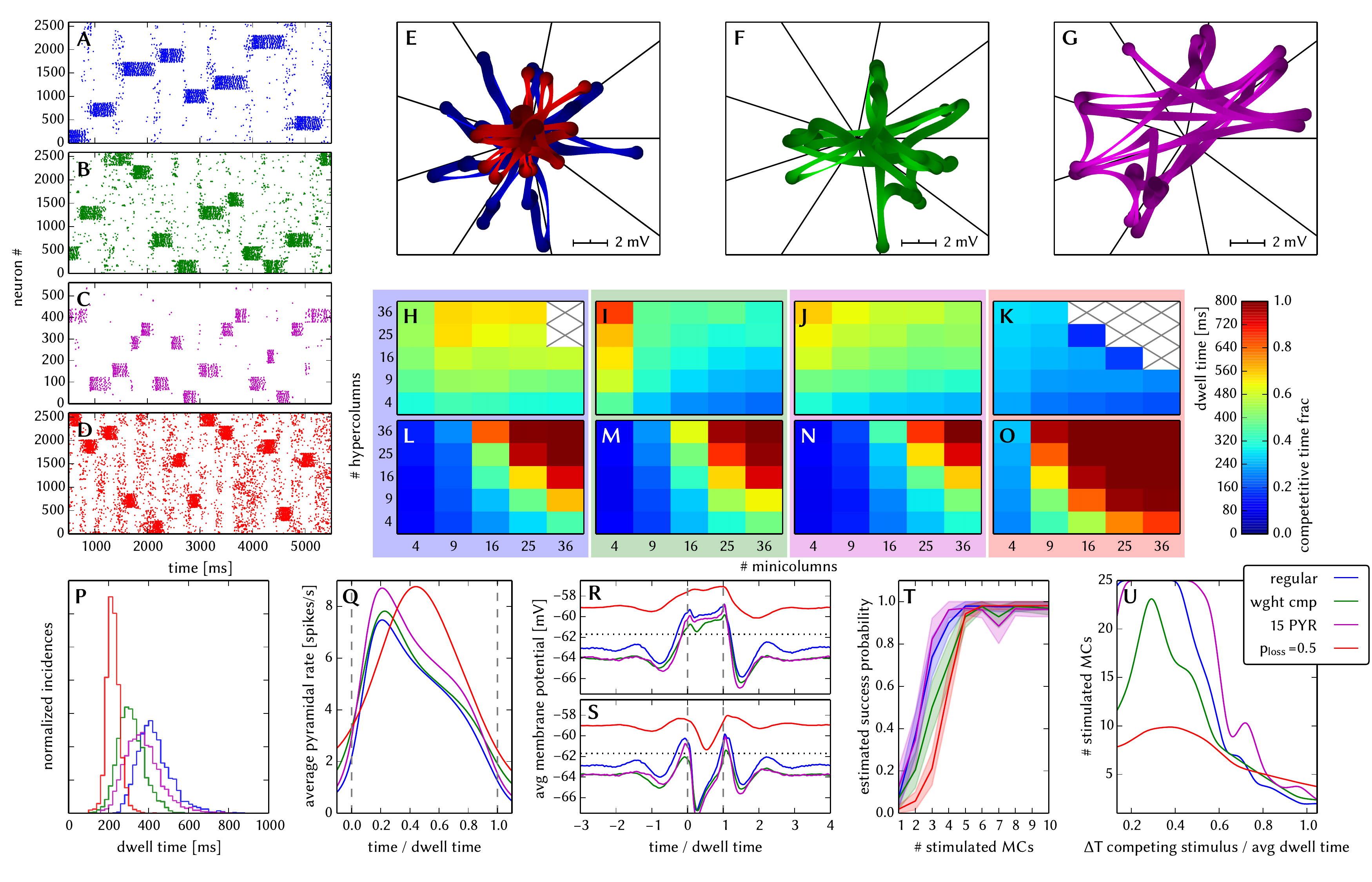}
    \caption{
	{\bf Compensation of homogeneous synaptic loss in the L2/3 model.}
    Unless explicitly stated otherwise, the default network model (9HC$\times$9MC) was used.
    Here, we use the following color code: blue for the original model, red for the distorted case (50\% synapse loss), green for the compensation via increased synaptic weights and purple for the compensation by scaling down the size of the PYR populations.
    (\textbf{A} - \textbf{D}) Raster plots of spiking activity.
                 The MCs are ordered such that those belonging to the same attractor (and \textit{not} those within the same HC) are grouped together.
                 Synapse loss weakens the interactions within and among MCs, causing shorter dwell times and longer competition times.
                 Both compensation methods successfuly counter these effects.
                 These phenomena can also be observed in subplots \tb{H}-\tb{P}.
    (\textbf{E} - \textbf{G}) Star plots of average PYR voltages from a sample of 5 PYR cells per MC.
                Synapse loss leads to a less pronounced difference between the average PYR membrane potential within and outside of active attractors.
                After compensation, the differences between UP and DOWN states become more pronounced again.
                These phenomena can also be observed in subplots \tb{R} and \tb{S}.
    (\textbf{H} - \textbf{K}) Average dwell time for various network sizes.
    (\textbf{L} - \textbf{O}) Fraction of time spent in competitive states (i.e. no active attractors) for various network sizes.
    (\textbf{P}) Distributions of dwell times.
    (\textbf{Q}) Average firing rate of PYR cells within an active period of their parent attractor.
    (\textbf{R}) Average voltage of PYR cells before, during and after their parent attractor is active (UP state).
    (\textbf{S}) Average voltage of PYR cells before, during and after an attractor they do not belong to is active (DOWN state).
                 For subplots \tb{Q}, \tb{R} and \tb{S}, the abscissa has been subdivided into multiples of the average attractor dwell time in the respective simulations.
                 In subplots \tb{R} and \tb{S} the dotted line indicates the leak potential $\Vrest$ of the PYR cells. 
    (\textbf{T}) Pattern completion in a 25HC$\times$25MC network.
                Estimated activation probability from 25 trials per abscissa value.
                Synapse loss shifts the curve to the right, i.e., more MCs need to be stimulated to achieve the same probability of activating their parent attractor.
                Both compensation methods restore the original behavior to a large extent.
    (\textbf{U}) Attentional blink in a 25HC$\times$25MC network: $p=0.5$ iso-probability contours, measured over 14 trials per ($\Delta T$, \#MCs) pair.
                Synapse loss suppresses attentional blink, as inhibition from active attractors becomes to weak to prevent the activation of other stimulated attractors.
                Compensation by increasing the weight of the remaining synapses alleviates this effect, but scaling down the PYR population sizes directly reduces the percentage of lost synapses and is therefore more effective in restoring attentional blink.
    \label{fig:synlosscompensation}
    }
\end{figure*}

Instead of compensating for synapse loss after its occurrence, it is also possible to circumvent it altogether after having estimated the expected synapse loss in a preliminary mapping run.
For the L2/3 model, this can be done without altering the number of functional units (i.e., the number of HCs and MCs) by changing the size of the PYR cell populations.
For this approach, however, the standard scaling rules (\Cref{sec:methods-l23-scaling}) need to be modified.
These rules are designed to keep the average number of inputs per neuron constant and would increase the total number of PYR-incident synapses by the same factor by which the PYR population is scaled.
This would inevitably lead to an increased number of shared inputs per PYR cell, with the immediate consequence of increased firing synchrony.
Instead, when reducing the PYR population size, we compensate for the reduced number of presynaptic partners by increasing relevant synaptic weights instead of connection probabilities.
This modified downscaling leads to a net reduction of the total number of synapses in the network, thereby potentially reducing synaptic loss between all populations.
\Cref{fig:synlosscompensation} shows the effects of scaling down the PYR population size until the total remaining number of synapses is equal to the realized number of synapses in the distorted case (50\% of the total number of synapses in the undistorted network).
More detailed plots of the effects of PYR population downscaling can be found in \Cref{fig:pyrscaling}.
The two presented compensation methods can also be combined to further improve the final result, as we show in \Cref{sec:l23alldistortions}.

\subsubsection{Synaptic weight noise}
\label{sec:l23noise}

\begin{figure*}
    \centering
    \includegraphics[width=\textwidth]{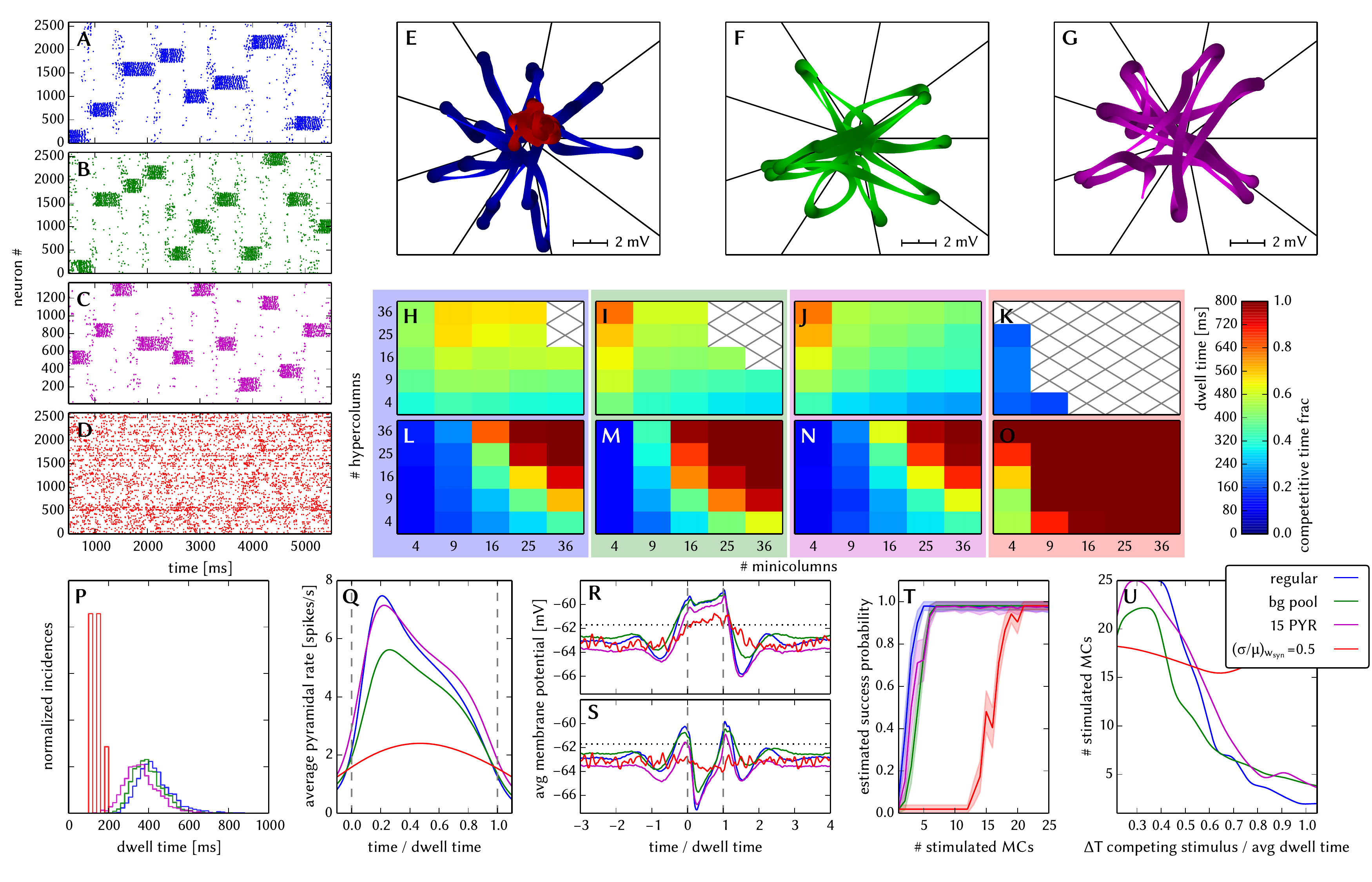}
    \caption{
	{\bf Compensation of synaptic weight noise in the L2/3 model.}
    Unless explicitly stated otherwise, the default network model (9HC$\times$9MC) was used.
    Here, we use the following color code: blue for the original model, red for the distorted case (50\% synaptic weight noise), green for the compensation via multiple background sources per PYR cell and purple for the same compensation method, but with scaled down PYR populations.
    Altogether, we note that the observed effects happen almost exclusively due to each PYR cell receiving background input via a single synapse.
    When compensated via the inclusion of multiple background sources, the network exhibits remarkable robustness towards synaptic weight noise.
    (\textbf{A} - \textbf{D}) Raster plots of spiking activity.
                 The MCs are ordered such that those belonging to the same attractor (and \textit{not} those within the same HC) are grouped together.
                 When each PYR cell has a single background source, high levels of synaptic weight noise cause some PYR cells to become completely silent, while others spike disproportionately often.
                 This can completely disrupt the stability of attractors, resulting in largely random spiking, with long competition times between the occasional appearance of weak, unstable attractors.
                 The inclusion of multiple background sources per PYR cell efficiently counters these effects.
                 This compensation strategy works just as well for downscaled PYR populations.
                 The phenomena described above can also be observed in subplots \tb{H}-\tb{P}.
    (\textbf{E} - \textbf{G}) Star plots of average PYR voltages from a sample of 5 PYR cells per MC.
                The disrupted attractor behavior and erratic PYR spiking result in weak fluctuations of average PYR voltages with essentially no clear UP or DOWN states.
                After compensation, the differences between UP and DOWN states become more pronounced again.
                These phenomena can also be observed in subplots \tb{R} and \tb{S}.
    (\textbf{H} - \textbf{K}) Average dwell time for various network sizes.
    (\textbf{L} - \textbf{O}) Fraction of time spent in competitive states (i.e. no active attractors) for various network sizes.
    (\textbf{P}) Distributions of dwell times.
    (\textbf{Q}) Average firing rate of PYR cells within an active period of their parent attractor.
    (\textbf{R}) Average voltage of PYR cells before, during and after their parent attractor is active (UP state).
    (\textbf{S}) Average voltage of PYR cells before, during and after an attractor they do not belong to is active (DOWN state).
                 For subplots \tb{Q}, \tb{R} and \tb{S}, the abscissa has been subdivided into multiples of the average attractor dwell time in the respective simulations.
                 In subplots \tb{R} and \tb{S} the dotted line indicates the leak potential $\Vrest$ of the PYR cells. 
    (\textbf{T}) Pattern completion in a 25HC$\times$25MC network.
                Estimated activation probability from 25 trials per abscissa value.
                Due to erratically firing PYR cells in the distorted network, much stronger stimulation is needed to guarantee the appearance of an attractor.
                Compensation restores the original behavior to a large extent.
    (\textbf{U}) attentional blink in a 25HC$\times$25MC network: $p=0.5$ iso-probability contours, measured over 14 trials per ($\Delta T$, \#MCs) pair.
                Due to the highly unstable attractors in the distorted network, attentional blink is completely suppressed.
                Compensation restores blink, but not to its original strength, due to the synaptic weight noise within the network itself.
    \label{fig:compensationnoise}
    }
\end{figure*}

One would not expect the synaptic weight noise to affect the L2/3 model strongly, as it should average out over a large number of connections between the constitutent populations.
It turns out that the surprisingly strong impact of synaptic weight noise is purely due to the implementation of background stimulus in this model and can therefore be easily countered.

\paragraph{Effects}

The relative deviation of the total synaptic conductance scales with $\expect{g}/\var{g} \sim 1/\sqrt{\nu_\mathrm{input}} \sim 1/\sqrt{N}$ (see \Cref{eqn:cond_var_single}), where $\nu_\mathrm{input}$ is the total input frequency and N the number of presynaptic neurons.
Therefore, interactions between large populations are not expected to be strongly affected by synaptic weight noise.

The only connections where an effect is expected are the RSNP$\rightarrow$PYR connections, because the presynaptic RSNP population consists of only 2 neurons per MC.
However, long-range inhibition also acts by means of a second-order mechanism, in which an active MC activates its counterpart in some other HC, which then in turn inhibits all other MCs in its home HC via BAS cells.
This mechanism masks much of how synaptic weight noise affects RSNP$\rightarrow$PYR connections.

Nevertheless, synaptic weight noise appears to have a strong effect on network dynamics (\Cref{fig:compensationnoise}, red curves).
The reason for that lies in the way the network is stimulated.
In the original model, each PYR cell receives input from a single Poisson source.
This is of course a computational simplification and represents diffuse noise arriving from many neurons within other cortical areas.
However, having only a single noise source connected by a single synapse to the target neuron makes the network highly sensitive to synaptic weight noise (see \Cref{sec:l23_fixed_pattern_noise}).

\paragraph{Compensation}
\label{sec:l23-noise_comp}

The compensation for this effect was done by increasing the number of independent noise sources per neuron, thereby reducing the statistically expected relative noise conductance variations per PYR cell.
The only limitation lies in the total number of available external spike sources and the bandwidth supplied by the off-wafer communication network (\Cref{sec:communication_infrastructure}).
Once this limit is reached, the number of noise inputs per PYR cell can still be increased even further if PYR cells are allowed to share noise sources.
Given a total number of available Poisson sources $N$ and a noise population size of $n$ sources per PYR cell, the average pairwise overlap between two such populations is $n^2/N$.
Therefore, as long as the average overlap remains small enough, the overlap-induced spike correlations will not affect the network dynamics.

In our example (\Cref{fig:compensationnoise}, green curves), we have chosen $n=100$, while the total number of Poisson sources is set at $N=5000$.
Note how this relatively simple compensation method efficiently restores most functionality criteria.
The most significant remaining differences can be seen in pattern completion and attentional blink (\tb{T}, \tb{U}) and appear mainly due to the affected RSNP$\rightarrow$PYR connections.

In addition to the investigation of synaptic weight noise on the default model, we repeated the same experiments for the model with reduced PYR population sizes (\Cref{fig:compensationnoise}, purple curves), which we have previously suggested as a compensation method for synaptic weight noise (\Cref{sec:l23-loss_comp}).
The fact that PYR population reduction does not affect the network functionality in the case of (compensated) synaptic weight noise is an early indicator for the compatibility of the suggested compensation methods when all distortion mechanisms are present (\Cref{sec:l23alldistortions}).

\subsubsection{Non-configurable axonal delays}
\label{sec:l23-delay}

\begin{figure}[tb]
		\centering		
		\includegraphics[width=\columnwidth]{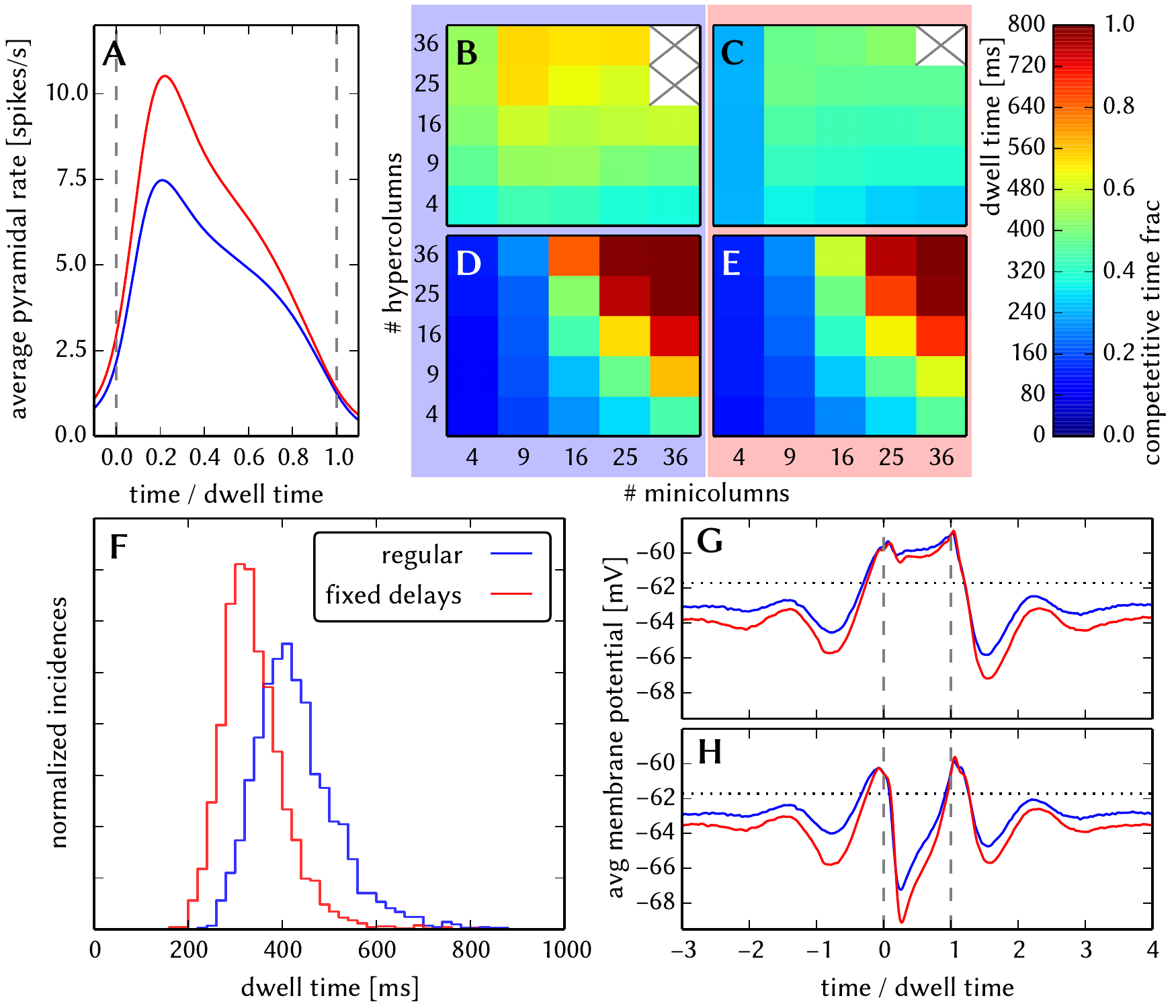}
		\caption{
						{\bf Effects of fixed axonal delays on the L2/3 model.}
                        Unless explicitly stated otherwise, the default network model (9HC$\times$9MC) was used.
                        Data from the regular and distorted models is depicted (or highlighted) in blue, and red, respectively.
			(\textbf{A}) Average firing rate of PYR cells within an active period of their parent attractor.
			(\textbf{B}, \textbf{C}) Average dwell time for various network sizes.
			(\textbf{D}, \textbf{E}) Fraction of time spent in competitive states (i.e. no active attractors) for various network sizes.
			(\textbf{F}) Distributions of dwell times.
			(\textbf{G}) Average voltage of PYR cells before, during and after their parent attractor is active (UP state).
			(\textbf{H}) Average voltage of PYR cells before, during and after an attractor they do not belong to is active (DOWN state).
                                     For subplots \tb{A}, \tb{G} and \tb{H}, the abscissa has been subdivided into multiples of the average attractor dwell time in the respective simulations.
                                     In subplots \tb{G} and \tb{H} the dotted line indicates the leak potential $\Vrest$ of the PYR cells.
		\label{fig:fixed_delays_L23}
	}
\end{figure}

In the original model, axonal delays between neurons are proportional to the distance between their home HCs.
At an axonal spike propagation velocity of 0.2 m/ms, the default (9HC$\times$9MC) network implements axonal delays distributed between 0.5 and 8 ms.
While PYR cells within an MC tend to spike synchronously in gamma waves, the distribution of axonal delays reduces synchronicity between spike volleys of different MCs.

Fixed delays, on the other hand, promote synchronicity, thereby inducing subtle changes to the network dynamics (\Cref{fig:fixed_delays_L23}).
The synchronous arrival of excitatory spike volleys causes PYR cells in active attractors to spike more often (\tb{A}).
Their higher firing rate in turn causes shorter attractor dwell times, due to their spike frequency adaptation mechanism (\tb{B}, \tb{C}, \tb{F}).
During an active attractor, the elevated firing rate of its constituent PYR cells causes a higher firing rate of the inhibitory interneurons belonging to all other attractors.
This, in turn, leads to a lower membrane potential for PYR cells during inactive periods of their parent attractor (\tb{G}, \tb{H}).
As these effects are not fundamentally disruptive and also difficult to counter without significantly changing other functional characteristics of the network, we chose not to design a compensation strategy for this distortion mechanism in the L2/3 network.

\subsubsection{Full simulation of combined distortion mechanisms}
\label{sec:l23alldistortions}

In a final step, we emulate the L2/3 model on the ESS~(\Cref{sec:ess}), and compensate simultaneously for all of the effects discussed above.
We first investigate how much synapse loss to expect for different network sizes, and then realize the network at two different scales in order to investigate all of the chosen functionality criteria.
The default network (9HC$\times$9MC) is used to analyze spontaneous attractors, while a large-scale model (25HC$\times$25MC) serves as the test substrate for pattern completion and pattern rivalry.

\begin{figure}
    \centering		
    \includegraphics{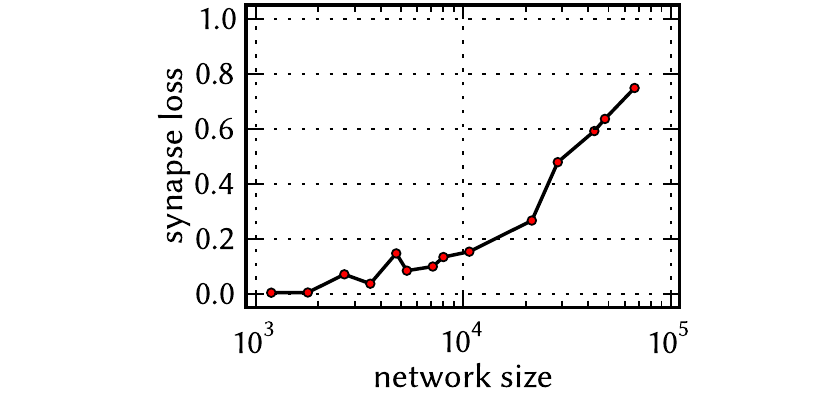}
	\caption{
		{\bf Synapse Losses after mapping the L2/3 model.}
    \label{figure:sweepL23}
	}
\end{figure}

\paragraph{Synapse loss}
The synapse loss after mapping the L2/3 model onto the BrainScaleS hardware is shown in \Cref{figure:sweepL23}
for different sizes, using the scaling rules defined in \cref{sec:methods-l23-scaling}.
Synapse loss starts to occur already at small sizes and increases rapidly above network sizes of \num{20000} neurons.
The jumps can be attributed to the different ratios between number of HCs and number of MCs per HC (\Cref{tab:l23_scaling_mapping}).

\begin{figure*}
    \centering		
	\includegraphics[width=\textwidth]{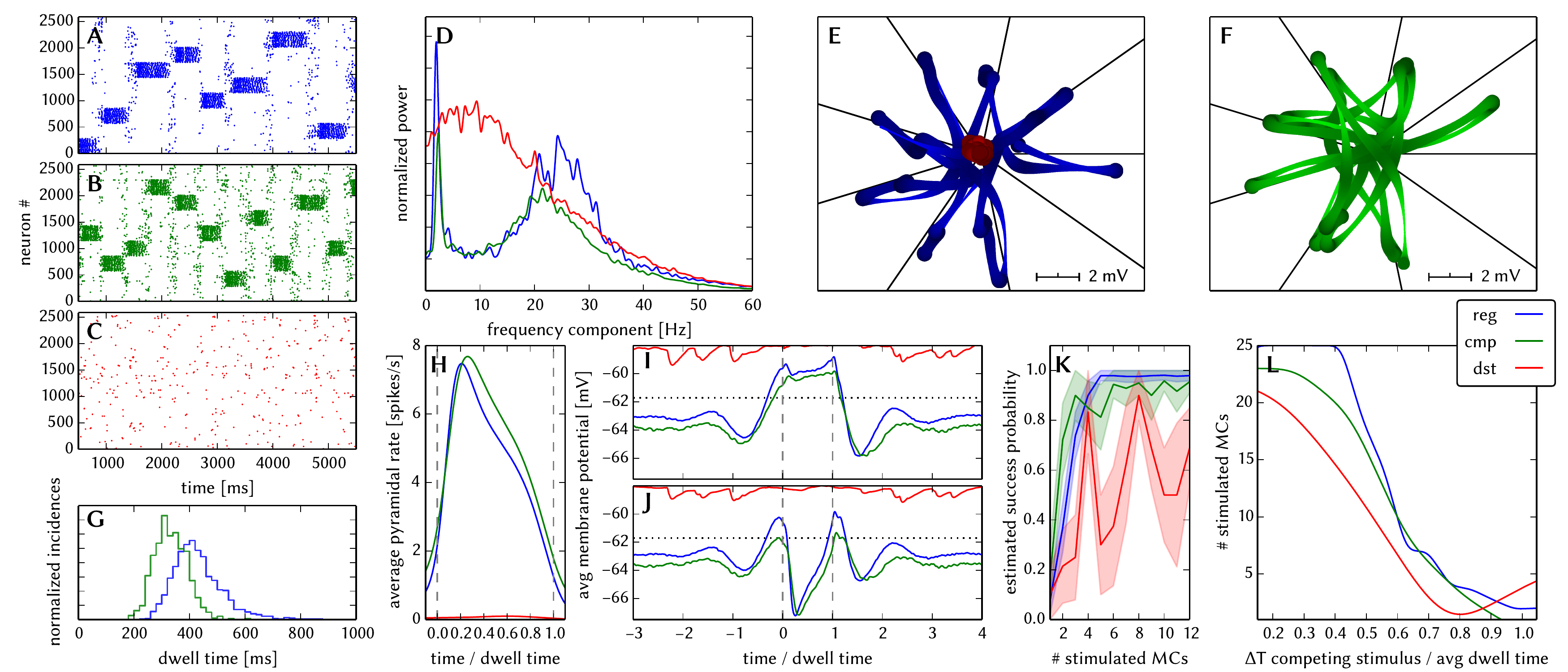}
        \caption{
			{\bf ESS emulation of the L2/3 model.}
            Unless explicitly stated otherwise, the default network model (9HC$\times$9MC) was used.
            Here, we use the following color code: blue for the original model, red for the distorted case on the ESS (with \SI{20}{\%} synaptic weight noise and $\approx\SI{20}{\%}$ synapse loss), and green for the compensated case on the ESS.
            (\textbf{A} - \textbf{C}) Raster plots of spiking activity.
            The MCs are ordered such that those belonging to the same attractor (and \textit{not} those within the same HC) are grouped together.
            A synapse loss of \SI{32}{\%} on the background synpases (see \Cref{table:l23-ess-synloss}) is the main reason for which no spontaneous attractors are evoked.
            For this reason, there are no red curves in \tb{G}, \tb{H}, \tb{I} and \tb{J}.
            Applying a weight compensation and increasing the number of background sources from 9 to 50 effectively restores the original behavior.
            (\textbf{D}) Power spectrum of global activity.
            Since no spontaneous attractors are evokes, neither attractor switching ($\sim\SI{3}{\Hz}$) nor gamma oscillations ($\sim\SI{25}{\Hz}$) can be observed.
            The spectrum of the distorted network complies with the asynchronous irregular firing observed in \tb{C}.
            Compensation restores both of the characteristic peaks in the spectrum.
            (\textbf{E} and \textbf{F}) Star plots of average PYR voltages from a sample of 5 PYR cells per MC.
            The disrupted attractor behavior results in a weak fluctuations of average PYR voltages with essentially no clear UP or DOWN states.
            After compensation, the differences between UP and DOWN states become more pronounced again.
            (\textbf{G}) Distributions of dwell times.
            The disrupted network effectively shows no spontaneous attractors.
            As expected from the software simulations, the dwell times remain, on average, slightly shorter after compensation.
            (\textbf{H}) Average firing rate of PYR cells within an active period of their parent attractor.
            The higher firing rates after compensation are caused by the fixed, short delays, which promote synchronous firing and therefore stronger mutual excitation among PYR cells.
            (\textbf{I}) Average voltage of PYR cells before, during and after their parent attractor is active (UP state).
            (\textbf{J}) Average voltage of PYR cells before, during and after an attractor they do not belong to is active (DOWN state).
            For subplots \tb{H}, \tb{I} and \tb{J}, the abscissa has been subdivided into multiples of the average attractor dwell time in the respective simulations.
            In subplots \tb{I} and \tb{J} the dotted line indicates the leak potential $\Vrest$ of the PYR cells. 
            (\textbf{K}) Pattern completion in a 25HC$\times$25MC network.
                        Estimated activation probability from 25 trials per abscissa value.
            (\textbf{L}) Attentional blink in a 25HC$\times$25MC network: $p=0.5$ iso-probability contours, measured over 14 trials per ($\Delta T$, \#MCs) pair.
            Since the distorted network showed no spontaneous attractors (\tb{C}), we used the average dwell time from the pattern completion experiment (\tb{K}) for normalization.
    		\label{figure:l23_ess}
            }
\end{figure*}

\paragraph{Small-scale model}
The default model (9HC$\times$9MC) can, in principle, be mapped onto the hardware without any synapse loss if the full wafer is available for use.
Nevertheless, in some scenarios, a full wafer might not be available, due to faulty components or part of its area being used for emulating other parts of a larger parent network.
We simulate this scenario by limiting the usable wafer area to 4 reticles (out of a total of 48 on the full wafer).
With the reduced available hardware size, the available pulse bandwidth of the off-wafer communication network decreases as well, such that diffusive background noise can not be modeled with one individual Poisson source per neuron.
Hence, each pyramidal neuron receives input from 9 out of 2430 background sources.
The total synapse loss for the given network setup amounts to \SI{22.2}{\%} and affects different projection types with varying strength (\Cref{table:l23-ess-synloss}).
Also external synapses are lost, since, in contrast to the synapse loss study (\Cref{sec:l23synloss}), they have not been prioritized in the mapping process in this case.
Additionally, we applied \SI{20}{\%} synaptic weight noise and simulated the network with a speedup factor of \num{12000}.
The behavior on the ESS is shown in \Cref{figure:l23_ess}.
The distorted network shows no spontaneous attractors (\tb{C}), which can be mainly attributed to the loss of over \SI{32}{\%} of the background synpases.
To recover the original network behavior, we first increased the number of background neurons per cell from 9 to 50 to compensate for synaptic weight noise,
and also scaled the weights by $1/(1-p_\mathrm{loss})$ for each projection type with extracted synapse loss values $p_\mathrm{loss}$ (\Cref{table:l23-ess-synloss}), following the synapse loss compensation method described in \Cref{sec:l23-loss_comp}.
Note that here the complete PyNN experiment is re-run: synaptic weights are scaled in the network definition leading to a new configuration of $g_\textrm{max}$ and the digital weights on the HICANNs (\Cref{sec:hicann}) after the mapping process.
These measures effectively restored the attractor characteristics of the network (\Cref{figure:l23_ess}).
The attractor dwell times remained a bit smaller than for the regular network (\tb{G}), which can be ascribed to the non-configurable delays (\Cref{sec:l23-delay}).

\begin{table}
	\centering
	\caption{{\bf Projection-wise synapse loss of the L2/3 model after the mapping process}}
\begin{tabular}{l|rr|rr}
\toprule
&\multicolumn{2}{c|}{\textbf{9HC$\times$9MC}}&\multicolumn{2}{c}{\textbf{25HC$\times$25MC}}\\
\textbf{projection} & \textbf{dist.} & \textbf{comp.} & \textbf{dist.} & \textbf{comp.} \\
\midrule
PYR $\to$ PYR (local) &	21.1 &	21.0 &	0.9 &	0.3\\
PYR $\to$ PYR (global) &	20.8 &	21.2 &	8.0 &	0.4\\
PYR $\to$ RSNP &	22.6 &	21.9 &	37.0 &	28.8\\
PYR $\to$ BAS &	8.2 &	7.6 &	15.0 &	0.2\\
BAS $\to$ PYR &	23.3 &	39.4 &	0.5 &	0.2\\
RSNP $\to$ PYR &	22.7 &	39.9 &	0.0 &	3.9\\
L4 $\to$ PYR &	44.1 &	45.4 &	15.5 &	2.3\\
background $\to$ PYR &	32.3 &	31.3 &	17.3 &	1.3\\
\midrule
total &	22.2 &	25.2 &	17.9 &	9.8\\
\bottomrule
\end{tabular}
\begin{flushleft}
	Projection-wise synapse loss in \% for the default (9HC$\times$9MC) and large-scale (25HC$\times$25MC) network. See text for the respective differences between the distorted (dist.) and compensated (comp.) networks.
\end{flushleft}

\label{table:l23-ess-synloss}
\end{table}

\paragraph{Large-scale model}
The ability of the network to perform pattern completion and exhibit pattern rivalry was tested on the ESS for the large-scale model with 25 HCs and 25 MCs per HC.
From the start, we use a background pool with \num{5000} Poisson sources and \num{100} sources per neuron to model the diffusive background noise,
as used for the compensation of synaptic weight noise (\Cref{sec:l23-noise_comp}).
As with the small-scale network, the synapse loss of \SI{17.9}{\%} shows significant heterogeneity (\Cref{table:l23-ess-synloss}), and affects mainly projections from PYR to inhibitory cells, but also connections from the background and L4 stimulus.
In contrast to the idealized case in \Cref{sec:l23-loss_effects}, where each synapse is deleted with a given probability, the synapse loss here happens for entire projections at the same time, i.e. all synapses between two populations are either realized completely or not at all.
We note that the realization of all PYR-RSNP synapses is a priori impossible, as each RSNP cell has $24\times24\times30=17280$ potential pre-synaptic neurons (cf. scaling rules in \Cref{sec:methods-l23-scaling}), which is more than the maximum possible number of pre-synaptic neurons per HICANN (14336, see \Cref{sec:hicann}).
The simulation results with \SI{20}{\%} synaptic weight noise for pattern completion and pattern rivalry are shown in \Cref{figure:l23_ess} \tb{K} and \tb{L} (red curves).
In both cases the network functionality is clearly impaired.
In particular, the ability of an active pattern to suppress other patterns is noticeably detoriated, which can be traced back to the loss of \SI{37}{\%} of PYR-RNSP connections.

In order to restore the functionality of the network we used a two-fold approach:
First, we attempted to reduce the binary loss of PYR-RSNP projections by reducing the number of PYR cells per MC from 30 to 20, which decreases the total number of neurons in the network, as well as the number of potential pre-synaptic neurons per RNSP cell.
The synapse loss was thereby reduced to \SI{28.8}{\%} for PYR-RSNP projections and was eliminated almost completely for all other projections (\Cref{table:l23-ess-synloss}).
Secondly, we compensated for the remaining synapse loss by scaling the synaptic weights as described in \Cref{sec:l23-loss_comp}.

After application of these compensation mechanisms, we were able to effectively restore the original functionality of the network.
Both pattern completion and attentional blink can be clearly observed.
The small remaining deviations from the default model can be attributed to the inhomogeneity of the synapse loss and the fixed delays on the wafer.

\subsection{Synfire chain with feed-forward inhibition}
\label{sec:synfire}

Our second benchmark network is a model of a series of consecutive
neuron groups with feed-forward inhibition, called \emph{synfire chain}
from here on \cite{kremkow2010functional}.
This network acts as a selective filter to a synchronous spike packet
that is applied to the first neuron group of the chain.
The behavior of the network is quantified by the dependence of the
filter properties on the strength and temporal width of the initial
pulse.
Our simulations show that synapse loss can be compensated in a
straightforward manner.
Further, the major impact of weight noise on the network functionality
stems from weight variations in background synapses, which can be
countered by modification of synaptic and neuronal parameters.
The effect of fixed axonal delays on the filtering properties of the
network can be countered only to a limited extent by modifying synaptic
time constants and the strength of local inhibition.
Simulations using the ESS show that the developed compensation
methods are applicable simultaneously.
Furthermore, they highlight some further sources of potential failure of pulse propagation that originate from bandwidth limitations in the off-wafer communication infrastructure.

\subsubsection{Architecture}

Feed-forward networks with a convergent-divergent connection scheme
provide an ideal substrate for the investigation of activity transport.
Insights have been gained regarding the influence of network
characteristics on its response to different types of stimulus
\cite{aertsen96,diesmann99,vogels05signal}.
Similar networks were also considered as computational entities rather
than purely as a medium for information transport
\cite{abeles2004modeling,schrader2010compositionality,kremkow2010gating}.
The behavior of this particular network has been shown to depend on the
connection density between consecutive groups, on the balance of
excitation and inhibition as well as on the presence and magnitude of
axonal delays in \cite{kremkow2010functional}.
This makes it sensitive to hardware-specific effects such as an
incomplete mapping of synaptic connectivity, the variation of synaptic
weights, bandwidth limitations which cause loss of individual spike events and
limited availability of adjustable axonal delays and jitter in the
spike timing that may be introduced by different hardware components.

\begin{figure}
    \centering
    \includegraphics{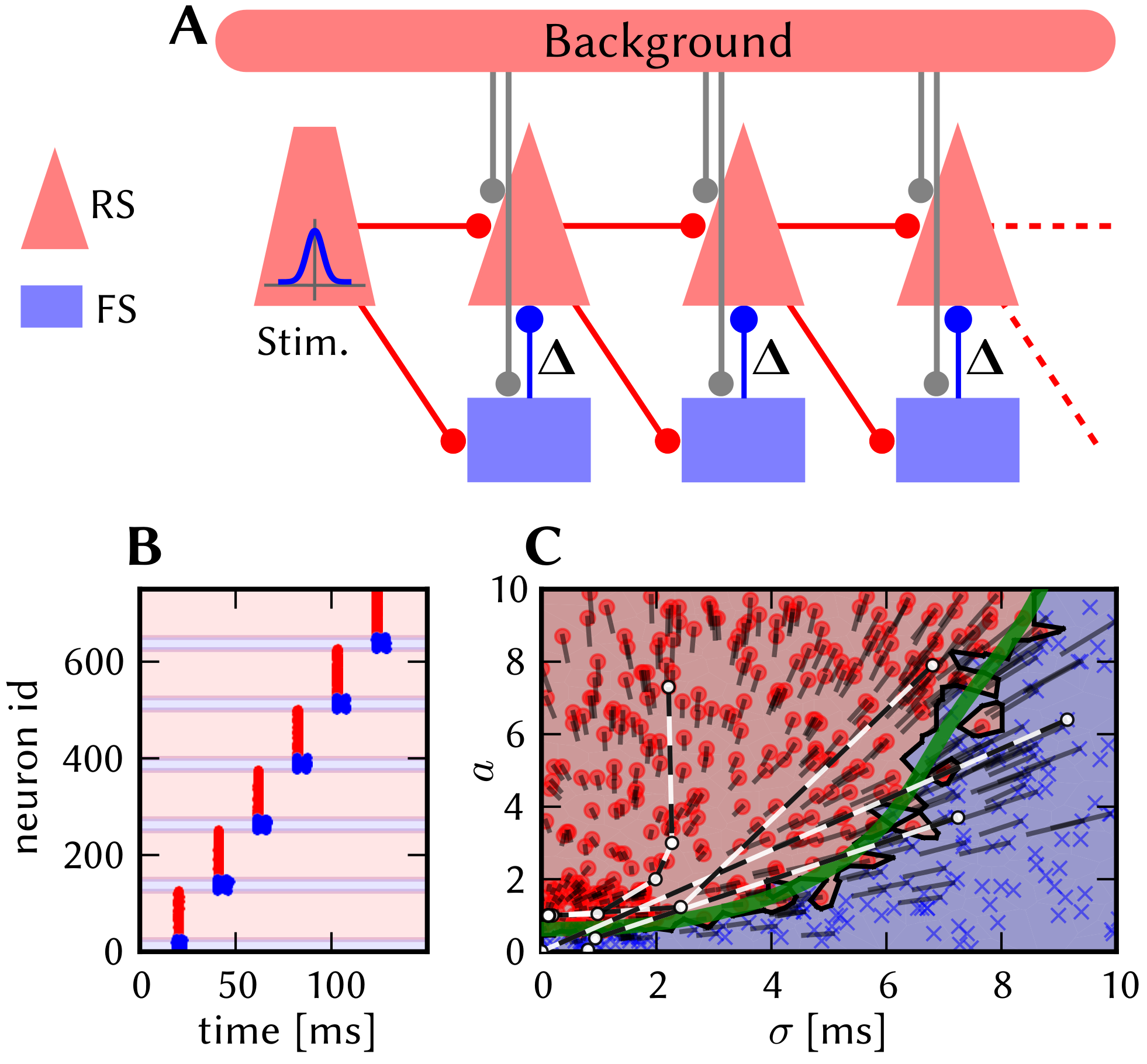}
    \caption{
		{\bf Synfire chain network.}
        (\textbf{A}) Connectivity of the synfire chain with
        feed-forward inhibition \cite{kremkow2010functional}.
        Excitatory projections are shown in red, inhibitory in blue.
        In the default realization the network consists of six consecutive groups.
        The local FS~$\to$~RS projection has an adjustable delay
        $\Updelta$, which affects the network dynamics.
        The intergroup delay is set to \SI{20}{\ms} for visualization
        purposes following the previous work; this has no influence on the filter
        properties because the delay of both intergroup projections is equal.
        The background stimulus is realized using random Gaussian
        current (original) and Poisson background spikes (adapted
        version for the hardware).
        The parameters for neurons and connections are given in
        \cref{tab:synfire-neuron-params} and
        \cref{tab:synfire_ffi_connectivity}.
        (\textbf{B}) Exemplary raster plot of the network behavior.
        The first group receives a pulse packet with $a=1$ and
        $\sigma_0=\SI{1}{\ms}$, which propagates as a
        synchronous spike volley along the chain.
        (\textbf{C}) Characterization of the network behavior in the
        $(\sigma, a$) state space.
        Each marker represents the initial stimulus parameters
        $(\sigma_0, a_0)$.
        The stimulus parameters were selected randomly from the region
        ($a_0 < 10$, $\sigma_0 \le \SI{10}{\ms}$).
        The region with ($a_0 < 2$, $\sigma_0 \le \SI{2}{\ms}$)
        was simulated more frequently to increase the
        resolution near the convergence points of the propagation.
        The marker color is linearly scaled with the activity in the last group,
        $a_6$, being blue for $a_6=0$ and red for $a_6=1$ and is set to red for $a_6 > 1$.
        To improve visibility, the background is colored according to the
        color of the nearest marker, red for $a_6 \ge 0.5$ and blue otherwise.
        Experiments where the RS group did not fire are marked as $\times$.
        The gray lines originating from each marker denote the direction towards
        the pulse volley parameters $(\sigma_1, a_1)$.
        The green line shows a fit to the separatrix between
        zero and nonzero activity at the last group of the synfire chain (see \cref{sec:ffi_criteria} for details).
        This approximation is used to compare the behavior of different
        modifications of the original network.
        The dashed black and white lines show four exemplary
        trajectories through the $(\sigma, a)$ state space.
    \label{fig:synfire_ffi_schema}
	}
\end{figure}
The feed-forward network comprises a series of successive neuron
groups, each group containing one excitatory and one inhibitory
population.
The excitatory population consists of 100 regular-spiking (RS), the
inhibitory of 25 fast-spiking (FS) cells.
Both cell types are modeled as LIF neurons with exponentially shaped
synaptic conductance without adaptation, as described in
\cref{sec:hicann}.
Both RS and FS neurons are parameterized using identical values
(\cref{tab:synfire-neuron-params}).

Each excitatory population projects to both populations of the
consecutive group while the inhibitory population projects to the
excitatory population in its local group
(\cref{fig:synfire_ffi_schema} \textbf{A}).
There are no recurrent connections within the RS or FS populations.
In the original publication \cite{kremkow2010functional}, each
neuron was stimulated independently by a Gaussian noise current.
Because the hardware system does not offer current stimulus for all
neurons, all neurons in the network received stimulus from
independent Poisson spike sources.
For Gaussian current stimulus, as well as in the diffusion limit of
Poisson stimulus (high input rates, low synaptic weights), the membrane
potential is stationary Gaussian, with an autocorrelation dominated by
the membrane time constant.
The only remaining differences are due to
the finite, but small, synaptic time constants.
The rate and synaptic weight of the background stimulus were adjusted to obtain similar values for
the mean and variance of the membrane potential, resulting in a
firing rate of \SI{2}{\kilo\hertz} with a synaptic weight
of \SI{1}{\nano\siemens}.

The initial synchronous stimulus pulse is emitted by a population of spike
sources, which has the same size and connection properties as a single RS
population within the network.
A temporally localized \emph{pulse packet} was used as a stimulus,
whereby each of the 100 spike sources emitted $a_0$ spikes that were
sampled from a Gaussian distribution with a common mean time and a given
standard deviation $\sigma_0$.
The variables $(\sigma_i, a_i)$ are later used to describe the
characteristics of the activity in the $i$th group of the chain, referring
to the temporal pulse width and number of spike pulses per neuron, respectively.

\subsubsection{Functionality criteria}
\label{sec:ffi_criteria}

The functionality of the feed-forward network is assessed by
examining the propagation of a synchronous pulse after the stimulus is applied to
the first group in the chain (\cref{fig:synfire_ffi_schema} \textbf{B}).
The propagation is quantified by applying initial stimuli of varying
strength $a_0 \in [0, 10]$ and temporal spread
$\sigma_0\in [\SI{0}{\ms}, \SI{10}{\ms}]$.
For each synfire group $i \in \{1, ..., 6\}$, the activation is determined by
setting $a_i$ to the number of emitted spikes divided by the number of neurons in the RS population.
$\sigma_i$ is the standard deviation of the spike pulse times.
Typically, the resulting ``trajectory'' in the $(\sigma, a)$ space 
(\cref{fig:synfire_ffi_schema})
is
attracted to one of two fixed points: either near $(\sigma=\SI{0}{\ms}, a=1)$, i.e., the
pulse packet propagates as a synchronous spike volley, and $(\SI{0}{\ms}, 0)$,
i.e., the propagation dies out (e.g., \cref{fig:ffi_loss_merged} \textbf{A}).

The network behavior is characterized by the separating line between
successful and extinguished propagation in the state space
$(\sigma, a)$ of the initial stimulus;
this line will be called \emph{separatrix} from here on.
The differentiation between extinguished and successful propagation
is defined by $a_6 \ge 0.5$ resp.\ $a_6 < 0.5$ in the last (6th) group.
This is justified because in the undistorted case, $a$ is clustered
around the values 0 and 1 (\Cref{fig:synfire_end_a}).
Due to the statistic nature of the connectivity, background stimulus
and pulse packet, the macroscopic parameters $\sigma$ and $a$ do not
fully determine the behavior of the system.
This means that in the reference simulation, there is a small region
around the separatrix where the probability of a stable pulse
propagation is neither close to zero nor to one.
Thus, in addition to the location of the separatrix
(\cref{sec:synfire_separatrix_fit}), the width of this region is taken
as a functionality criterion.

The background stimulus is adjusted such that the spontaneous firing
rate in the network is below $0.1$ Hz, in accordance with
\cite{kremkow2010functional}.
In cases in which distortion mechanisms induce a much
stronger background firing, the spike trains are filtered before the
analysis by removing spikes which appear not to be within a spike
volley (\Cref{sec:synfire_spike_filter}).

\subsubsection{Synapse loss}
\label{sec:synfire_synloss}

Homogeneous synapse loss affects the strength of excitatory and
inhibitory projections equally on average.
Additionally, the number of incoming spikes seen by a single neuron
varies as synapses are removed probabilistically, in contrast to the
undistorted model with a fixed number of incoming connections for each
neuron type (\cref{tab:synfire_ffi_connectivity}).
Synapse loss was applied to all internal connections as well as to the
connection from the synchronized stimulus population to the first
group in the network; background stimulus was not affected (cf.\ \cref{dist:synapseloss}).
\begin{figure}
    \centering
    \includegraphics{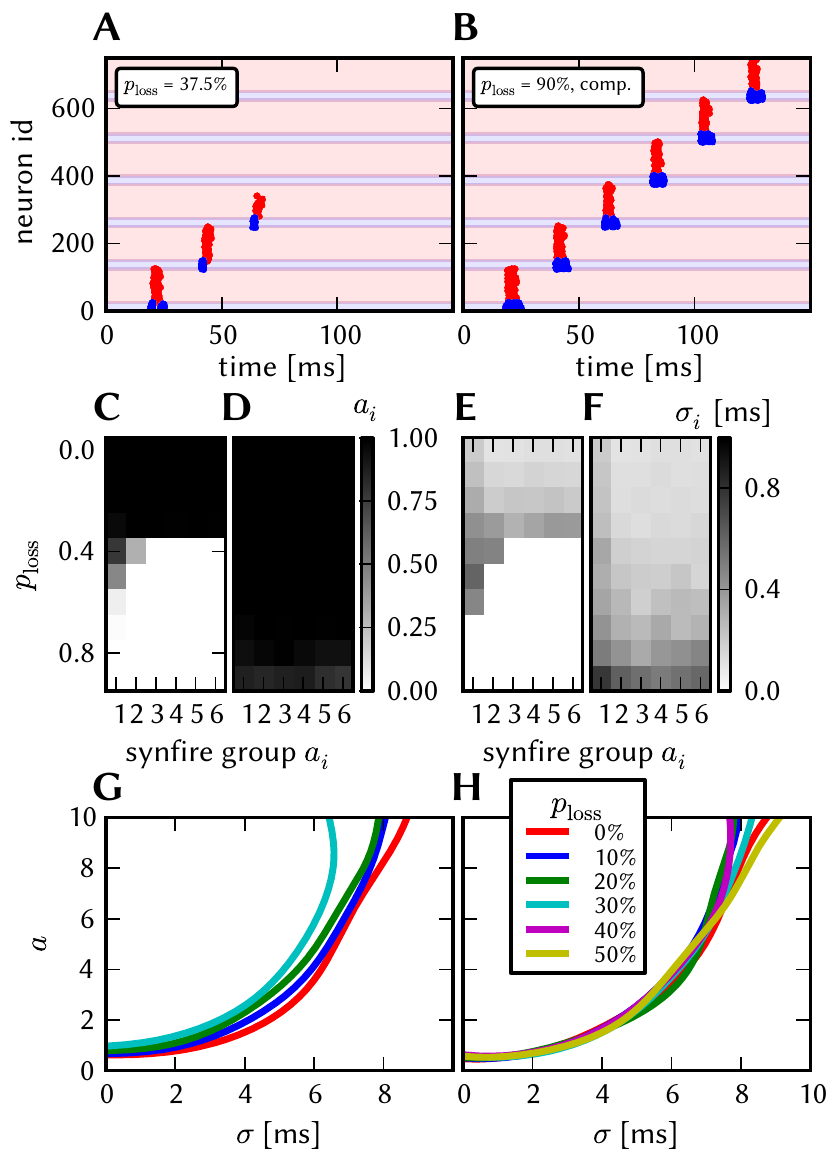}
    \caption{
		{\bf Effect and compensation of synapse loss for the synfire chain network.}
        (\textbf{A}) Synfire network with 37.5\% synapse loss applied
        to all connections within the network.
        External connections (synchronous stimulus and background) are not
        affected.
        (\textbf{B}) Raster plot with active compensation at 90\% synapse loss.
        (\textbf{C}) Activation $a_i$ in each group $i$ with varying values of
        synapse loss.
        (\textbf{D}) $a_i$ as in \textbf{C} but with active compensation.
        (\textbf{E}) Pulse width $\sigma_i$ in each group $i$ with varying
        values of synapse loss.
        (\textbf{F}) Pulse width as in \textbf{E} but with active compensation.
        (\textbf{G}) Comparison of approximated separatrix locations for synapse 
        loss values from 0\% to 50\%.
        The lines for 40\% and 50\% are missing because no stable region exists.
        (\textbf{H}) Approximated separatrix locations with active
        compensation.
    \label{fig:ffi_loss_merged}
	}
\end{figure}
\Cref{fig:ffi_loss_merged}~\textbf{A} shows a single experiment
with synapse loss of \SI{37.5}{\percent}, contrasting with the undistorted case
(\cref{fig:synfire_ffi_schema} \textbf{A}).
Above a certain value of synapse loss, the signal fails to propagate
to the last group.
As shown in \cref{fig:ffi_loss_merged} \textbf{C} and \textbf{E} for one stimulus
parameter set, successful propagation stops at a synapse loss value
between 30\% and 40\%.
The pulse width increases with rising synapse loss due to the
increasing variation of synaptic conductance for individual neurons
(\textbf{E}).
The effect is reversed by increasing all synaptic weights in the
network by a factor of $1 / (1 - p_{\mathrm{loss}})$, with
$p_{\mathrm{loss}}$ being the probability of synapse loss.
This compensation strategy can effectively counter synapse loss of up to \SI{90}{\percent}
(\textbf{B}, \textbf{D}) and the pulse width increase is shifted to
larger values of synapse loss (\textbf{F}).  The distortion mechanism has only a
minor effect on the $a$-value of the separatrix in the depicted region
(\textbf{G}).
However, the location of the separatrix at $\sigma_0 = 0$ rises with
synapse loss until it reaches the fixed point at approx.\
$(\SI{0.1}{\ms}, 1)$, at which point a bifurcation occurs and the the
attractor region for $(\SI{0.1}{\ms}, 1)$ disappears 
(as described in \cite{diesmann2001state} for the case of varying weights).
In the compensated case, the separatrix locations are identical with
the undistorted case within the measurement precision.

\begin{figure}[tbp]
    \centering
    \includegraphics{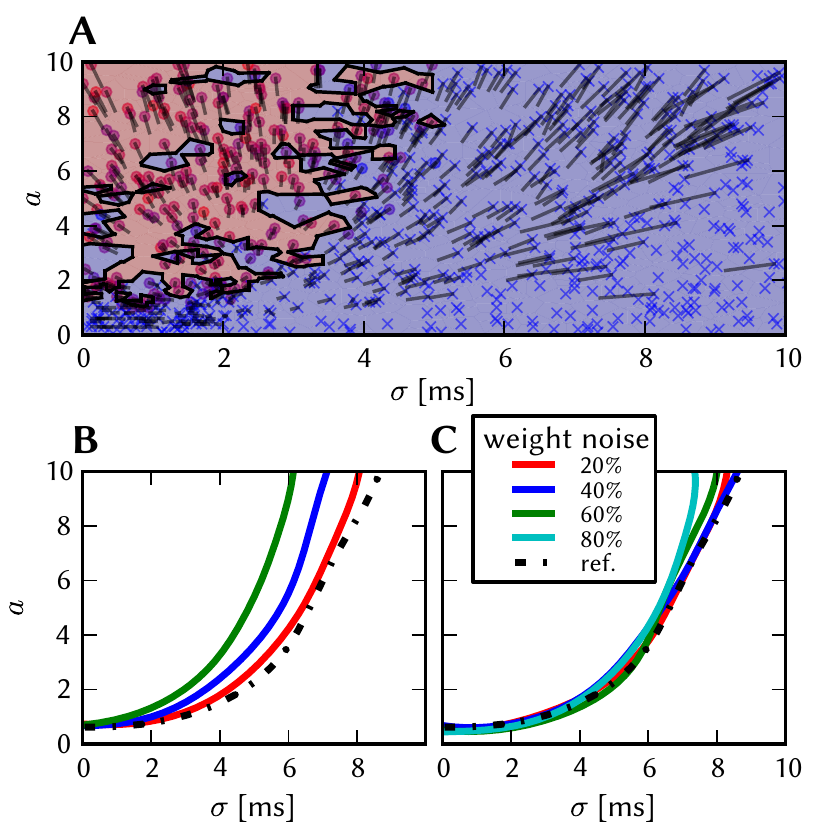}
    \caption{
		{\bf Effect of synaptic weight noise on the synfire chain model.}
        The spike data for all three plots was filtered to remove spontaneous
        spikes in individual neurons, which stem from weight increase
        in some background synapses due to weight noise.
        (The filter parameters were $T=\SI{10}{\ms}, N=25$ cf.\ \cref{sec:synfire_spike_filter})
        (\textbf{A}) State space at 80\% weight noise.
        The set of inputs that evokes activity in the last group is patchy
        as a consequence of the distortion mechanism.
        In the compensated case the separation is sharp again, as shown in
        \cref{fig:ffi_weight_noise_comp}.
        (\textbf{B}) Approximate separatrix locations for smaller values
        of weight noise.
        (\textbf{C}) Approximate separatrix locations for the
        compensated case.
    \label{fig:ffi_weight_noise}
	}
\end{figure}

\subsubsection{Synaptic weight noise} 
\label{sec:synfire_weight_noise}
The effect of synaptic weight noise is shown in
\cref{fig:ffi_weight_noise}.
Similarly to the effect of synapse loss, the region of stable propagation shrinks
(\textbf{B}); additionally, the border between the regions of stable
and extinguished propagations becomes less sharp (\textbf{A}).
This is caused by two effects: Varying strength of the background
stimulus, and varying strength of the synaptic connections within the
network.
The first effect is significant because the background stimulus to
each neuron is provided through a single synapse.
Thus, the effective resting potential of each neuron is shifted, significantly
changing its excitability and, in some cases, inducing spontaneous
activity.
One possibility of countering this effect is to utilize several synapses for background stimulus
thereby averaging out the effect of individual strong or weak synapses, as has been done
in the case of the L2/3 model in \cref{sec:l23-noise_comp}.
Here, a different method was employed: The resting potential $\Vrest$
was raised while simultaneously lowering the synaptic weight from the
background stimulus.
The parameters were chosen in such a way that the mean and variance of
the distribution of membrane voltages in each neuron population was
kept at the value of the undistorted network:

\begin{align}
    \langle V \rangle &\approx w_0 \cdot \langle K \rangle + \Vrest \quad \mathrm{and}\\
    \var{V} &\approx w_0^2 \var{K} + \var{w} \left( \langle K \rangle ^2 + \var{K}\right) \; , \label{eq:synfire_var_V}
\end{align}
where $K(t) = \sum_{\mathrm{spk}\;j} \kappa(t - t_j)$ represents the effect of the background stimulus, $\kappa$ being the PSP kernel, and $\var{w} = w_0^2 \sigma^2$ appears due to synaptic weight noise.
In the distorted case, the width of this distribution is a combined
effect of the random background stimulus and the weight
variation, while in the original case it originates from the stochasticity of the
stimulus only.
In the undistorted case, $\var{w}$ is 0, and only the first term contributes to $\var{V}$.
With increasing $\sigma^2$, the contribution of the second term to
$\var{V}$ increases, which is compensated by changing $w_0$
accordingly, keeping $\var{V}$ at the original level.
This, in turn, changes $\langle V \rangle$, which is compensated by a change of $\Vrest$.

The effect of synaptic weight noise within the network itself is
less significant compared to its impact on the noise stimulus.
\cref{fig:ffi_weight_noise} \textbf{C} shows
that removing the effect of background stimulus noise alone is
sufficient to counteract synaptic noise values of up to 50\%.

\begin{figure}
    \centering
    \includegraphics{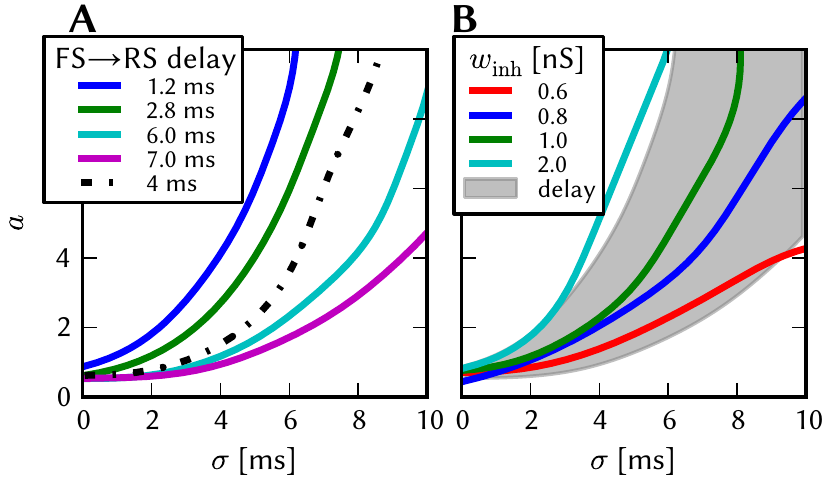}
    \caption{
		{\bf Delays in the synfire chain model.}
        \textbf{(A)} Reproduction of \cite{kremkow2010functional}, fig. 4c.
        The location of the separatrix is modified by
        changing the axonal delay of local inhibition.
        For a value of \SI{0.4}{\ms}, no stable region is present.
        \textbf{(B)} The location of the separatrix is modified by varying weights for synapses taking part in local
        inhibition.
        The axonal delay of local inhibition was fixed at \SI{1.5}{\ms} and the inhibitory time constant was increased by a factor of 3.
        The gray region shows the range of the separatrix location for
        delay values from \SI{1.2}{\ms} to
        \SI{7}{\ms} (the range in plot \textbf{A}) as reference.
    \label{fig:ffi_delay}
	}
\end{figure}

\subsubsection{Non-configurable axonal delays}
\label{sec:synfire_delays}

\Cref{fig:ffi_delay} \textbf{A} shows the effect of varying axonal
delays between the inhibitory and excitatory population of a single
synfire group.
As was shown in \cite{kremkow2010functional}, the delay can be
employed to control the position of the separatrix between stable and
unstable propagation.
Because the axonal delay is not configurable for on-wafer connections,
a different method is required to regain the ability to control the
separatrix.
While \cref{sec:synfire_synloss} and \cref{sec:synfire_weight_noise}
show that synaptic weight noise and synapse loss can influence the
location of the separatrix, a method is required that is independent
of those distortion mechanisms.
\cite{diesmann02} shows that several parameters, including group size
and noise level, can modify the separatrix location, albeit for a
model without feed-forward inhibition.
Here, we investigate to which extent parameter modification can
reproduce the effect of variable delays.
For very short delays (in this case, \SI{0.1}{\ms}, not shown),
stable propagation does not occur, because the onset of local inhibition
is nearly synchronous with the onset of external excitation.
This effect was countered by increasing the synaptic time constant and
simultaneously decreasing the synaptic weight for local inhibition,
thus extending the duration of inhibition that acts on the RS
population.
The inhibitory synaptic time constant was increased by a factor of 3
while simultaneously reducing the synaptic weight of the inhibitory
projection.
\cref{fig:ffi_delay} \textbf{B} shows the result of the
compensation for \SI{1.5}{\ms} local inhibition delay.
For both values of axonal delay, the location of the separatrix can be
controlled by changing the weight of inhibition.
However, its shape differs from the delay-induced case because of
the modified delay mechanism of inhibition.
Reduction of the weight beyond a certain point is not possible, as
balanced inhibition is required for network functionality
\cite{kremkow2010functional}.
It is important to note that this kind of compensation is specific to the
state space region which is examined, and that it can not be extended
to arbitrarily large delays.

\subsubsection{Full simulation of combined distortion mechanisms}
\label{sec:synfire-ess}
At last, we simulate the synfire chain with the ESS and compensate simultaneously for all the causes of distortions addressed above.
Before running ESS simulations, we have verified the compatibility of the proposed compensation strategies for different distortion mechanisms in software simulations dealing with the simultaneous incidence of synaptic weight noise, synapse loss and non-configurable axonal delays (\Cref{sec:synfire_appendix_all}).
We proceed with a quantification of synapse loss after mapping the synfire chain for different network sizes to the hardware.
For the ESS simulations we limit the model to very few hardware resources to artificially generate synapse loss, such that all of the above distortion mechanisms are present.
Additional hardware simulations investigating the influence of spike loss and jitter on the network functionality are provided in \Cref{hardware_synfire:jitter}.

\begin{figure}[ht!]
    \centering
\includegraphics{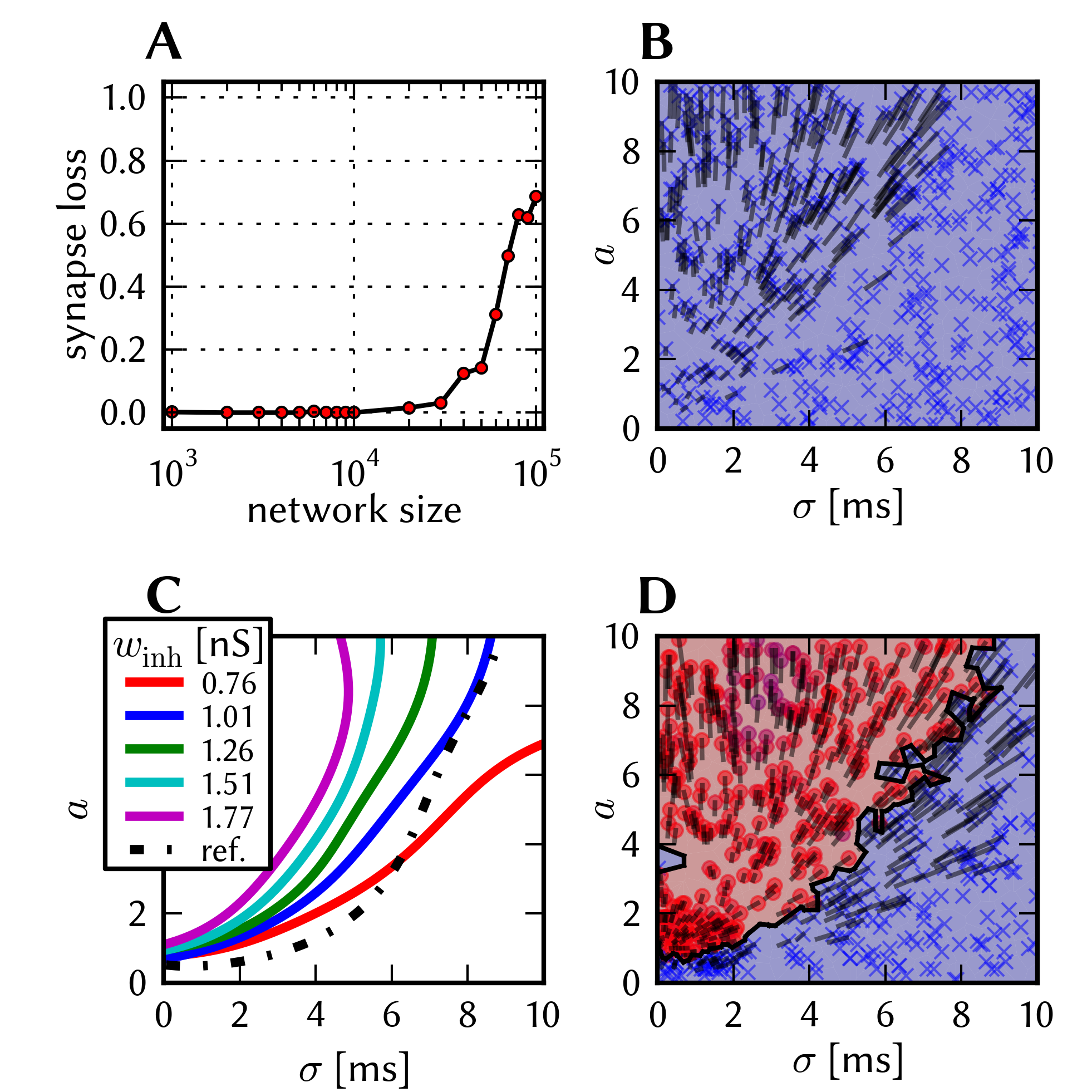}
    \caption{
		{\bf Distorted and compensated simulations of the feedforward synfire chain on the ESS:}
		(\textbf{A}) Synapse loss after mapping the model with different numbers of neurons onto the BrainScaleS System.
		(\textbf{B}) ($\sigma$,$a$) state space on the ESS with default parameters, \SI{20}{\%} weight noise, and \SI{27.4}{\%} synapse loss.
		(\textbf{C}) After compensation for all distortion mechanisms, different separatrices are possible by setting different values of the inhibitory weight.
		(\textbf{D}) Compensated state space belonging to the blue separatrix in \textbf{C}.
    \label{fig:ffi_ess_all_comp}
    }
\end{figure}

\paragraph{Synapse loss}
We mapped the synfire chain at different network sizes onto the BrainScaleS wafer-scale hardware in order to quantify the synapse loss (\Cref{fig:ffi_ess_all_comp} \textbf{A}).
For this purpose we developed network scaling rules that depend on the number and the size of the synfire groups (\Cref{sec:method_synfire_scaling}).
Due to its modular structure and feed-forward connectivity scheme, there is no synapse loss for networks with up to \num{10000} neurons.
However, for network sizes above \num{30000} neurons, the ratio of lost synapses increases abruptly.
With increasing network size more neurons have to be mapped onto one HICANN thereby reducing the number of hardware synapses per neuron.
Moreover, as the group size grows with the network size (cf. \cref{tab:synfire_scaling}), also the number of pre-synaptic neurons for all neurons mapped onto one HICANN increases, so that the maximum number of inputs to a HICANN, i.e. the synapse drivers, becomes a limiting constraint.
The combination of both factors unavoidably leads to synapse loss.

\paragraph{Distorted and compensated simulation}
For the ESS simulation, we applied the following modifications to the benchmark network: originally, each cell in the network receives Poisson background stimulus from an individual source with 2000 \Hz.
Because the off-wafer pulse routing network does not support such high bandwidths (cf. \cref{sec:communication_infrastructure}), we reduce the total number of background sources from 750 to 192 and let each neuron receive input from 8 sources, while decreasing the Poisson rate by a factor of 8, using the same mechanism as for the compensation of synaptic weight noise in the L2/3 model (cf. \cref{sec:l23-noise_comp}).
For the same reason, the network was emulated with a speedup factor of \num{5000} compared to biological real-time, whereby the effective bandwidth for stimulation and recording is doubled with respect to the normal operation with a speedup of \num{10000}.
As seen before, no synapse loss occurs for small networks.
However, as discussed for the L2/3 model in \Cref{sec:l23alldistortions}, one can consider situations where only a small part of the wafer is available for experiments, or where some neurons or synaptic elements are defective or missing a calibration.
Therefore, in order to generate synapse loss in the feed-forward network, we limited the network to only 8 out of 48 reticles of the wafer and furthermore declare half of the synapse drivers as not available.
This resulted in a total synapse loss of \SI{27.4}{\%}.
As with the L2/3 model, the synapse loss was not homogeneous but depended strongly on the projection type (\Cref{tab:synfire-ess-synloss}).

\begin{table}
	\caption{{\bf Projection-wise synapse loss of the synfire chain model after the mapping process}
	}
    \centering
    \begin{tabular}[h]{l c} 
        \toprule 
		\textbf{projection} & \textbf{synapse loss [\%]} \\ 
        \midrule 
        Pulse Packet $\to$ RS$_{ 0 } $ & 21.3\\
        Pulse Packet $\to$ FS$_{ 0 } $ & 12.7\\ 
        RS$_n$ $\to$ RS$_{ n + 1 } $ & 32.4\\
        RS$_n$ $\to$ FS$_{ n + 1 } $ & 32.0\\ 
        FS$_n$ $\to$ RS$_{ n } $ &  20.8\\ 
        Poisson background $\to$ ALL & 0\\
        \midrule
        total & 27.4\\
        \bottomrule
    \end{tabular}
    \label{tab:synfire-ess-synloss}
\end{table}

We simulated the synfire chain with default neuron and synapse parameters on the ESS with \SI{20}{\%} synaptic weight noise and the above synapse loss.
The $(\sigma,a)$ state space (\cref{fig:ffi_ess_all_comp} \textbf{B}) shows no stable point of propagation.
This can be mainly attributed to the small and non-configurable axonal delays which are in the range of \SIrange{0.6}{1.1}{\milli\second} for the chosen speedup factor of \num{5000}.

In order to recover the original behavior, we applied the previously developed compensation methods described in \Cref{sec:synfire_synloss,sec:synfire_weight_noise,sec:synfire_delays}.
Synapse loss was compensated separately for each projection type using \cref{tab:synfire-ess-synloss}.
For synaptic weight noise effectively two compensation methods were applied, as, by using 8 Poisson sources per neuron instead of one, the effect of weight variations is already reduced.
Therefore, this fact was considered in the implementation of the second compensation method that scales the synaptic weight and shifts the resting potential $\Vrest$ to keep the mean and variance of the membrane voltage constant (\Cref{sec:synfire_weight_noise}), by replacing $\var{w}$ with $\frac{1}{8}\var{w}$ in \cref{eq:synfire_var_V}.
We were able to compensate for all distortion mechanisms while still maintaining control over the position of the separatrix (\Cref{fig:ffi_ess_all_comp}~\textbf{C}).

However, we encountered some abnormalities as can be seen in \Cref{fig:ffi_ess_all_comp}~\textbf{D} showing the $(\sigma,a)$ state space for one of the separatrices:
For $\sigma\approx\SI{3}{\ms}$ and $a>7$  one can recognize a purple region indicating that not all RS cells of the last group spiked.
Actually, spikes occurred for all RS cells in the simulated hardware network, but not all spikes were recorded because they were lost in the off-wafer communication network (\Cref{sec:communication_infrastructure}).
For very small $\sigma_0$ an additional effect can appear:
input bandwidth limitations can result in very dense pulse volleys not being propagated through the synfire chain, as can be seen e.g. for the blue point with $\sigma_0=\SI{0.02}{\ms}$ and $a_0=3.3$ in the left of \textbf{D}.
In that particular case the large majority of input spikes were lost in the off-wafer communication network so that they did not even reach the first synfire group.
We remark that this effect only appeared for $\sigma_0$ smaller than \SI{0.1}{\ms}.

\subsection{Self-sustained asynchronous irregular activity}
\label{model:ai}
Our third benchmark is a cortically inspired network with random,
distance-dependent connectivity which displays self-sustained
asynchronous and irregular firing (short: ``AI network'').
We define functionality measures on several levels of abstraction,
starting from single network observables such as the network firing rate,
the correlation coefficient and the coefficient of variation, the
properties of the power spectrum of the network activity, up to global
behavior such as the dependence of network dynamics on the internal synaptic weights $\Gi$
and $\Ge$.
We test two compensation strategies based on a mean field approach and
on iterative modification of individual neuron parameters.
While the first method offers a way to control the mean firing rate in
the presence of synapse loss, the second is applicable to 
synapse loss and fixed-pattern weight noise simultaneously, in contrast
to the other presented compensation methods.
Non-configurable axonal delays do not significantly affect the
network functionality because the intrinsic hardware delay is
approximately equal to the delay utilized in the model.
A scaling method for the network size is introduced and the effectivity of the second
compensation method was demonstrated using the ESS on a large network
with mapping-induced synapse loss and imposed fixed-pattern
synapse noise.

\subsubsection{Architecture}
\label{sec:ai-model-description}

\begin{figure}[t]
    \centering
    \includegraphics{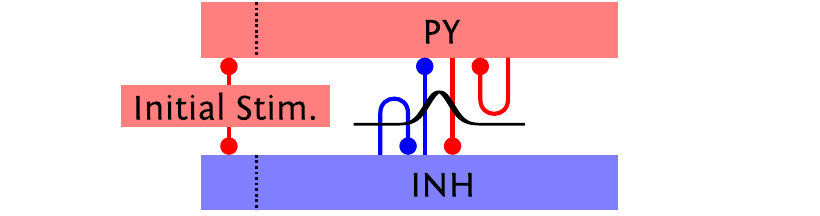}
    \caption{
		{\bf Schematic of the connectivity of the random cortical network.}
        Excitatory PY and inhibitory INH neurons are connected
        randomly with a spatial, Gaussian connection probability
        profile.
        The connection properties are given in \cref{sec:methods-ai}.
        A small part of the network is stimulated in the beginning of
        the experiment.
    \label{fig:ai_schematic}
    }
\end{figure}

Self-sustained states in spiking neural networks are known to be exquisitely sensitive to the correlation dynamics generated by recurrent activity \cite{kumar08highconductance,elboustani2009}.
Because of this sensitivity, a model of self-sustained activity within the asynchronous-irregular regime can provide a strong comparison between hardware and software platforms, by requiring the hardware network to reproduce the low firing, weakly correlated, and highly irregular dynamics of this state.
Notably, it is often observed that this activity regime provides a good match to the dynamics observed experimentally in the awake, activated cortex \cite{destexhe1999impact, brunel_jcns2000, destexhe03hcs}.
Additionally, one can note that the self-sustained activity regime provides an interesting test of the BrainScaleS hardware system, as in this state, the model network is not driven by external Poisson input, but has dynamics dominated by internally generated  noise \cite{destexhe2006}, beyond the initial brief Poisson stimulation to a small percentage of the network.

The self-sustained regime constitutes an attractor of a dynamical system \cite{amit97model}.
Networks based on this principle have been implemented in neuromorphic VLSI hardware \cite{giulioni2012}.

Here, we used a reduced model based on that published in \cite{destexhe09self}.
Neurons in the network followed the AdEx equations \labelcref{eqn:adex1,eqn:adex2,eqn:adex3} with parameters as in \cite{muller2012}, modeling regular spiking pyramidal cells (PY) with spike frequency adaptation \cite{connors90intrinsic} and fast spiking inhibitory cells (INH) with relatively little spike frequency adaptation.
Instead of explicitly modeling the thalamocortical or corticocortical networks, as in the previous work, we have chosen to modify the model, simplifying it to a single two-dimensional toroidal sheet and adding local connections and conduction delays.
The addition of local connectivity follows the experimental observation that horizontal connections in neocortex project, for the most part, to their immediate surroundings \cite{hellwig2000}, while the choice of linear conduction delays reflects electrophysiological estimates of conduction velocity in these unmyelinated horizontal fibers, in the range of 0.1 to \SI{0.5}{\metre\per\second} \cite{hirsch1991,murakoshi1993,bringuier1999, gonzalez-burgos2000, telfeian2003}.
Propagation delays are known to add richness to the spatiotemporal dynamics of neural network models \cite{roxin2005}, and in this case are observed to expand the region in the 2D space spanned by the excitatory and inhibitory conductances that supports self-sustained activity, albeit only slightly.

\Cref{fig:ai_schematic} shows a schematic of the AI network with its distance-dependent connectivity.
A small part of the neurons is stimulated at the beginning of the experiment.
Depending on its parameters, the network is able to sustain asynchronous irregular firing activity.
The details about the architecture and the parameters used are given in \cref{sec:methods-ai}.
 
\subsubsection{Functionality criteria}
\label{sec:ai-criteria}

\begin{figure*}[htb]
    \centering
	\includegraphics{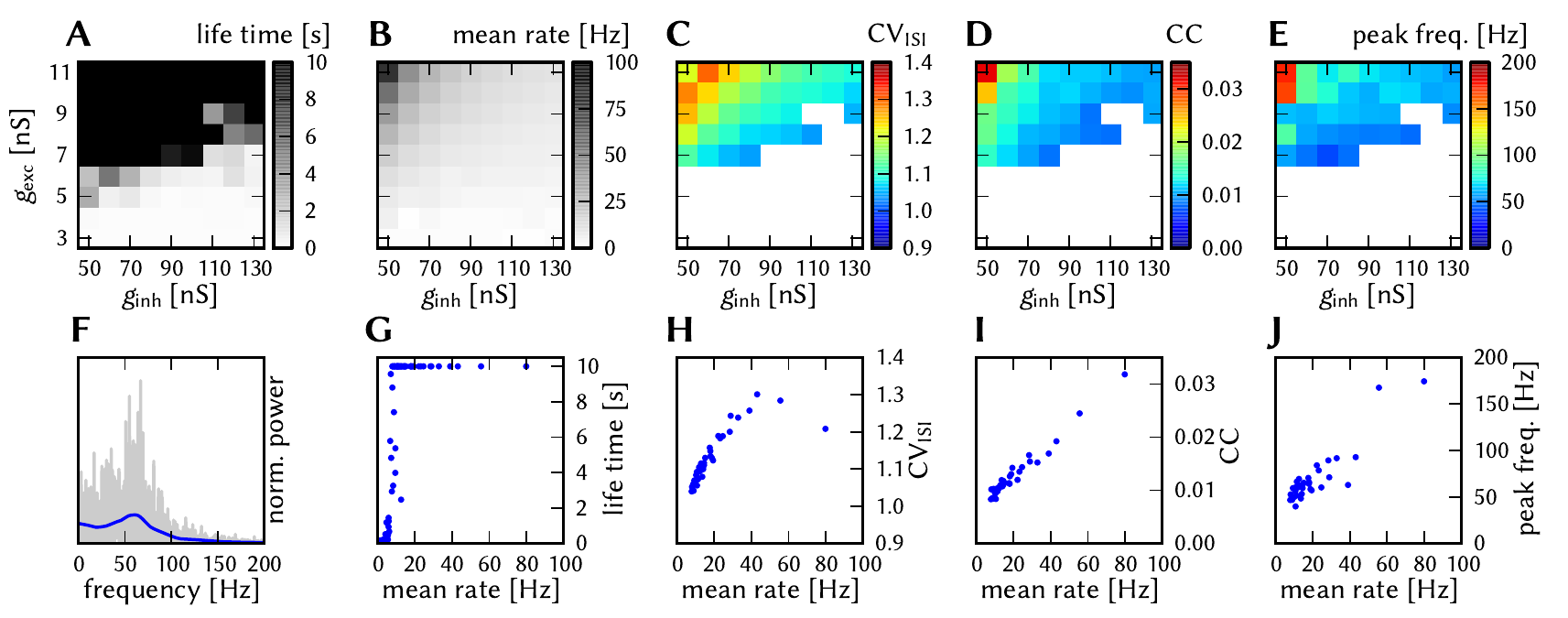}
    \caption{
		{\bf Behavior of the undistorted AI network.}
		On the top: survival time (\textbf{A}), 
		mean firing rate (\textbf{B}), 
		coefficient of variance $\cvisi$ (\textbf{C}), 
		coefficient of correlation CC (\textbf{D}) and 
		position of peak in power spectrum of global activity (\textbf{E}) in the parameter space for $\Ge$ and $\Gi$ for the default network with 3920 neurons without any distortions.
		(\textbf{F}) Power spectrum of the global pyramidal activity for the state ($\Ge=9\nano\siemens$, $\Gi=90\nano\siemens$).
		The population activity was binned with a time of $1\ms$, the raw spectrum is shown in gray, the blue curve shows a Gauss-filtered ($\sigma=\SI{5}{\Hz}$) version for better visualization.
		The position of the peak in the filtered version was used for (\textbf{E}).
		In (\textbf{G} - \textbf{J}) the dependence of single criteria on the mean firing rate is shown:
		survival time (\textbf{G}),
		$\cvisi$ (\textbf{H}),
		CC (\textbf{I}),
		position of peak in power spectrum (\textbf{J}).
		In the last three plots only surviving states of the ($\Ge$, $\Gi$) space were considered.
    \label{fig:ai-statespace-undistorted}
        }
\end{figure*}

The global functionality criterion for this network consists of the ability to sustain activity in an asynchronous and irregular activity regime.
The activity is considered self-sustained upon persistence to the end of the chosen simulation period.
The activity characteristics are quantified for the pyramidal cells using the mean and variance of the firing rates, the irregularity of individual spike trains ($\cvisi$, \cref{eq:cv-isi}), the synchrony via the correlation coefficient (CC, \cref{eq:cc}) and the power spectrum (see, e.g. 3.1.4 in \cite{rieke1997spikes}) of the excitatory activity.
The implementation details are given in \cref{sec:methods-ai-criteria}.

These criteria were evaluated for a range of excitatory and inhibitory synaptic weights $\Ge$ and $\Gi$ for the default network consisting of 3920 neurons.
\Cref{fig:ai-statespace-undistorted} (\textbf{A}) shows the region in the ($\Ge$, $\Gi$) parameter space that allows self-sustained activity, which is achieved at pyramidal firing rates above \SI{8}{\Hz}~(\textbf{G}).

The coefficient of variation of the firing rates across neurons ($\cvrate$)
is small ($<0.2$, see the \SI{0}{\%} weight noise data in \cref{fig:ai-weight_distortion} \textbf{B}),
as all neurons have identical numbers of afferent synapses with identical weights in each network realization.
In addition to the parameter space plots in the top row of \cref{fig:ai-statespace-undistorted}, we plot the other criteria against the mean firing rate in the bottom row and
recognize the latter as the principal property of each state that mostly determines all other criteria.

The activity is irregular ($\cvisi > 1$) across all states (\textbf{C}) and is mainly determined by the network firing rate: the $\cvisi$ first increases with the firing rate, then saturates and decreases for rates higher than \SI{50}{\Hz}~(\textbf{H}).
Over the entire parameter space, the spike trains of the pyramidal cells are only weakly correlated, with a CC between 0.01 and 0.03.

The average CC increases with the firing rate, which can be attributed to local areas in which neurons synchronize over short time periods.
At last, we look at the power spectrum of the global pyramidal activity,
exemplarily for the (\SI{9}{\nano\siemens}, \SI{90}{\nano\siemens}) state in (\textbf{F}). 
 As a comparison for further studies we follow \cite{brunel_jcns2000} and use the position of the non-zero peak in the power-spectrum, which is shown for each ($\Ge$, $\Gi$) point (\textbf{E}) and as a function of the firing rate (\textbf{J}): The position of the power spectrum peak frequency (\cref{sec:methods-ai-criteria}) increases linearly with the mean firing rate.

\subsubsection{Non-configurable axonal delays}
\label{sec:ai-delay-distortion}
For the analysis of the effects of non-configurable delays we repeated the ($\Ge$, $\Gi$) sweep with all axonal delays set to \SI{1.5}{\ms}, cf.\ \cref{sec:investigated_distortions}.
This distortion mechanism did not affect any of the functionality criteria, as each neuron still received synaptic input comparable to the reference case.
One might expect an influence on the power spectrum of global activity as we switched from distance-dependent delays to a globally constant delay of \SI{1.5}{\ms} as it changes the temporal correlation of the effect of a neuron on all of its efferents.
In fact, the power spectra did not change significantly, which can be explained as follows:
In the reference case, the \emph{average} of all distance-dependent delays in the network amounts to \SI{1.55}{\ms} (cf. \cref{fig:ai-delay-histogram}), which is close to the constant delay value of \SI{1.5}{\ms} we use to model the non-configurable delays on the hardware.
In this particular case, the hardware delay matches the average delay in the network such that no distortion is introduced.
Accordingly, parameter space sweeps on the ESS yielded the same results.

In \cref{sec:methods_ai_delay_sims} we provide further simulations on the influence of the distribution of delays on the behavior of the network, showing that 
the effect of the distance-dependent delays is small and that it is mostly the average delay which matters.
In our case, this delay exactly corresponds to the average delays on the wafer when running at a speedup of \num{10000} compared to biological real-time, such that there is no need for a compensation here.

We note that for variants of this benchmark, where the average network delay is higher or lower than \SI{1.5}{\ms}, there exists a simple but effective compensation strategy by just modifying the speedup of the emulation on the hardware, such that the average network delay is directly mapped onto the hardware delay.
We can assume a modified experiment where the average delay amounts to \SI{3}{\ms}.
By choosing a speedup of \num{20000}, this delay can be directly mapped to the \SI{150}{\ns} average delay on the hardware.
Such a change of emulation speed is not arbitrary, as one has to make sure that the neural dynamics can still be emulated at the chosen speed (cf. supported parameter ranges in \cref{table:hardware_parameter_ranges}).
Furthermore, the reduced bandwidth for the pulse communication, especially for external stimulation, must be considered.
While this is no issue for this self-sustaining kind of network, these conditions must be also fulfilled for potential other networks that are interconnected to the AI network.

\subsubsection{Synaptic weight noise}
\begin{figure*}[!htb]
    \centering
	\includegraphics{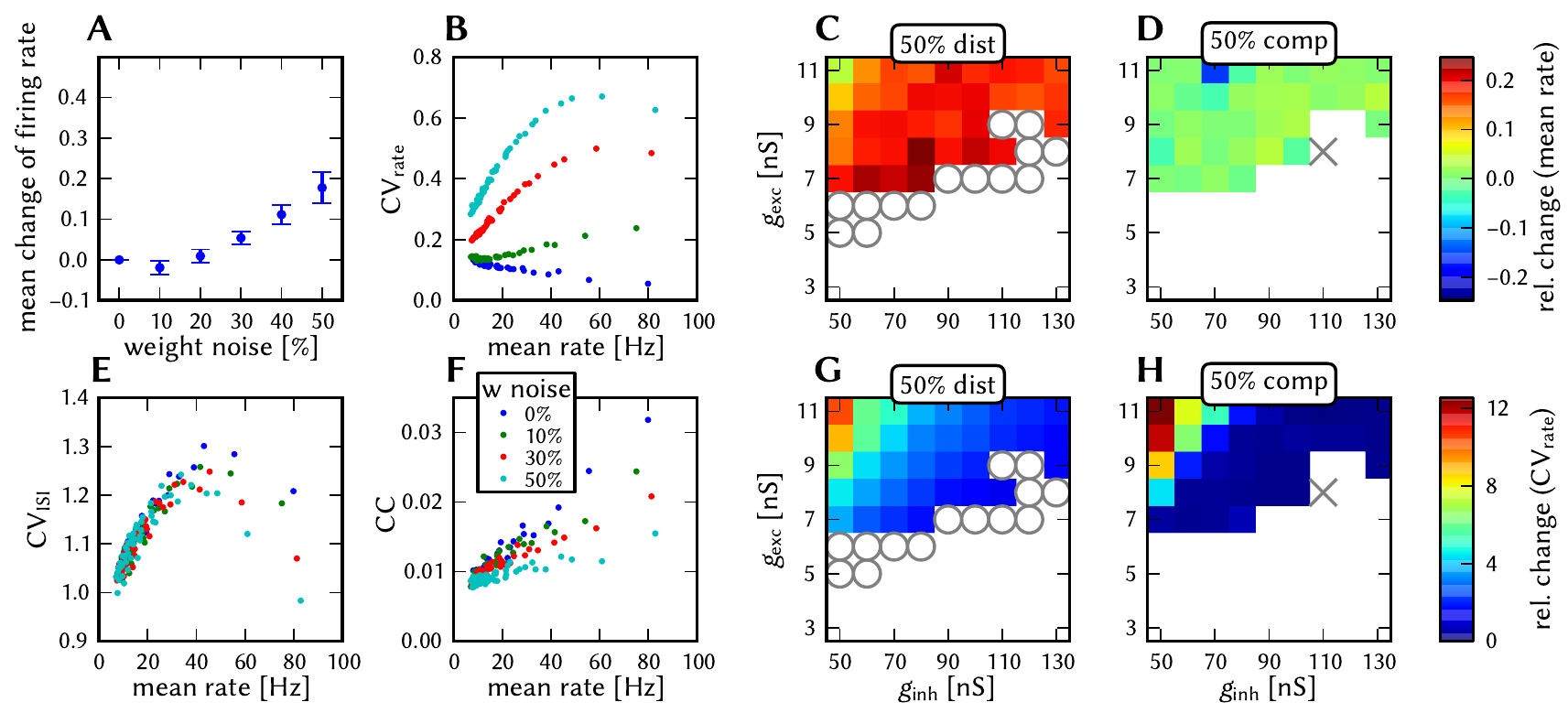}
    \caption{
		{\bf Effect and compensation of synapse weight noise in the AI network:}
		(\textbf{A}) Relative change of the firing rate with respect to the undistorted network averaged over all sustained states for varying synapse weight noise.
		(\textbf{B}) $\cvrate$ as a function of mean rate for every survived state for varying synapse weight noise.
		(\textbf{C} and \textbf{D}) Relative change of the firing rate with respect to the undistorted for each state for \SI{50}{\%} synapse weight noise(\textbf{C}) and compensated (\textbf{D}).
		(\textbf{E}) $\cvisi$ as a function of mean rate for varying synapse weight noise.
		(\textbf{F}) CC as a function of mean rate for varying synapse weight noise.
		(\textbf{G} and \textbf{H}) Relative change of $\cvrate$ with respect to the undistorted for each state for \SI{50}{\%} synapse weight noise(\textbf{G}) and compensated (\textbf{H}).
		In (\textbf{C} and \textbf{D}) and  (\textbf{G} and \textbf{H}): A cross marks a state that was sustained in the undistorted but not sustained in the compared case. A circle marks a state that was not sustained in the original but sustained in the compared case.
    \label{fig:ai-weight_distortion}
        }
\end{figure*}
The effects of synaptic weight noise between \SI{10}{\%} and \SI{50}{\%} (cf. \Cref{sec:dist-weights}) on the AI network are shown in \Cref{fig:ai-weight_distortion}:
The region of self-sustained states in the ($\Ge$, $\Gi$) space is increased by this distortion mechanism, cf. the circles in (\textbf{C}) marking states that survived with \SI{50}{\%} synaptic weight noise but not in the undistorted case.
The firing rate increases with the degree of noise~(\textbf{A}): the change is the stronger the lower $\Ge$ and diminishes for states with an already high firing rate in the undistorted case~(\textbf{C}).
Synaptic weight noise leads to an increase of the variation of firing rates ($\cvrate$), with the change being stronger for high population firing rates~(\textbf{B}).
The $\cvisi$ as a function of firing rates remains unchanged for low rates, but decreases for higher firing rates in proportion to the noise level~(\textbf{E}).
Furthermore, weight noise introduces randomness into the network, thereby reducing synchrony: The pairwise correlation between neurons decreases linearly with the amount of weight noise~(\textbf{F}).
The power spectrum of the global activity is not affected by this distortion mechanism.

\subsubsection{Synapse loss}
\begin{figure*}[!htb]
    \centering
	\includegraphics{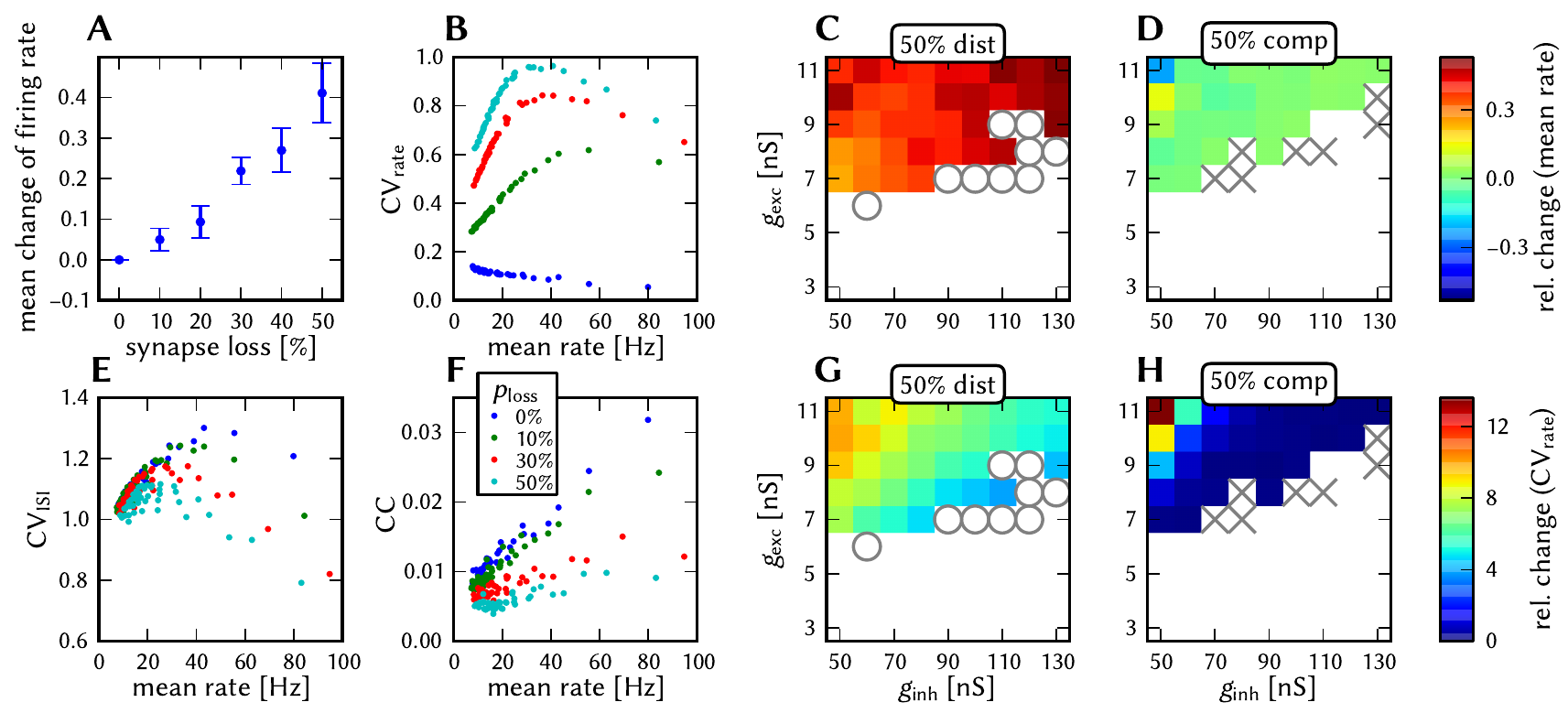}
    \caption{
		{\bf Effect and compensation of synapse loss in the AI network:}
		(\textbf{A}) Relative change of the firing rate with respect to the undistorted network averaged over all sustained states for varying synapse loss.
		(\textbf{B}) $\cvrate$ as a function of mean rate for every survived state for varying synapse loss.
		(\textbf{C}, \textbf{D}) Relative change of the firing rate with respect to the undistorted case for each state for \SI{50}{\%} synapse loss (\textbf{C}) and compensated (\textbf{D}).
		(\textbf{E}) $\cvisi$ as a function of mean rate for varying synapse loss.
		(\textbf{F}) CC as a function of mean rate for varying synapse loss.
		(\textbf{G}, \textbf{H}) Relative change of $\cvrate$ with respect to the undistorted case for each state for \SI{50}{\%} synapse loss (\textbf{G}) and compensated (\textbf{H}).
		In \textbf{C}, \textbf{D}, \textbf{G} and \textbf{H}: A cross marks a state that was sustained in the undistorted but not sustained in the compared case. A circle marks a state, that was not sustained in the original but sustained in the compared case.
    \label{fig:ai-loss_distortion2}
        }
\end{figure*}

Synapse loss has a similar influence on the network behavior as synaptic weight noise:
\Cref{fig:ai-loss_distortion2} shows the results of the $\Ge$-$\Gi$ sweeps for synapse loss values between \SI{10}{\%} and {50}{\%} (cf. \Cref{dist:synapseloss}).
The region of sustained states increases with synapse loss but not as strongly as for weight noise (\textbf{C}).
The firing rate increases with synapse loss (\textbf{A}):
Compared to the change caused by synaptic weight noise, however, the effect is much stronger for synapse loss.
The same holds for the variance of the firing rates across the pyramidal neurons, which again increases with synapse loss, as can be seen in (\textbf{B}).
Note that the $\cvrate$ first increases with the mean rate, then reaches a maximum and finally saddles for high rates.
We remark that for high synapse loss, some neurons did not fire at all.
Both the irregularity and the correlation of firing decrease with increasing synapse loss, leaving the network still in an asynchronous irregular state (\textbf{E} and \textbf{F}).
Synapse loss shows no effect on the power spectrum of global pyramidal activity.

\subsubsection{Compensation strategies}
\label{sec:ai_comp_strat}
The hardware-induced distortions on the AI network analyzed in the previous sections leave two major criteria that need to be recovered:
The population firing rate and the variation of firing rates across the population.
We consider the other effects (change of CC, $\cvisi$, peak frequency in power spectrum) as minor because they are mainly determined by the mean rate and discard them in the following.

One apparent approach for recovering the original firing rate is to change the strengths of the synaptic weights $\Ge$ and $\Gi$.
Considering the conducted ($\Ge, \Gi$) parameter space sweeps, we could simply select the distorted state that best matches the criteria of the undistorted reference.
However, this method requires to scan $\Ge$ and $\Gi$ over a wide range to finally get to the desired result.
Preferably, one wants to have a compensation method that can be applied to a single experiment and works without huge parameter sweeps.

\paragraph{Mean field compensation for rate change}
~
\begin{figure}[ht]
    \includegraphics{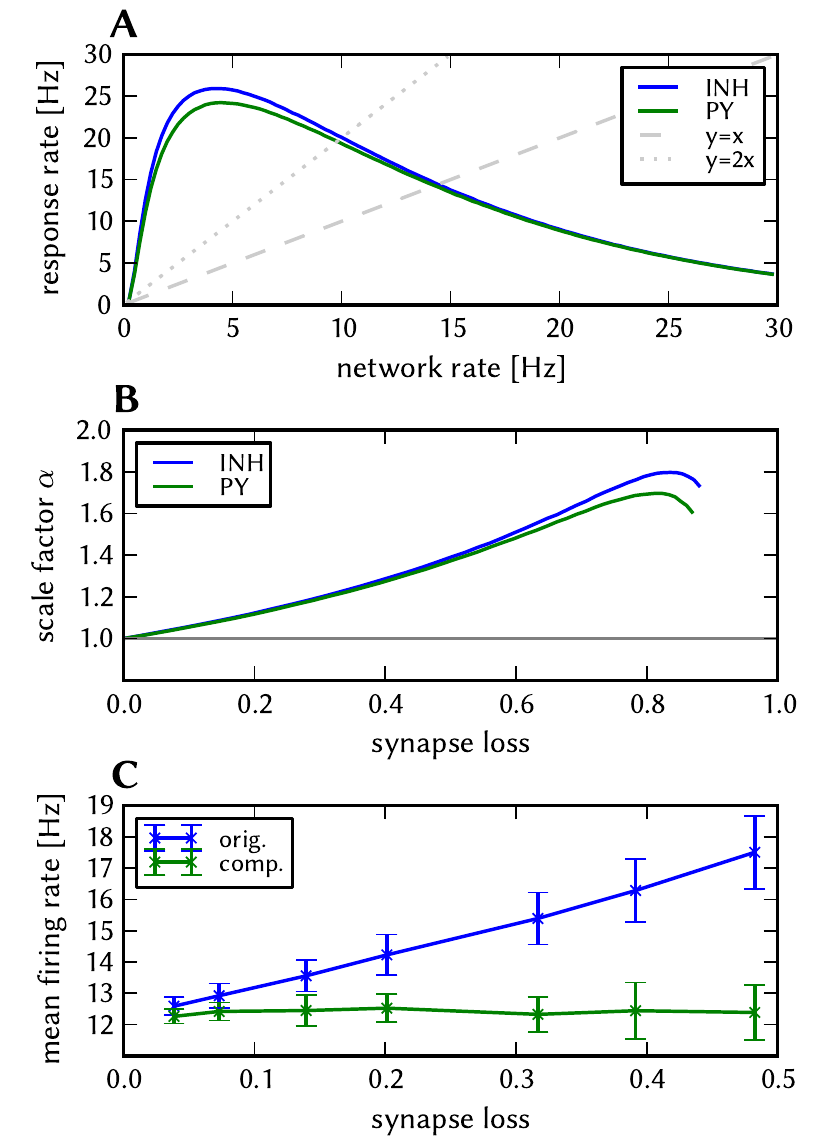}
    \caption{
		{\bf Mean-field-based compensation method for the AI network.}
        (\textbf{A}) Mean firing rate of a single PY and INH neuron given a poisson stimulus by the external network with a given rate.
        (\textbf{B}) Compensation factor $\alpha$ calculated from the data in \textbf{A}.
        (\textbf{C}) Compensation applied to the self-sustained network (with parameters $\Gi = \SI{90}{\nano\siemens}$, $\Ge = \SI{9}{\nano\siemens}$).
		The error bars denote the standard deviation of mean firing rates across all neurons.
        ``orig.'' marks the original network without compensation, in
        ``comp.'' the neuron parameters were modified according to the compensation factor.
        The scaling of internal delays had only minimal effect on the firing rate (not shown)
    \label{fig:mean_firing_rate}
	}
\end{figure}
The mean firing rate in the network rises with an increasing synapse
loss value.
This effect can be understood using a mean-field approach (see, e.g. \cite{kumar08highconductance}) in which the response rate of a single
neuron's firing rate is assumed to be a function of the mean network
firing rate.

\begin{align}
    \nu_i = f\left(\nu_{\mathrm{in,exc}}, \nu_{\mathrm{in,inh}}\right)
\end{align}
With this ansatz, which is similar to the approach in
\cite{brunel_jcns2000} where the afferent neurons are replaced by
independent Poisson processes with equal instantaneous rate in a
sparse random network, the mean firing rate in a self-sustained state
can be calculated as a stable, self-consistent solution of the
gain function being equal to the firing rate of a single neuron:

\begin{align}
    \hat \nu (p_\mathrm{loss}) = f\left(N_{\mathrm{exc}} (1 - p_\mathrm{loss}) \hat \nu, N_{\mathrm{inh}} (1 - p_\mathrm{loss}) \hat \nu \right)
	\label{eq:ai-stability-condition}
\end{align}
Here, $N_{\mathrm{exc}}$ and $N_{\mathrm{inh}}$ are the number of
pre-synaptic connections of a given neuron, and $p_\mathrm{loss}$ is the modeled
synapse loss value.
\Cref{fig:mean_firing_rate} \textbf{A} shows the gain function (right-hand side of \cref{eq:ai-stability-condition})
of PY and INH neurons for $p_\mathrm{loss}=0$ yielding the stable solution $\tilde \nu(0)\approx\SI{14}{\Hz}$ as the intersection of the $y=x$ diagonal and the gain function.
Analogously, the solution for $p_\mathrm{loss}=0.5$ can be determined as the intersection with the $y=2x$ line (considering $\nu_{\mathrm{in}}(p_\mathrm{loss}) = p_\mathrm{loss}\cdot\nu_{\mathrm{in}}$).
The result
justifies the assumption of the mean firing rate of inhibitory and
excitatory neurons being equal for $p_\mathrm{loss} < 0.5$.

The parameter change that is necessary to restore the original mean
firing rate can be calculated using the following relationship for the
time scaling of the solution of a differential equation:

\begin{align}
    \dot {\mathbf{x}}(t) &= \mathbf{F}(\mathbf{x}, t) \\
    \mathbf{y}(t) &:= \mathbf{x}(\alpha t) \\
    \dot {\mathbf{y}}(t) &= \alpha \dot {\mathbf{x}}(\alpha t) = \alpha {\mathbf{F}}(\mathbf{y}(t), \alpha t) \\
    &\Rightarrow \tilde {\mathbf{F}}(\mathbf{x}, t) := \alpha \mathbf{F}(\mathbf{x}, \alpha t)
\end{align}
Assuming that $\mathbf{x}$ is the state of the dynamic variables within a
network, $\mathbf{y}$ describes a network which follows the same time
dependence with the dynamics scaled by the factor $\alpha$ in time.
As the given random cortical network shows self-sustained behavior,
the transition from $\mathbf{F}$ to $\mathbf{\tilde F}$ requires only the modification
of internal network parameters, because there is no external input (which
would have to also be modified otherwise).
In particular, the transition encompasses scaling $\taum$,
$\tausyn_{x}$, $\tauref$, $\tau_{\textrm{w}}$ and the
synaptic delays by $\alpha$, while leaving the conductance jump after
each presynaptic PSP unchanged.
$\alpha$ is calculated from the measured gain function (cf.\ \cref{fig:mean_firing_rate}) via 

\begin{align}
    \alpha = \frac {\tilde \nu(p_\mathrm{loss})} {\tilde \nu(0)}
\end{align}

The resulting firing rate with and without compensation is shown in
\cref{fig:mean_firing_rate} \textbf{C}.
The results also show that the variance of the
firing rates across neurons
grows with rising synapse loss due to the increasing difference
in connectivity within the networks.
An extension of the mean-field-based compensation to this kind of
inhomogeneous connectivity would be impractical, as it requires
knowledge of the actual network realization (which is available only
after the mapping step) and the measurement
of \Cref{fig:mean_firing_rate} \textbf{A} for all occuring counts of
presynaptic inhibitory and excitatory neurons.
Thus, a different method is considered in \cref{sec:iterative_compensation}.

In conclusion, this method can be applied when the actual synapse
loss value and the mean response function of a single neuron is known.
It only depends on the single neuron response properties; the
amount of synapse loss has to be known a priori, but not the complete
network dynamics.
The method depends on the ability to modify synaptic delays according
to the scaling rule.
However, for the given network, this scaling has only a minimal effect
on the mean firing rate.

\paragraph{Iterative compensation}
\label{sec:iterative_compensation}
The iterative compensation method aims at reducing two distortion effects: the change of the mean firing rate of the pyramidal neurons and its variance across neurons, which are both apparent for synapse loss and synaptic weight noise.
It relies on the controlability of the hardware neuron parameters allowing to fine tune the AdEx parameters for every individual neuron (\Cref{sec:hicann}).
The iterative compensation functions as follows: We start with the results of the reference and the distorted network.
From the reference simulation we extract the target mean rate $\nu^\mathrm{tgt}$ of the neurons in a population.
For each neuron in the distorted network, we compare its actual firing rate against $\nu^\mathrm{tgt}$, and modify the excitability of the neuron in proportion to the difference between target and measured firing rate.
The distorted network with modified neuron parameters is then simulated and the output is compared again to the reference network.
This iterative compensation step is repeated until the characteristics of the last step approximately match those of the reference simulation. 
In our simulations, we modified the spike initiation threshold $\Vthresh$, with its change $\Delta \Vthresh=c_\mathrm{comp}(\nu^\mathrm{tgt}-\nu^\mathrm{act})$ being proportional to the difference between the actual and the target rate.
We found that, when choosing the compensation factor $c_\mathrm{comp}$ appropriately, 10 iterations are sufficient to restore the mean and variance of the firing rates in the undistorted network.
While the compensated mean rate exactly corresponds to $\nu^\mathrm{tgt}$, the compensated $\cvrate$ is higher than in the reference network, but reliably below the $1.2$-fold of the reference value.
The iterative compensation applied in the following is described in detail in \cref{sec:methods-ai-iterative-comp}.
We remark that the proposed iterative compensation requires a controllable, deterministic mapping, which guarantees that in each iteration the neurons and synapses are always mapped onto the same hardware elements.
Furthermore, the complete compensation process needs to be repeated for each network instance.
In fact, we perform a calibration of the apparent permanent causes of distortion (fixed-pattern noise and synapse loss) similar to \cite{pfeil2013} in order to reduce their effects.
Hence, whenever we change the random seed that is used to generate the probabilistic connectivity between the neurons, the iterative compensation needs to be run anew.
Thus, a reference from a non-distorted simulation or, e.g., from theory is needed.
However, once obtained, the result of the compensation can be used for long-running simulations or as part of a larger compound network.

\subsubsection{Results of iterative compensation}
\paragraph{Synaptic weight noise}
In order to verify the iterative compensation strategy we applied it to the distorted parameter space with \SI{50}{\%} synaptic weight noise.
Note that, here and in \cref{sec:ai-ess}, weight noise was implemented persistently, being always the same in all iterations, representing the case where fixed-pattern noise, and not trial-to-trial variability, determines the synaptic weight noise (cf. \cref{sec:dist-weights}).
Accordingly, the following findings are not applicable to the opposite case.
The results of the iterative compensation are shown in \cref{fig:ai-weight_distortion},
which displays the relative difference of the mean and variance
of the firing rates with respect to the reference simulation in \textbf{D} and \textbf{H}.
The region of sustained activity in the ($\Ge$, $\Gi$) parameter space of the compensated network matches the one of the reference simulation very well.
The mean and variance of firing rates could be successfully recovered for most of the states; with the exception of states with a mean rate higher than \SI{25}{\Hz},
where both criteria still differ notably from the reference after 10 iterations (upper left regions in the parameter spaces).
We expect that the performance of the iterative compensation for those states could be further improved by tuning the compensation factor $c_\mathrm{comp}$ (\Cref{sec:methods-ai-iterative-comp}) for high firing rates.
The other criteria such as $\cvisi$ and peak frequency could be fully recovered, following the assumption made earlier, that those criteria mainly depend on the firing rate.
However, the coefficient of pairwise cross-correlation (CC) of the compensated networks is lower than in the reference simulation, i.e., the randomness introduced by the synaptic weight noise is still effective.

\paragraph{Synapse loss}
The results of the application of the iterative compensation strategy to the ($\Ge$, $\Gi$) parameter space with \SI{50}{\%} synapse loss are shown in \cref{fig:ai-loss_distortion2} (\textbf{D} and \textbf{H}), displaying
the relative difference of mean and variance of firing rates.
The compensation was not as effective as for synaptic weight noise:
Some states with a low base firing rate were unstable (marked with a cross), i.e. the network did not survive until the end of simulation.
As before, the mean and variance of firing rates can be successfully restored for low and medium base firing rates.
Again, for high firing rates, the iterative compensation only performed moderately (upper left regions in the parameter spaces \textbf{D} and \textbf{H}).
The other criteria show the same behavior as in the weight noise compensation, i.e. the peak frequency and $\cvisi$ are in good match with the reference while the pairwise correlation (CC) decreased due to the randomness introduced by the synapse loss.
We repeated the iterative compensation for the parameter space with \SI{30}{\%} synapse loss: The results (not shown) are comparable to the \SI{50}{\%} case, but exhibit fewer unstable states, i.e., there were more combinations of $\Ge$ and $\Gi$ whose compensated network survived. 

\paragraph{Conclusion}
We conclude that the iterative compensation of distorted networks works for both synapse loss and fixed-pattern synaptic weight noise.
The compensation also works when both are present at the same time, see \cref{sec:methods_ai_further_sims} for details.
While there seems to be no limit for weight noise, compensation of synapse-loss induced distortions is only possible up to a certain degree, as the network tends to become less stable with fewer synapses involved.

\subsubsection{Full simulation of combined distortion mechanisms}
\label{sec:ai-ess}
\begin{figure}[htb]
    \centering		
	\includegraphics{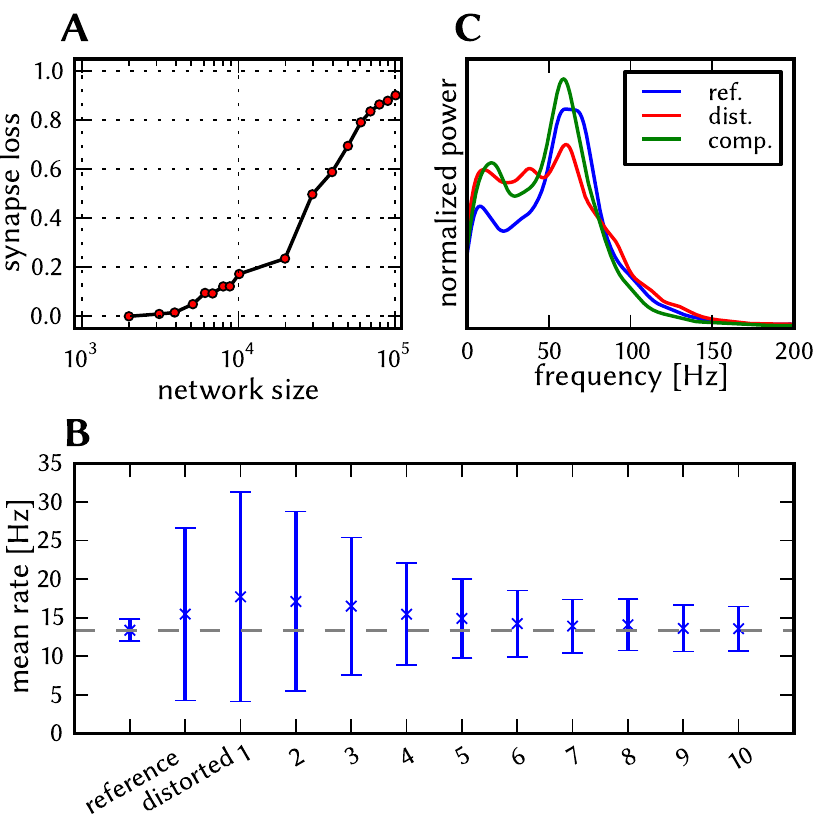}
    \caption{
	{\bf AI network on the ESS:}
	(\textbf{A}) Synapse loss after mapping the network with different sizes onto the BrainScaleS system
	(\textbf{B}) Iterative compensation of the large-scale network with \num{22445} neurons on the ESS: evolution of mean and standard deviation of firing rates for 10 iterations
	(\textbf{C}) Gauss-filtered power spectrum of global activity of the pyramidal neurons in the large-scale network. Reference spectrum shown in blue (simulated with Nest), distorted and compensated spectra in red resp. green, both simulated with the ESS. 
    \label{fig:ai-ess}
	}
\end{figure}
In a last step the iterative compensation method designed for the AI network was tested in ESS simulations.
Like for the other two models we forced distortions to test the developed compensation strategies.
Therefore, we scaled up the network such that a significant fraction of synapses was lost during the mapping process.
This large-scale network was then emulated on the ESS and compared to the undistorted reference simulation with NEST.
Afterwards, we applied the compensation strategy developed in the previous section to restore the original behavior of the AI network.

\paragraph{Synapse loss}
Mapping such homogeneous networks that lack any modularity represents the worst-case scenario for the mapping process, as they have little room for optimization.
In \cref{fig:ai-ess}~\textbf{A} the relative synapse loss is plotted for various network sizes using the scaling method described in \cref{sec:methods-ai-scaling}.
One can see that already for low numbers of neurons some synapse loss occurs, although there are sufficient hardware synapses and synapse drivers: due to the sparseness of the on-wafer routing switches some routing buses don't find a free switch to connect to its respective target HICANNs, such that synapses are lost.
A kink in the graph
of the synapse loss can be seen at around \num{20000} neurons, where at least 64 neurons are mapped onto one HICANN (cf. \cref{table:size_and_connectivity}). 
In such a network with random connectivity it is merely possible to find 64 neurons whose pool of pre-synaptic neurons is smaller than \num{14336}, which is the maximum number of pre-synaptic neurons per HICANN, such that synapse loss must occur.
Recall that there is a maximum of \num{14336} pre-synaptic neurons for all neurons mapped onto one HICANN.
As the connectivity in the AI network is probabilistic, the chance to find groups of 64 neurons whose pool of pre-synaptic neurons is smaller than \num{14336} is close to zero.

\paragraph{Large-scale network}
In order to produce a demanding scenario, we scaled the model to a size of \num{22445} neurons (\Cref{sec:methods-ai-scaling}).
The size was chosen such that the network almost occupies an entire wafer, while mapping up to 64 neurons onto one HICANN.
This large-scale network has a total of approximately 5.6 million synapses.
The statistics of the reference simulation can be found in \cref{table:ai-ess-results} and are in accordance with the scaling behavior investigated in the Supplement, \cref{sec:methods_ai_scaling_sims}.

\paragraph{Distorted network}
In the above scenario, \SI{28.1}{\%} of synapses were lost during the mapping process (for projection-wise numbers see \cref{table:ai-ess-synloss}).
We remark that the synapse loss at this size is higher than during the synapse loss sweep in \cref{fig:ai-ess}~\textbf{A}, as
 we used a sequence of mapping algorithms that guarantees a balance between synapse loss of excitatory and inhibitory connections.
Still, there were slightly more inhibitory connections lost than excitatory ones (\Cref{table:ai-ess-synloss}).
Additionally, we applied a fixed-pattern noise of \SI{20}{\%} to the synaptic weights in the ESS simulation.
The result of the latter can be found in \cref{table:ai-ess-results}: the network still survived until the end of the simulation, but the firing rate and its variance increased compared to the reference simulation, which complies with the prediction of the distortion analysis.

\begin{table}
	\caption{{\bf Statistics of the large-scale AI network}}
    \centering
\begin{tabular}{lccc}
\toprule
\textbf{criteria} & \textbf{ref.} & \textbf{dist.} & \textbf{comp.} \\
\midrule
	Rate [\Hz] &13.4 & 15.5 & 13.6\\
	$\cvrate$ &0.107 & 0.726 & 0.212\\
	$\cvisi$  &1.12 & 1.11 & 1.09\\
	CC &0.00103 & 0.0011 & 0.00166\\
	Peak Frequency[\Hz] &60.3 & 60.7 & 59.0\\
\bottomrule
\end{tabular}
    \flushleft{Reference (ref.) simulated with NEST, distorted (dist.) and compensated (comp.) with the ESS.}
    \label{table:ai-ess-results}
\end{table}

\begin{table}
	\caption{{\bf Projection-wise synapse loss of the large-scale AI network after the mapping process.}}
	\centering
	\begin{tabular}{lccc}
		\toprule
		\textbf{projection} & \textbf{synapse loss [\%]} \\
		\midrule
		PY $\rightarrow$ PY & 26.9 \\
		PY $\rightarrow$ INH & 28.1 \\
		INH $\rightarrow$ PY & 31.1 \\
		INH $\rightarrow$ INH & 33.4 \\
		STIM $\rightarrow$ PY & 77.5 \\
		STIM $\rightarrow$ INH & 89.4 \\
		\midrule
		total & 28.1 \\
		\bottomrule
	\end{tabular}
	\flushleft{PY: excitatory pyramidal neurons. INH: fast spiking inhibitory cells, STIM: external Poisson sources for initial stimulation}
    \label{table:ai-ess-synloss}
\end{table}

\paragraph{Compensated network}
We then used the iterative compensation method from \cref{sec:ai_comp_strat} to compensate the abovementioned distortions and repeated the ESS simulation with the modified network.
The evolution of the firing rates over 10 iterations is shown in \cref{fig:ai-ess}~\textbf{B}:
One can clearly see how, step by step, the firing rate approaches the target rate and that at the same time the variance of firing rates decreases.
The statistics of the final iteration are listed in \cref{table:ai-ess-results}:
It was possible to fully recover the target mean rate.
The variation of firing across neurons ($\cvrate$) was significantly reduced from $0.726$ to $0.212$ but was still twice as large as in the reference network.
The other functionality criteria match the reference simulation very well (\Cref{table:ai-ess-results}), as does the power spectrum of global activity in \cref{fig:ai-ess}~\textbf{C}.

\section{Conclusions}
\label{sec:conclusions}

In this study, we have presented a systematic comparison between neural network simulations carried out with ideal software models and a specific implementation of a neuromorphic computing system.
The results for the neuromorphic system were obtained with a detailed simulation of the hardware architecture.
The core concept is, essentially, a functionalist one: neural networks are defined in terms of functional measures on muliple scales, from individual neuron behavior up to network dynamics.
The various neuron and synapse parameters are then tuned to achieve the target performance in terms of these measures.

The comparison was based on three cortically inspired benchmark networks: a layer 2/3 columnar architecture, a model of a synfire chain with feed-forward inhibition and a random network with self-sustained, irregular firing activity.
We have chosen these specific network architectures for two reasons.
First of all, they implement very different, but widely acknowledged computational paradigms and activity regimes found in neocortex: winner-take-all modules, spike-correlation-based computation, self-sustained activity and asynchronous irregular firing.
Secondly, due to their diverse properties and structure, they pose an array of challenges for their hardware emulation, being affected differently by the studied hardware-specific distortion mechanisms.

All three networks were exposed to the same set of hardware constraints and a detailed comparison with the ideal software model was carried out.
The agreement was quantified by looking at several chosen microscopic and and macroscopic observables on both the cell and network level, which we dubbed ``functionality criteria''.
These criteria were chosen individually for each network and were aimed at covering all of the relevant aspects discussed in the original studies of the chosen models.

Several hardware constraint categories have been studied: the dynamics of the embedded neuron and synapse models, limited parameter ranges, synapse loss due to limited hardware resources, synaptic weight noise due to fixed-pattern and trial-to-trial variations, and the lack of configurable axonal delays.
The final three effects were studied in most detail, as they are expected to affect essentially every hardware-emulated model.
The investigated distortion mechanisms were studied both individually, as well as combined, similarly to the way they would occur on a real hardware substrate.
As expected, above certain magnitudes of the hardware-specific distortion mechanisms, substantial deviations of the functionality criteria were observed.

For each of the three network models and for each type of distortion mechanism, several compensation strategies were discussed, with the goal of tuning the hardware implementation towards maximum agreement with the ideal software model.
With the proposed compensation strategies, we have shown that it is possible to considerably reduce, and in some cases even eliminate the effects of the hardware-induced distortions.
We therefore regard this study as an exemplary workflow and a toolbox for neuromorphic modelers, from which they can pick the most suitable strategy and eventually tune it towards their particular needs.

In addition to the investigated mechanisms, several other sources of distortions are routinely observed on neuromorphic hardware.
A (certainly not exhaustive) list might include mismatch of neuron and synapse parameters, shared parameter values (i.e., not individually configurable for each neuron or synapse) or limited parameter programming resolution.
These mechanisms are highly back-end-specific and therefore difficult to generalize.
However, although they are likely to pose individual challenges by themselves, some of their ultimate effects on the target network functionality can be alleviated with the compensation strategies proposed here.

Our proposed strategies aim at neuromorphic implementations that compete in terms of network functionality with conventional computers but offer major potential advantages in terms of power comsumption, simulation speed and fault tolerance of the used hardware components.
If implemented successfully, such neuromorphic systems would serve as fast and efficient simulation engines for computational neuroscience.
Their potential advantages would then more than make up for the overhead imposed by the requirement of compensation.

From this point of view, hardware-induced distortions are considered a nuisance, as they hinder precise and reproducible computation.
In an alternative approach, one might consider the performance of the system itself at some computational task as the ``fitness function'' to be maximized.
In this context, some particular architecture of an embedded model, together with an associated target behavior, would then become less relevant.
Instead, one would design the network structure specifically for the neuromorphic substrate or include training algorithms that are suitable for such an inherently imperfect back-end.
The use of particular, ``ideal'' software models as benchmarks might then given up altogether in favor of a more hardware-oriented, stand-alone approach.
Here, too, the proposed compensation strategies can be actively embedded in the design of the models or their training algorithms.

The hardware architecture used for our studies is, indeed, suited for both approaches.
It will be an important aspect of future research with neuromorphic systems to develop procedures that tolerate or even actively embrace the temporal and spatial imperfections inherent to all electronic circuits.
These questions need to be addressed by both model and hardware developers, in a common effort to determine which architectural aspects are important for the studied computational problems, both from a biological and a machine learning perspective.






\section*{Data Availability}
The authors confirm that all data underlying the findings are fully available without restriction.
The three benchmark models, the performed simulations, as well as the analysis and compensation
methods are fully described in the manuscript and the supporting information.
For the original L2/3 network with detailed neuron and synapse models, we provide the complete simulation data at:\\
http://brainscales.kip.uni-heidelberg.de/\\ largePublicContent/plos\textunderscore one\textunderscore 2014\textunderscore fit\textunderscore data.tar.gz \\
The executable system specification of the BrainScaleS wafer-scale
neuromorphic hardware as used for the simulations in this article is provided on a Linux
live-system available at:\\
http://brainscales.kip.uni-heidelberg.de/\\ largePublicContent/plos\textunderscore one\textunderscore 2014\textunderscore ess\textunderscore live\textunderscore system.iso
\section*{Acknowledgments}
We would like to thank Eric M\"uller for his invaluable help with the software infrastructure; Jens Kremkow for his support with the synfire chain model; Mitja Kleider, Christoph Koke, Dominik Schmidt and Sebastian Schmitt for providing measurements from the BrainScaleS neuromorphic system, as well as the necessary framework.
This research was supported by EU grant \#269921 (BrainScaleS) and the Manfred St\"ark Foundation.



\bibliographystyle{spbasic}

\begin{thebibliography}{98}
\providecommand{\natexlab}[1]{#1}
\providecommand{\url}[1]{{#1}}
\providecommand{\urlprefix}{URL }
\expandafter\ifx\csname urlstyle\endcsname\relax
  \providecommand{\doi}[1]{DOI~\discretionary{}{}{}#1}\else
  \providecommand{\doi}{DOI~\discretionary{}{}{}\begingroup
  \urlstyle{rm}\Url}\fi
\providecommand{\eprint}[2][]{\url{#2}}

\bibitem[{Abeles et~al(2004)Abeles, Hayon, and Lehmann}]{abeles2004modeling}
Abeles M, Hayon G, Lehmann D (2004) Modeling compositionality by dynamic
  binding of synfire chains.  \emph{ Journal of computational neuroscience }
  17(2):179--201

\bibitem[{Aertsen et~al(1996)Aertsen, Diesmann, and Gewaltig}]{aertsen96}
Aertsen A, Diesmann M, Gewaltig MO (1996) {Propagation of synchronous spiking
  activity in feedforward neural networks.}  \emph{ J Physiol Paris }
  90(3-4):243--247

\bibitem[{Amit and Brunel(1997)}]{amit97model}
Amit DJ, Brunel N (1997) Model of global spontaneous activity and local
  structured activity during delay periods in the cerebral cortex.  \emph{
  Cereb Cortex }  7(3):237--52

\bibitem[{Berge and H{\"a}fliger(2007)}]{berge_iscas07}
Berge HKO, H{\"a}fliger P (2007) High-speed serial {AER} on {FPGA}. In: ISCAS,
  IEEE, pp 857--860

\bibitem[{Bergman et~al(2008)Bergman, Borkar, Campbell, Carlson, Dally,
  Denneau, Franzon, Harrod, Hill, Hiller et~al}]{bergman2008exascale}
Bergman K, Borkar S, Campbell D, Carlson W, Dally W, Denneau M, Franzon P,
  Harrod W, Hill K, Hiller J, et~al (2008) Exascale computing study: Technology
  challenges in achieving exascale systems.  \emph{ Defense Advanced Research
  Projects Agency Information Processing Techniques Office (DARPA IPTO), Tech
  Rep }  15

\bibitem[{Bi and Poo(1998)}]{poo98stdp}
Bi GQ, Poo MM (1998) {Synaptic modifications in cultured hippocampal neurons:
  dependence on spike timing, synaptic strength, and postsynaptic cell type.}
  \emph{ The Journal of neuroscience : the official journal of the Society for
  Neuroscience }  18(24):10,464--10,472

\bibitem[{Bill et~al(2010)Bill, Schuch, Br\"uderle, Schemmel, Maass, and
  Meier}]{bill2010compensating}
Bill J, Schuch K, Br\"uderle D, Schemmel J, Maass W, Meier K (2010)
  Compensating inhomogeneities of neuromorphic {VLSI} devices via short-term
  synaptic plasticity.  \emph{ Front Comp Neurosci }  4(129)

\bibitem[{Bontorin et~al(2007)Bontorin, Renaud, Garenne, Alvado, Le~Masson, and
  Tomas}]{bontorin2007realtime}
Bontorin G, Renaud S, Garenne A, Alvado L, Le~Masson G, Tomas J (2007) A
  real-time closed-loop setup for hybrid neural networks. In: Proceedings of
  the 29th Annual International Conference of the IEEE Engineering in Medicine
  and Biology Society (EMBS2007)

\bibitem[{Brette and Gerstner(2005)}]{brette_05}
Brette R, Gerstner W (2005) Adaptive exponential integrate-and-fire model as an
  effective description of neuronal activity.  \emph{ J Neurophysiol }  94:3637
  -- 3642

\bibitem[{Brette et~al(2007)Brette, Rudolph, Carnevale, Hines, Beeman, Bower,
  Diesmann, Morrison, Goodman, {Harris Jr}, Zirpe, Natschlager, Pecevski,
  Ermentrout, Djurfeldt, Lansner, Rochel, Vieville, Muller, Davison, Boustani,
  and Destexhe}]{brette2007simulation}
Brette R, Rudolph M, Carnevale T, Hines M, Beeman D, Bower JM, Diesmann M,
  Morrison A, Goodman PH, {Harris Jr} FC, Zirpe M, Natschlager T, Pecevski D,
  Ermentrout B, Djurfeldt M, Lansner A, Rochel O, Vieville T, Muller E, Davison
  AP, Boustani SE, Destexhe A (2007) Simulation of networks of spiking neurons:
  A review of tools and strategies.  \emph{ Journal of Computational
  Neuroscience }  23(3):349--398

\bibitem[{Bringuier et~al(1999)Bringuier, Chavane, Glaeser, and
  Frégnac}]{bringuier1999}
Bringuier V, Chavane F, Glaeser L, Frégnac Y (1999) Horizontal propagation of
  visual activity in the synaptic integration field of area 17 neurons.  \emph{
  Science }  283(5402):695--699

\bibitem[{Br{\"u}derle et~al(2009)Br{\"u}derle, M{\"u}ller, Davison, Muller,
  Schemmel, and Meier}]{bruederle09establishing}
Br{\"u}derle D, M{\"u}ller E, Davison A, Muller E, Schemmel J, Meier K (2009)
  Establishing a novel modeling tool: {A} python-based interface for a
  neuromorphic hardware system.  \emph{ Front Neuroinform }  3(17)

\bibitem[{Br{\"u}derle et~al(2011)Br{\"u}derle, Petrovici, Vogginger, Ehrlich,
  Pfeil, Millner, Gr{\"u}bl, Wendt, M{\"u}ller, Schwartz, de~Oliveira, Jeltsch,
  Fieres, Schilling, M{\"u}ller, Breitwieser, Petkov, Muller, Davison,
  Krishnamurthy, Kremkow, Lundqvist, Muller, Partzsch, Scholze, Z{\"u}hl, Mayr,
  Destexhe, Diesmann, Potjans, Lansner, Sch{\"u}ffny, Schemmel, and
  Meier}]{bruederle_biolcybern2010}
Br{\"u}derle D, Petrovici M, Vogginger B, Ehrlich M, Pfeil T, Millner S,
  Gr{\"u}bl A, Wendt K, M{\"u}ller E, Schwartz MO, de~Oliveira D, Jeltsch S,
  Fieres J, Schilling M, M{\"u}ller P, Breitwieser O, Petkov V, Muller L,
  Davison A, Krishnamurthy P, Kremkow J, Lundqvist M, Muller E, Partzsch J,
  Scholze S, Z{\"u}hl L, Mayr C, Destexhe A, Diesmann M, Potjans T, Lansner A,
  Sch{\"u}ffny R, Schemmel J, Meier K (2011) A comprehensive workflow for
  general-purpose neural modeling with highly configurable neuromorphic
  hardware systems.  \emph{ Biological Cybernetics }  104:263--296

\bibitem[{Brunel(2000)}]{brunel_jcns2000}
Brunel N (2000) Dynamics of sparsely connected networks of excitatory and
  inhibitory spiking neurons.  \emph{ Journal of Computational Neuroscience }
  8(3):183--208

\bibitem[{Buxhoeveden and Casanova(2002)}]{buxhoeveden2002minicolumn}
Buxhoeveden D, Casanova M (2002) The minicolumn and evolution of the brain.
  \emph{ Brain Behav Evol }  60:125--151

\bibitem[{Connors and Gutnick(1990)}]{connors90intrinsic}
Connors B, Gutnick M (1990) Intrinsic firing patterns of diverse neocortical
  neurons.  \emph{ Trends Neurosci }  13:99--104

\bibitem[{Cossart et~al(2003)Cossart, Aronov, and Yuste}]{cossart03attractor}
Cossart R, Aronov D, Yuste R (2003) Attractor dynamics of network up states in
  the neocortex.  \emph{ Nature }  423:238--283

\bibitem[{Costas-Santos et~al(2007)Costas-Santos, Serrano-Gotarredona,
  Serrano-Gotarredona, and Linares-Barranco}]{serrano_tcas2007}
Costas-Santos J, Serrano-Gotarredona T, Serrano-Gotarredona R, Linares-Barranco
  B (2007) A spatial contrast retina with on-chip calibration for neuromorphic
  spike-based {AER} vision systems.  \emph{ IEEE Transactions on Circuits and
  Systems }  54(7):1444--1458

\bibitem[{Davison et~al(2008)Davison, Br\"{u}derle, Eppler, Kremkow, Muller,
  Pecevski, Perrinet, and Yger}]{davison08pynn}
Davison AP, Br\"{u}derle D, Eppler J, Kremkow J, Muller E, Pecevski D, Perrinet
  L, Yger P (2008) {PyNN}: a common interface for neuronal network simulators.
  \emph{ Front Neuroinform }  2(11)

\bibitem[{Delbr\"{u}ck and Liu(2004)}]{delbrueck04silicon}
Delbr\"{u}ck T, Liu SC (2004) A silicon early visual system as a model animal.
  \emph{ Vision Res }  44(17):2083--2089

\bibitem[{Destexhe(2009)}]{destexhe09self}
Destexhe A (2009) Self-sustained asynchronous irregular states and {U}p/{D}own
  states in thalamic, cortical and thalamocortical networks of nonlinear
  integrate-and-fire neurons.  \emph{ Journal of Computational Neuroscience }
  3:493 -- 506

\bibitem[{Destexhe and Contreras(2006)}]{destexhe2006}
Destexhe A, Contreras D (2006) Neuronal computations with stochastic network
  states.  \emph{ Science }  314(5796):85--90

\bibitem[{Destexhe and Pare(1999)}]{destexhe1999impact}
Destexhe A, Pare D (1999) {Impact of Network Activity on the Integrative
  Properties of Neocortical Pyramidal Neurons In Vivo}.  \emph{ J Neurophysiol
  }  81(4):1531--1547

\bibitem[{Destexhe et~al(2003)Destexhe, Rudolph, and Pare}]{destexhe03hcs}
Destexhe A, Rudolph M, Pare D (2003) The high-conductance state of neocortical
  neurons in vivo.  \emph{ Nature Reviews Neuroscience }  4:739--751

\bibitem[{Diesmann(2002)}]{diesmann02}
Diesmann M (2002) Conditions for stable propagation of synchronous spiking in
  cortical neural networks: Single neuron dynamics and network properties. PhD
  thesis, Ruhr-Universit{\"a}t Bochum

\bibitem[{Diesmann and Gewaltig(2002)}]{diesmann01nest}
Diesmann M, Gewaltig MO (2002) {NEST}: An environment for neural systems
  simulations. In: Plesser T, Macho V (eds) Forschung und wisschenschaftliches
  Rechnen, Beitr{\"a}ge zum Heinz-Billing-Preis 2001, GWDG-Bericht, vol~58,
  Ges. f{\"u}r Wiss. Datenverarbeitung, G{\"o}ttingen, pp 43--70

\bibitem[{Diesmann et~al(1999)Diesmann, Gewaltig, and Aertsen}]{diesmann99}
Diesmann M, Gewaltig MO, Aertsen A (1999) Stable propagation of synchronous
  spiking in cortical neural networks.  \emph{ Nature }  402:529--533

\bibitem[{Diesmann et~al(2001)Diesmann, Gewaltig, Rotter, and
  Aertsen}]{diesmann2001state}
Diesmann M, Gewaltig MO, Rotter S, Aertsen A (2001) State space analysis of
  synchronous spiking in cortical neural networks.  \emph{ Neurocomputing }
  38:565--571

\bibitem[{Djurfeldt et~al(2008)Djurfeldt, Lundqvist, Johansson, Rehn, Ekeberg,
  and Lansner}]{djurfeldt2008bluegene}
Djurfeldt M, Lundqvist M, Johansson C, Rehn M, Ekeberg O, Lansner A (2008)
  Brain-scale simulation of the neocortex on the ibm blue gene/l supercomputer.
   \emph{ IBM Journal of Research and Development }  52(1.2):31--41

\bibitem[{Ehrlich et~al(2007)Ehrlich, Mayr, Eisenreich, Henker, Srowig,
  Gr\"ubl, Schemmel, and Sch\"uffny}]{ehrlich_ssd07}
Ehrlich M, Mayr C, Eisenreich H, Henker S, Srowig A, Gr\"ubl A, Schemmel J,
  Sch\"uffny R (2007) Wafer-scale {VLSI} implementations of pulse coupled
  neural networks. In: Proceedings of the International Conference on Sensors,
  Circuits and Instrumentation Systems (SSD-07)

\bibitem[{Ehrlich et~al(2010)Ehrlich, Wendt, Z\"uhl, Sch\"uffny, Br\"uderle,
  M\"uller, and Vogginger}]{ehrlich2010anniip}
Ehrlich M, Wendt K, Z\"uhl L, Sch\"uffny R, Br\"uderle D, M\"uller E, Vogginger
  B (2010) A software framework for mapping neural networks to a wafer-scale
  neuromorphic hardware system. In: Proceedings of the Artificial Neural
  Networks and Intelligent Information Processing Conference (ANNIIP) 2010, pp
  43--52

\bibitem[{El~Boustani and Destexhe(2009)}]{elboustani2009}
El~Boustani S, Destexhe A (2009) A master equation formalism for macroscopic
  modeling of asynchronous irregular activity states.  \emph{ Neural
  Computation }  21(1):46--100

\bibitem[{Eppler et~al(2008)Eppler, Helias, Muller, Diesmann, and
  Gewaltig}]{eppler2008}
Eppler JM, Helias M, Muller E, Diesmann M, Gewaltig MO (2008) {Py{NEST}}: a
  convenient interface to the {NEST} simulator.  \emph{ Front Neuroinform }
  2(12)

\bibitem[{Fieres et~al(2008)Fieres, Schemmel, and Meier}]{fieres_ijcnn2008}
Fieres J, Schemmel J, Meier K (2008) Realizing biological spiking network
  models in a configurable wafer-scale hardware system. In: Proceedings of the
  2008 International Joint Conference on Neural Networks (IJCNN)

\bibitem[{Furber et~al(2012)Furber, Lester, Plana, Garside, Painkras, Temple,
  and Brown}]{furber2012}
Furber SB, Lester DR, Plana LA, Garside JD, Painkras E, Temple S, Brown AD
  (2012) Overview of the {SpiNNaker} system architecture.  \emph{ IEEE
  Transactions on Computers }  99(PrePrints)

\bibitem[{Galluppi et~al(2010)Galluppi, Rast, Davies, and
  Furber}]{galluppi2010}
Galluppi F, Rast A, Davies S, Furber S (2010) A general-purpose model
  translation system for a universal neural chip. In: Wong K, Mendis B,
  Bouzerdoum A (eds) Neural Information Processing. Theory and Algorithms,
  Lecture Notes in Computer Science, vol 6443, Springer Berlin / Heidelberg, pp
  58--65

\bibitem[{Giulioni et~al(2012)Giulioni, Camilleri, Mattia, Dante, Braun, and
  Del~Giudice}]{giulioni2012}
Giulioni M, Camilleri P, Mattia M, Dante V, Braun J, Del~Giudice P (2012)
  Robust working memory in an asynchronously spiking neural network realized in
  neuromorphic vlsi.  \emph{ Frontiers in Neuroscience }  5(149)

\bibitem[{González-Burgos et~al(2000)González-Burgos, Barrionuevo, and
  Lewis}]{gonzalez-burgos2000}
González-Burgos G, Barrionuevo G, Lewis DA (2000) Horizontal synaptic
  connections in monkey prefrontal cortex: An in vitro electrophysiological
  study.  \emph{ Cerebral Cortex }  10(1):82--92

\bibitem[{H\"afliger(2007)}]{hafliger2007adaptive}
H\"afliger P (2007) Adaptive {WTA} with an analog {VLSI} neuromorphic learning
  chip.  \emph{ IEEE Transactions on Neural Networks }  18(2):551--72

\bibitem[{Hartmann et~al(2010)Hartmann, Schiefer, Scholze, Partzsch, Mayr,
  Henker, and Schuffny}]{hartmann2010}
Hartmann S, Schiefer S, Scholze S, Partzsch J, Mayr C, Henker S, Schuffny R
  (2010) Highly integrated packet-based aer communication infrastructure with
  3gevent/s throughput. In: Electronics, Circuits, and Systems (ICECS), 2010
  17th IEEE International Conference on, pp 950--953

\bibitem[{Hasler and Marr(2013)}]{hasler2013}
Hasler J, Marr HB (2013) Finding a roadmap to achieve large neuromorphic
  hardware systems.  \emph{ Frontiers in Neuroscience }  7(118)

\bibitem[{Helias et~al(2012)Helias, Kunkel, Masumoto, Igarashi, Eppler, Ishii,
  Fukai, Morrison, and Diesmann}]{helias2012supercomputers}
Helias M, Kunkel S, Masumoto G, Igarashi J, Eppler JM, Ishii S, Fukai T,
  Morrison A, Diesmann M (2012) Supercomputers ready for use as discovery
  machines for neuroscience.  \emph{ Frontiers in Neuroinformatics }  6(26)

\bibitem[{Hellwig(2000)}]{hellwig2000}
Hellwig B (2000) A quantitative analysis of the local connectivity between
  pyramidal neurons in layers 2/3 of the rat visual cortex.  \emph{ Biological
  Cybernetics }  82:111--121

\bibitem[{Hines and Carnevale(2003)}]{hines03neuron}
Hines M, Carnevale N (2003) The NEURON simulation environment., M.A. Arbib, pp
  769--773

\bibitem[{Hines et~al(2009)Hines, Davison, and Muller}]{hines09}
Hines ML, Davison AP, Muller E (2009) {NEURON and Python}.  \emph{ Front
  Neuroinform }

\bibitem[{Hirsch and Gilbert(1991)}]{hirsch1991}
Hirsch J, Gilbert C (1991) Synaptic physiology of horizontal connections in the
  cat's visual cortex.  \emph{ The Journal of Neuroscience }  11(6):1800--1809

\bibitem[{Indiveri(2008)}]{indiveri2008_vlsi}
Indiveri G (2008) Neuromorphic vlsi models of selective attention: From single
  chip vision sensors to multi-chip systems.  \emph{ Sensors }  8(9):5352--5375

\bibitem[{Indiveri et~al(2006)Indiveri, Chicca, and Douglas}]{indiveri_tnn2006}
Indiveri G, Chicca E, Douglas R (2006) A {VLSI} array of low-power spiking
  neurons and bistable synapses with spike-timing dependent plasticity.  \emph{
  IEEE Transactions on Neural Networks }  17(1):211--221

\bibitem[{Indiveri et~al(2009)Indiveri, Chicca, and
  Douglas}]{indiveri2009artificial}
Indiveri G, Chicca E, Douglas R (2009) Artificial cognitive systems: From
  {VLSI} networks of spiking neurons to neuromorphic cognition.  \emph{
  Cognitive Computation }  1(2):119--127

\bibitem[{Indiveri et~al(2011)Indiveri, Linares-Barranco, Hamilton, van Schaik,
  Etienne-Cummings, Delbruck, Liu, Dudek, H{\"a}fliger, Renaud, Schemmel,
  Cauwenberghs, Arthur, Hynna, Folowosele, Saighi, Serrano-Gotarredona,
  Wijekoon, Wang, and Boahen}]{indiveri2011}
Indiveri G, Linares-Barranco B, Hamilton TJ, van Schaik A, Etienne-Cummings R,
  Delbruck T, Liu SC, Dudek P, H{\"a}fliger P, Renaud S, Schemmel J,
  Cauwenberghs G, Arthur J, Hynna K, Folowosele F, Saighi S,
  Serrano-Gotarredona T, Wijekoon J, Wang Y, Boahen K (2011) Neuromorphic
  silicon neuron circuits.  \emph{ Frontiers in Neuroscience }  5(0)

\bibitem[{Kampa et~al(2006)Kampa, Letzkus, and Stuart}]{kampa04cortical}
Kampa BM, Letzkus JJ, Stuart GJ (2006) Cortical feed-forward networks for
  binding different streams of sensory information.  \emph{ Nature Neuroscience
  }  9(12):1472--1473

\bibitem[{Kremkow et~al(2010{\natexlab{a}})Kremkow, Aertsen, and
  Kumar}]{kremkow2010gating}
Kremkow J, Aertsen A, Kumar A (2010{\natexlab{a}}) Gating of signal propagation
  in spiking neural networks by balanced and correlated excitation and
  inhibition.  \emph{ The Journal of neuroscience }  30(47):15,760--15,768

\bibitem[{Kremkow et~al(2010{\natexlab{b}})Kremkow, Perrinet, Masson, and
  Aertsen}]{kremkow2010functional}
Kremkow J, Perrinet L, Masson G, Aertsen A (2010{\natexlab{b}}) Functional
  consequences of correlated excitatory and inhibitory conductances in cortical
  networks.  \emph{ J Comput Neurosci }  28:579--594

\bibitem[{Kumar et~al(2008)Kumar, Schrader, Aertsen, and
  Rotter}]{kumar08highconductance}
Kumar A, Schrader S, Aertsen A, Rotter S (2008) The high-conductance state of
  cortical networks.  \emph{ Neural Computation }  20(1):1--43

\bibitem[{Laing and Lord(2009)}]{laing2009stochastic}
Laing C, Lord GJ (2009) \emph{Stochastic Methods in Neuroscience}. Oxford
  University Press

\bibitem[{Lande et~al(1996)Lande, Ranjbar, Ismail, and Berg}]{lande96}
Lande T, Ranjbar H, Ismail M, Berg Y (1996) An analog floating-gate memory in a
  standard digital technology. In: Microelectronics for Neural Networks, 1996.,
  Proceedings of Fifth International Conference on, pp 271 --276

\bibitem[{Lewis et~al(2000)Lewis, Etienne-Cummings, Cohen, and
  Hartmann}]{lewis00toward}
Lewis MA, Etienne-Cummings R, Cohen AH, Hartmann M (2000) Toward biomorphic
  control using custom a{VLSI} chips. In: Proceedings of the International
  conference on robotics and automation, IEEE Press

\bibitem[{Lundqvist et~al(2006)Lundqvist, Rehn, Djurfeldt, and
  Lansner}]{lundqvist2006attractor}
Lundqvist M, Rehn M, Djurfeldt M, Lansner A (2006) Attractor dynamics in a
  modular network of neocortex.  \emph{ Network: Computation in Neural Systems
  }  17:3:253--276

\bibitem[{Lundqvist et~al(2010)Lundqvist, Compte, and
  Lansner}]{lundqvist2010bistable}
Lundqvist M, Compte A, Lansner A (2010) Bistable, irregular firing and
  population oscillations in a modular attractor memory network.  \emph{ PLoS
  Comput Biol }  6(6)

\bibitem[{Markram et~al(1998)Markram, Gupta, Uziel, Wang, and
  Tsodyks}]{markram98information}
Markram H, Gupta A, Uziel A, Wang Y, Tsodyks M (1998) Information processing
  with frequency-dependent synaptic connections.  \emph{ Neurobiol Learn Mem }
  70(1-2):101--112

\bibitem[{Markram et~al(2004)Markram, Toledo-Rodriguez, Wang, Gupta,
  Silberberg, and Wu}]{markram04interneurons}
Markram H, Toledo-Rodriguez M, Wang Y, Gupta A, Silberberg G, Wu C (2004)
  Interneurons of the neocortical inhibitory system.  \emph{ Nat Rev Neurosci }
   5(10):793--807

\bibitem[{McDonnell et~al(2014)McDonnell, Boahen, Ijspeert, and
  Sejnowski}]{ieee2014specialissue}
McDonnell MD, Boahen K, Ijspeert A, Sejnowski TJ (eds) (2014) Engineering
  Intelligent Electronic Systems Based on Computational Neuroscience,
  Proceedings of the IEEE, vol 102: 5

\bibitem[{Mead(1989)}]{mead89analog}
Mead CA (1989) \emph{Analog {VLSI} and Neural Systems}. Addison Wesley,
  Reading, MA

\bibitem[{Mead(1990)}]{mead90neuromorphic}
Mead CA (1990) Neuromorphic electronic systems.  \emph{ Proceedings of the IEEE
  }  78:1629--1636

\bibitem[{Mead and Mahowald(1988)}]{mead88silicon}
Mead CA, Mahowald MA (1988) A silicon model of early visual processing.  \emph{
  Neural Networks }  1(1):91--97

\bibitem[{Merolla and Boahen(2006)}]{merolla2006dynamic}
Merolla PA, Boahen K (2006) Dynamic computation in a recurrent network of
  heterogeneous silicon neurons. In: Proceedings of the 2006 IEEE International
  Symposium on Circuits and Systems (ISCAS 2006)

\bibitem[{Millner et~al(2010)Millner, Gr\"{u}bl, Meier, Schemmel, and
  Schwartz}]{millner10}
Millner S, Gr\"{u}bl A, Meier K, Schemmel J, Schwartz MO (2010) A {VLSI}
  implementation of the adaptive exponential integrate-and-fire neuron model.
  In: Lafferty J, Williams CKI, Shawe-Taylor J, Zemel R, Culotta A (eds)
  Advances in Neural Information Processing Systems 23, pp 1642--1650

\bibitem[{Mitra et~al(2009)Mitra, Fusi, and Indiveri}]{mitra09realtime}
Mitra S, Fusi S, Indiveri G (2009) Real-time classification of complex patterns
  using spike-based learning in neuromorphic {VLSI}.  \emph{ {IEEE}
  Transactions on Biomedical Circuits and Systems }  3:(1):32--42

\bibitem[{Morrison et~al(2005)Morrison, Mehring, Geisel, Aertsen, and
  Diesmann}]{morrison05distributed}
Morrison A, Mehring C, Geisel T, Aertsen A, Diesmann M (2005) Advancing the
  boundaries of high connectivity network simulation with distributed
  computing.  \emph{ Neural Comput }  17(8):1776--1801

\bibitem[{Morrison et~al(2008)Morrison, Diesmann, and
  Gerstner}]{morrison08_stdp}
Morrison A, Diesmann M, Gerstner W (2008) Phenomenological models of synaptic
  plasticity based on spike timing.  \emph{ Biological Cybernetics }
  98(6):459--478

\bibitem[{Mountcastle(1997)}]{mountcastle1997columnar}
Mountcastle VB (1997) The columnar organization of the neocortex.  \emph{ Brain
  }  120(4):701--722

\bibitem[{Muller and Destexhe(2012)}]{muller2012}
Muller L, Destexhe A (2012) Propagating waves in thalamus, cortex and the
  thalamocortical system: Experiments and models.  \emph{ Journal of
  Physiology-Paris }  106(5–6):222 -- 238

\bibitem[{Murakoshi et~al(1993)Murakoshi, Guo, and Ichinose}]{murakoshi1993}
Murakoshi T, Guo JZ, Ichinose T (1993) Electrophysiological identification of
  horizontal synaptic connections in rat visual cortex in vitro.  \emph{
  Neuroscience Letters }  163(2):211 -- 214

\bibitem[{Netter and Franceschini(2002)}]{netter02arobotic}
Netter T, Franceschini N (2002) A robotic aircraft that follows terrain using a
  neuromorphic eye. In: Conf. Intelligent Robots and System, pp 129--134

\bibitem[{Perin et~al(2011)Perin, Berger, and Markram}]{perin2011synaptic}
Perin R, Berger TK, Markram H (2011) A synaptic organizing principle for
  cortical neuronal groups.  \emph{ PNAS }  pp 5419--5424

\bibitem[{Peters and Sethares(1997)}]{peters1997organization}
Peters A, Sethares C (1997) The organization of double bouquet cells in monkey
  striate cortex.  \emph{ Journal of Neurocytology }  26(12):779 -- 797

\bibitem[{Pfeil et~al(2012)Pfeil, Potjans, Schrader, Potjans, Schemmel,
  Diesmann, and Meier}]{Pfeil12_90}
Pfeil T, Potjans TC, Schrader S, Potjans W, Schemmel J, Diesmann M, Meier K
  (2012) Is a 4-bit synaptic weight resolution enough? - constraints on
  enabling spike-timing dependent plasticity in neuromorphic hardware.  \emph{
  Frontiers in Neuroscience }  6(90)

\bibitem[{Pfeil et~al(2013)Pfeil, Gr\"ubl, Jeltsch, M\"uller, M\"uller,
  Petrovici, Schmuker, Br\"uderle, Schemmel, and Meier}]{pfeil2013}
Pfeil T, Gr\"ubl A, Jeltsch S, M\"uller E, M\"uller P, Petrovici MA, Schmuker
  M, Br\"uderle D, Schemmel J, Meier K (2013) Six networks on a universal
  neuromorphic computing substrate.  \emph{ Frontiers in Neuroscience }  7:11

\bibitem[{Renaud et~al(2007)Renaud, Tomas, Bornat, Daouzli, and
  Saighi}]{renaud2007neuromimetic}
Renaud S, Tomas J, Bornat Y, Daouzli A, Saighi S (2007) Neuromimetic {IC}s with
  analog cores: an alternative for simulating spiking neural networks. In:
  Proceedings of the 2007 IEEE Symposium on Circuits and Systems (ISCAS2007)

\bibitem[{Rieke et~al(1997)Rieke, Warland, de~Ruyter~van Steveninck, and
  Bialek}]{rieke1997spikes}
Rieke F, Warland D, de~Ruyter~van Steveninck R, Bialek W (1997) \emph{Spikes -
  Exploring the neural code.} MIT Press, Cambridge, MA.

\bibitem[{Rocke et~al(2008)Rocke, McGinley, Maher, Morgan, and
  Harkin}]{rocke2008fpaanetworks}
Rocke P, McGinley B, Maher J, Morgan F, Harkin J (2008) Investigating the
  suitability of fpaas for evolved hardware spiking neural networks. In: Hornby
  G, Sekanina L, Haddow P (eds) Evolvable Systems: From Biology to Hardware,
  Lecture Notes in Computer Science, vol 5216, Springer Berlin / Heidelberg, pp
  118--129

\bibitem[{Roxin et~al(2005)Roxin, Brunel, and Hansel}]{roxin2005}
Roxin A, Brunel N, Hansel D (2005) Role of delays in shaping spatiotemporal
  dynamics of neuronal activity in large networks.  \emph{ Phys Rev Lett }
  94:238,103

\bibitem[{Schemmel et~al(2006)Schemmel, Gr\"ubl, Meier, and
  Muller}]{schemmel_ijcnn06}
Schemmel J, Gr\"ubl A, Meier K, Muller E (2006) Implementing synaptic
  plasticity in a {VLSI} spiking neural network model. In: Proceedings of the
  2006 International Joint Conference on Neural Networks (IJCNN), IEEE Press

\bibitem[{Schemmel et~al(2007)Schemmel, Br\"uderle, Meier, and
  Ostendorf}]{schemmel_iscas07}
Schemmel J, Br\"uderle D, Meier K, Ostendorf B (2007) Modeling synaptic
  plasticity within networks of highly accelerated {I}\&{F} neurons. In:
  Proceedings of the 2007 IEEE International Symposium on Circuits and Systems
  (ISCAS), IEEE Press, pp 3367--3370

\bibitem[{Schemmel et~al(2008)Schemmel, Fieres, and Meier}]{schemmel_ijcnn2008}
Schemmel J, Fieres J, Meier K (2008) Wafer-scale integration of analog neural
  networks. In: Proceedings of the 2008 International Joint Conference on
  Neural Networks (IJCNN)

\bibitem[{Schemmel et~al(2010)Schemmel, Br\"uderle, Gr\"ubl, Hock, Meier, and
  Millner}]{schemmel_iscas2010}
Schemmel J, Br\"uderle D, Gr\"ubl A, Hock M, Meier K, Millner S (2010) A
  wafer-scale neuromorphic hardware system for large-scale neural modeling. In:
  Proceedings of the 2010 IEEE International Symposium on Circuits and Systems
  (ISCAS), pp 1947--1950

\bibitem[{Scholze et~al(2010)Scholze, Henker, Partzsch, Mayr, and
  Schuffny}]{scholze2010heap}
Scholze S, Henker S, Partzsch J, Mayr C, Schuffny R (2010) Optimized queue
  based communication in vlsi using a weakly ordered binary heap. In: Mixed
  Design of Integrated Circuits and Systems (MIXDES), 2010 Proceedings of the
  17th International Conference, pp 316 --320

\bibitem[{Scholze et~al(2011{\natexlab{a}})Scholze, Eisenreich, H\"oppner,
  Ellguth, Henker, Ander, H\"anzsche, Partzsch, Mayr, and
  Sch\"uffny}]{scholze11a}
Scholze S, Eisenreich H, H\"oppner S, Ellguth G, Henker S, Ander M, H\"anzsche
  S, Partzsch J, Mayr C, Sch\"uffny R (2011{\natexlab{a}}) A 32 {GB}it/s
  communication {SoC} for a waferscale neuromorphic system.  \emph{
  Integration, the VLSI Journal }

\bibitem[{Scholze et~al(2011{\natexlab{b}})Scholze, Schiefer, Partzsch,
  Hartmann, Mayr, H\"oppner, Eisenreich, Henker, Vogginger, and
  Sch\"uffny}]{scholze11b}
Scholze S, Schiefer S, Partzsch J, Hartmann S, Mayr CG, H\"oppner S, Eisenreich
  H, Henker S, Vogginger B, Sch\"uffny R (2011{\natexlab{b}}) {VLSI}
  implementation of a 2.8{GE}vent/s packet based {AER} interface with routing
  and event sorting functionality.  \emph{ Frontiers in Neuromorphic
  Engineering }  5(117):1--13

\bibitem[{Schrader et~al(2010)Schrader, Diesmann, and
  Morrison}]{schrader2010compositionality}
Schrader S, Diesmann M, Morrison A (2010) A compositionality machine realized
  by a hierarchic architecture of synfire chains.  \emph{ Frontiers in
  Computational Neuroscience }  4

\bibitem[{Serrano-Gotarredona et~al(2006)Serrano-Gotarredona, Oster,
  Lichtsteiner, Linares-Barranco, Paz-Vicente, G\'omez-Rodr\'iguez, Riis,
  Delbr\"uck, Liu, Zahnd, Whatley, Douglas, H\"afliger, Jimenez-Moreno, Civit,
  Serrano-Gotarredona, Acosta-Jim\'enez, and
  Linares-Barranco}]{serrano_nips2005}
Serrano-Gotarredona R, Oster M, Lichtsteiner P, Linares-Barranco A, Paz-Vicente
  R, G\'omez-Rodr\'iguez F, Riis HK, Delbr\"uck T, Liu SC, Zahnd S, Whatley AM,
  Douglas RJ, H\"afliger P, Jimenez-Moreno G, Civit A, Serrano-Gotarredona T,
  Acosta-Jim\'enez A, Linares-Barranco B (2006) {AER} building blocks for
  multi-layer multi-chip neuromorphic vision systems. In: Weiss Y,
  Sch\"{o}lkopf B, Platt J (eds) Advances in Neural Information Processing
  Systems 18, MIT Press, Cambridge, MA, pp 1217--1224

\bibitem[{Serrano-Gotarredona et~al(2009)Serrano-Gotarredona, Oster,
  Lichtsteiner, Linares-Barranco, Paz-Vicente, Gomez-Rodriguez, Camunas-Mesa,
  Berner, Rivas-Perez, Delbruck, Liu, Douglas, Hafliger, Jimenez-Moreno,
  Ballcels, Serrano-Gotarredona, Acosta-Jimenez, and
  Linares-Barranco}]{serrano2009caviar}
Serrano-Gotarredona R, Oster M, Lichtsteiner P, Linares-Barranco A, Paz-Vicente
  R, Gomez-Rodriguez F, Camunas-Mesa L, Berner R, Rivas-Perez M, Delbruck T,
  Liu SC, Douglas R, Hafliger P, Jimenez-Moreno G, Ballcels A,
  Serrano-Gotarredona T, Acosta-Jimenez A, Linares-Barranco B (2009) Caviar: A
  45k neuron, 5m synapse, 12g connects/s aer hardware
  sensory--processing--learning--actuating system for high-speed visual object
  recognition and tracking.  \emph{ Neural Networks, IEEE Transactions on }
  20(9):1417--1438

\bibitem[{Song et~al(2005)Song, Sjöström, Reigl, Nelson, and
  Chklovskii}]{song05nonrandom}
Song S, Sjöström PJ, Reigl M, Nelson S, Chklovskii DB (2005) Highly nonrandom
  features of synaptic connectivity in cortical circuits.  \emph{ PLOS Biology
  }  3(3):517--519

\bibitem[{Telfeian and Connors(2003)}]{telfeian2003}
Telfeian AE, Connors BW (2003) Widely integrative properties of layer 5
  pyramidal cells support a role for processing of extralaminar synaptic inputs
  in rat neocortex.  \emph{ Neuroscience Letters }  343(2):121 -- 124

\bibitem[{Thomson et~al(2002)Thomson, West, Wang, and
  Bannister}]{thomson02synaptic}
Thomson AM, West DC, Wang Y, Bannister AP (2002) Synaptic connections and small
  circuits involving excitatory and inhibitory neurons in layers 2-5 of adult
  rat and cat neocortex: triple intracellular recordings and biocytin labelling
  in vitro.  \emph{ Cerebral Cortex }  12:936--953

\bibitem[{Vogels and Abbott(2005)}]{vogels05signal}
Vogels TP, Abbott LF (2005) Signal propagation and logic gating in networks of
  integrate-and-fire neurons.  \emph{ J Neurosci }  25(46):10,786--95

\bibitem[{Vogelstein et~al(2007)Vogelstein, Mallik, Vogelstein, and
  Cauwenberghs}]{vogelstein2007reconfigurable}
Vogelstein RJ, Mallik U, Vogelstein JT, Cauwenberghs G (2007) Dynamically
  reconfigurable silicon array of spiking neuron with conductance-based
  synapses.  \emph{ IEEE Transactions on Neural Networks }  18:253--265

\bibitem[{Zou et~al(2006)Zou, Bornat, Tomas, Renaud, and
  Destexhe}]{zou2006analogneurons}
Zou Q, Bornat Y, Tomas J, Renaud S, Destexhe A (2006) Real-time simulations of
  networks of hodgkin-huxley neurons using analog circuits.  \emph{
  Neurocomputing }  69:1137--1140

\end{thebibliography}



\newcounter{pagetmp}
\setcounter{pagetmp}{\value{page}}


\begin{appendices}

\setcounter{section}{0}
\renewcommand{\thesection}{S\arabic{section}}
\renewcommand{\theequation}{S\arabic{section}.\arabic{equation}}
\renewcommand{\thefigure}{S\arabic{section}.\arabic{figure}}
\renewcommand{\thetable}{S\arabic{section}.\arabic{table}}

\clearpage
\section{Neuromorphic hardware}
\setcounter{page}{1}
\setcounter{table}{0}
\setcounter{equation}{0}
\setcounter{figure}{0}

\begin{table*}[b]
    \caption{
		{\bf Parameter ranges of the BrainScaleS wafer-scale hardware}
	}
    \centering
    \begin{tabular*}{\textwidth}{l l c c c p{0.3\textwidth}}
    \toprule
        Description & Name & Min & Max & Unit & Comment \\
        \midrule
                \multicolumn{6}{l}{\textbf{Neuron (Adaptive Exponential Integrate\&Fire)}}\\
        \midrule
        Absolute refractory period    & $\tauref$     & 0.16   & 10.0  & \ms           & \\
        Spike detection potential     & $\Vspike$     & -125.0 & 45.0  & \mV           & \\
        Reset potential               & $\Er$         & -125.0 & 45.0  & \mV           & \\
        Leakage reversal potential    & $\Vrest$      & -125.0 & 45.0  & \mV           & \\
        Membrane time constant        &$\taum$        & 9      & 105   & \ms           & \\
        Adaptation coupling param     & $a$           & 0      & 10.0  & \nano\siemens & adaptation can be fully disabled\\
        Spike triggered adapt. param  & $b$           & 0      & 86    & \pA           &  \\
        Adaptation time constant      & $\tauw$       & 20.0   & 780.0 & \ms           &  \\
        Threshold slope factor         & $\adexdeltaT$ & 0.4    & 3.0   & \mV           &\multirow{2}{0.3\textwidth}{exponential spike generation can be fully disabled}\\
        Spike initiation threshold    & $\Vthresh$    & -125.0 & 45.0  & \mV           & \\
        Excitatory reversal potential & $\Ereve$      & -125.0 & 45.0  & \mV           & \\
        Inhibitory reversal potential & $\Erevi$      & -125.0 & 45.0  & \mV           & \\
        Exc. synaptic time constant   & $\tausyne$    & 1.0    & 100.0 & \ms           & \\
        Inh. synaptic time constant   & $\tausyni$    & 1.0    & 100.0 & \ms           & \\
        \midrule
                \multicolumn{6}{l}{\textbf{Synapses}}\\
        \midrule
	Weight                  & $\wsyn$          & 0 & 0.300 & \micro\siemens & 4-bit resolution \\ 
	Axonal delay (on-wafer) & $\mathrm{delay}$ & 1.2     & 2.2    & \ms            & not configurable \\
        \midrule
                \multicolumn{6}{l}{\textbf{Short Term Plasticity}}\\
        \midrule
        Utilization of synaptic efficacy & $\tsoU$                   & 0.11 & 0.47  &     & possible values: $[\frac{1}{9},\frac{3}{11}, \frac{5}{13},\frac{7}{15}]$\\
        Recovery time constant           & $\taurec$   & 40.0 & 900.0 & \ms &\multirow{3}{0.3\textwidth}{One of the two time constants has to be set to $0.0$. Available range depends on $\tsoU$ (maximum range given).} \\
        Facilitation time constant       & $\taufacil$ & 35.0 & 200.0 & \ms & \\
         & & & & & \\
         & & & & & \\
        \midrule
                \multicolumn{6}{l}{\textbf{Stimulus}}\\
        \midrule
        External spike sources & $\nu$ & 0.0 & 4000 & \Hz & cf. \cite{scholze11b} \\ 
        \bottomrule
    \end{tabular*}
	\flushleft{
        All ranges correspond to a membrane capacitance of $\Cm=\SI{0.2}{\nano\farad}$ and a hardware speedup of $10^4$ compared to real time.
        It is possible to choose an arbitrary value for $\Cm$, but then the ranges of parameters $a$, $b$ and of the synaptic weights are multiplied by $\frac{\Cm}{\SI{0.2}{\nano\farad}}$.
        }
    \label{table:hardware_parameter_ranges}
\end{table*}
\subsection{Short-term plasticity}
\label{sec:methods-hw-stp}
As mentioned in \cref{sec:hicann}, the hardware short-term plasticity mechanism is an implementation of the phenomenological model by \cite{markram98information}.
We first describe the hardware STP model and then provide the translation between the original and the hardware model. 

\paragraph{Model description}
Unlike the theoretical model \cite{markram98information}, which allows the occurrence of both depression and facilitation at the same time, the hardware implementation does not allow their simultaneous activation.
The ongoing pre-synaptic activity is tracked with a time-varying active partition $I$ with $0 \leq I \leq 1$, which decays exponentially to zero with time constant $\tau_{\mathrm{stdf}}$.
Following a pre-synaptic spike, $I$ is increased by a fixed fraction $U_{\mathrm{SE}}(1-I)$, resulting in the following dynamics for the active partition:
\begin{equation}
I_{n+1} = \left[I_{n} + U_{\mathrm{SE}}(1-I_{n})\right]\exp\left(-\frac{\Delta t}{\tau_\mathrm{stdf}}\right) \quad , \label{eqn:TSO}
\end{equation}
with $\Delta t$ being the time interval between the $n$th and $(n+1)$st afferent spike.

This active partition can be used to model depressing or facilitating synapses as follows:

\begin{align}
    w_{\mathrm{STP}}^{\mathrm{depression}} & = 1 - \lambda \cdot I \label{eqn:hw_stp1} \\
    w_{\mathrm{STP}}^{\mathrm{facilitation}} & = 1 + \lambda \cdot ( I - \beta) \quad . \label{eqn:hw_stp2}
\end{align}
Here, $w_{\mathrm{STP}}^{x}$ corresponds to a multiplicative factor to the static synaptic weight, with $\lambda$ and $\beta$ being configurable variables, and $x$ denotes the mode being either $\mathrm{depression}$ or $\mathrm{facilitation}$.

According to \cref{eq:tm_stp} the n-th effective synaptic weight is then given by

\begin{align}
	\wsyn_{n} & = w_{\mathrm{static}}w_{\mathrm{STP}}^{x} \quad .
\end{align}
Due to a technical limitation, the change of synaptic weights by STP can not be larger than the static weight, such that $0\leq w_{\mathrm{STP}}^{x} \leq 2$.
We refer to \cite{schemmel_ijcnn2008} for details of the hardware implementation of STP and to \cite{bill2010compensating} for neural network experiments on neuromorphic hardware using this STP model.

\paragraph{Transformation from original model}
The original model by \cite{markram98information} (\Cref{eq:tm_stp}) can be translated to the hardware model (\Cref{eqn:TSO,eqn:hw_stp1,eqn:hw_stp2}) when one of the two time constants ($\taurec$ or $\taufacil$) is equal to zero.

For depression only ($\taufacil=0$), the $n$th synaptic weight is given by (cf. \cref{eq:tm_stp}):

\begin{align}
	\wsyn_{n} & = \wsyn_\mathrm{max}R_{n}\tsoU \quad .
\end{align}
The time course of $R$ can be exactly represented by $(1-I)$ if the scaling factor $\lambda$ of the short-term plasticity mechanism is set to 1.
Additionally, the static synaptic weight $w_{\mathrm{static}}$ has to be adapted such that the applied synaptic weights are equal, giving us the following transformation:
	$\tau_\mathrm{stdf}  = \taurec$, 
	$U_\mathrm{SE}  = \tsoU$,
	$\lambda = 1$ and
	$w_{\mathrm{static}} = \wsyn_\mathrm{max}\tsoU$.

For facilition only ($\taurec=0$), the recovered partition remains fully available all the time ($R=1=\mathrm{const}$) and only the utilization varies with time.
Thus the $n$th synaptic weight is given by:

\begin{align}
	\wsyn_{n} & = \wsyn_\mathrm{max}\tsou_n \quad .
\end{align}
The time course of $\tsou$ now has to be emulated by the right-hand side of \cref{eqn:hw_stp2}; more precisely, we use $I$ to represent the course of $\tsou-\tsoU$.
Additionally we set $U_\mathrm{SE}=\tsoU$ and $\tau_\mathrm{stdf}=\taufacil$, and level the limits for the synaptic weights.
In the original model, $\tsou$ is always between $\tsoU$ and 1, while for the hardware model the STP factor is limited to values between 0 and 2 due to technical reasons.
By setting $\lambda=1$ and considering that $I$ is always within 0 and 1, the supplied range for $w_{\mathrm{STP}}^{\mathrm{facilitation}}$ is $[1-\beta,2-\beta]$.
In order to match the range of applied weights of both models, we need to solve the following system of equations:

\begin{align*}
	(1-\beta)\cdot  w_{\mathrm{static}} & = \tsoU \cdot \wsyn_\mathrm{max} \\
	(2-\beta)\cdot  w_{\mathrm{static}} & = 1 \cdot \wsyn_\mathrm{max} \quad .
\end{align*}
Solving for $w_{\mathrm{static}}$ and $\beta$ yields

\begin{align*}
  w_{\mathrm{static}} & = (1-\tsoU) \cdot \wsyn_\mathrm{max} \\
  \beta & = \frac{1-2\tsoU}{1-\tsoU} \quad .
\end{align*}
\subsection{Parameter ranges}

Here, we provide a full list of available parameter ranges for the BSS waferscale platform in \cref{table:hardware_parameter_ranges}.
As mentionend in \cref{sec:hicann}, one has the choice between two different capacitances in the hardware neuron.
The parameter ranges specified in \cref{table:hardware_parameter_ranges} correspond to the big capacitance (\SI{2.6}{\pico\farad}).
When using the small capacitance (\SI{0.4}{\pico\farad}) some parameter ranges change:
the limits of $\taum$ are multiplied by $\frac{0.4}{2.6}$, the ranges for $a$, $b$, and the synaptic weight are divided by $\frac{0.4}{2.6}$.
The ranges for electric potentials of the AdEx model ($\Vspike$, $\Er$, $\Vrest$, $\Vthresh$, $\Ereve$ and $\Erevi$) result from the following transformation from biological to hardware voltages (cf. \cref{sec:software}):
\begin{align}
	V_\mathrm{hardware} = \alpha_V\cdot V_\mathrm{bio} + V_\mathrm{shift} \quad, \label{eq:voltage_scaling}
\end{align}
with $\alpha_V=10$ and $V_\mathrm{shift}=\SI{1300}{\milli\volt}$.

In \cref{table:size_and_connectivity} we show how the tradeoff between total neuron number and maximum fan-in per neuron is realized on this device.

\begin{table}[t]
	\caption{{\bf List of typical usage scenarios of the wafer-scale hardware system}}
    \centering
    \begin{tabular}{r r r r}
    \toprule
        Nr of Neurons & Synapses/& DenMems/& Neurons/ \\
         & Neuron & Neuron& HICANN\\
        \midrule
		\num{196608}& 224         & 1 & 512\\
		\num{98304} & 448         & 2 & 256 \\
		\num{49152} & 896         & 4 & 128 \\
		\num{24576} & \num{1792}  & 8 & 64 \\
		\num{12288} & \num{3584}  & 16 & 32 \\
		\num{6144}  & \num{7168}  & 32 & 16 \\
		\num{3072}  & \num{14336} & 64 & 8 \\
        \bottomrule
    \end{tabular}
	\flushleft{One can either opt for many neurons with few synapses or for fewer neurons but a higher connection density.}
    \label{table:size_and_connectivity}
\end{table}

\subsection{Parameter Variation Measurements}
\label{sec:methods-hw-measurements}

\begin{figure*}[h!]
    \centering
    \includegraphics[width=\textwidth]{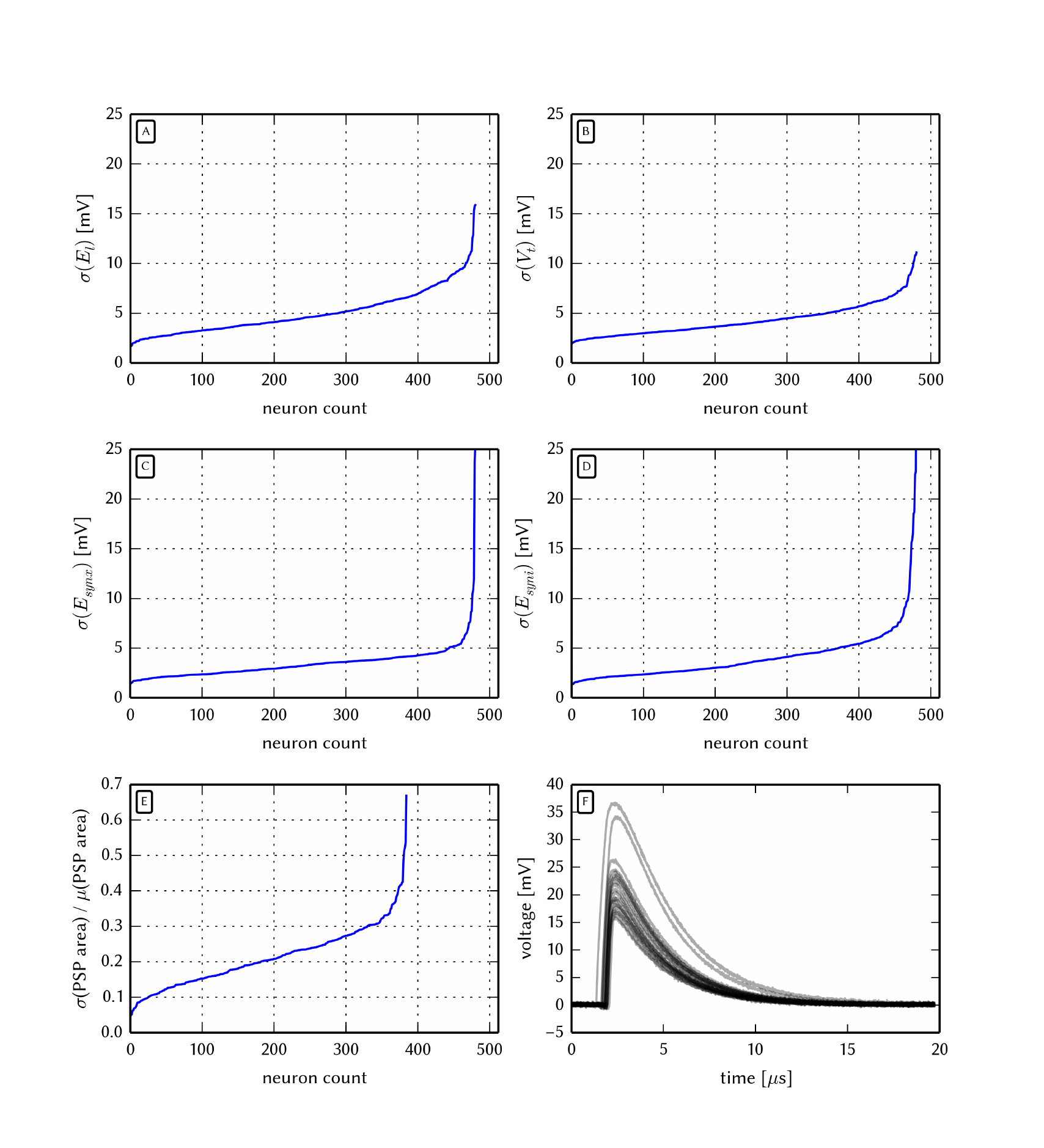}
    \caption{
(A-D) Cumulative distribution of trial-to-trial variation for selected parameters.
Each graph shows the number of neurons on one chip with a standard deviation of the measured value that is less than the value shown on the ordinate.
All values are given in hardware units.
In order to obtain values in the biological domain (\Cref{sec:software}), the voltages must be divided by the conversion factor of $\alpha_V=10$ (cf. \cref{eq:voltage_scaling}).
The standard deviation was estimated from 30 measurements for each neuron.
(A) Leakage potential
(B) Threshold potential
(C, D) Excitatory and inhibitory reversal potential
(E) Relative variation of the PSP integral.
The standard deviation was estimated from 20 trials per neuron.
Neurons were omitted from the measurements when an initial sweep over the available parameter range did not include the required PSP integral of \SI{8e-9}{\volt\second}.
(F) Example PSP traces for a randomly chosen neuron from the measurement in (E).
In order to minimize readout noise, each trace is an average over 400 individual PSPs which were evoked in short succession without rewriting floating gate parameters.
As the re-write variation is the main source of trial-to-trial variability (\cref{sec:hicann}), the variation within the 400 samples is much smaller than the trial-to-trial variation that is shown in figures (E) and (F).
}
\label{fig:methods-hw-trialtotrial}
\end{figure*}

\Cref{fig:methods-hw-trialtotrial} shows variation measurements on HICANN chips.
These measurements allow us to estimate the amount of variation that is present in the circuits (\cref{sec:hicann,sec:dist-weights}).

The measurements are conducted on a single-chip prototype system (plots \tb{A}-\tb{D}) and on one chip on a prototype wafer system (plots \tb{E} and \tb{F}).
Some neurons (on the right-hand-side of the plots) had been previously labeled non-functional and blacklisted, therefore showing no data points.
They will also be omitted during system operation.
Additionally, neurons that exhibit a larger variation than a chosen threshold can be blacklisted as well, reducing the total number of available neurons, but also limiting the magnitude of parameter noise.
This effect is not explicitly included in the ESS simulations in the main text, but it is conceptually covered by some of the experiments, where the network is restricted to only a small fraction of the wafer (\Cref{sec:l23alldistortions}), or where additionally parts of the synapses are declared as not available (\Cref{sec:synfire-ess}).

From the measurements in \cref{fig:methods-hw-trialtotrial}, we can e.g. estimate the variation of the voltages 
$\Vspike$, $\Vrest$, $\Ereve$ and $\Erevi$ in the biological domain: For all, the vast majority of neurons has a trial-to-trial variation below \SI{10}{\milli\volt} on the hardware, which corresponds to \SI{1}{\milli\volt} in the biological when using a voltage scaling factor $\alpha_V=10$ (cf. \cref{eq:voltage_scaling}).

\cleardoublepage

\clearpage
\section{Cortical layer 2/3 attractor memory}
\setcounter{page}{1}
\setcounter{table}{0}
\setcounter{equation}{0}
\setcounter{figure}{0}

\subsection{Original model parameters}
\label{sec:l23_original_params_appendix}

In \Cref{table:l23neuronparams}, \Cref{tab:l23synapseparams}, \Cref{tab:l23structure} and \Cref{tab:l23connectivity}, we summarize the parameters and characteristics of the original model, as found in \cite{lundqvist2006attractor}.
These have served as the basis for the model fit, for which the parameters can be found in the next subsection.

\begin{table*}
    \caption{
		{\bf Original neuron parameters}
        }
    \begin{tabular}{l c c c l}
    \toprule
        Parameter & PYR & RSNP & BAS & Unit \\
        \midrule
        $\gext$ & 0.082 & 0.15 & 0.15 & $\mu$S/mm$^2$ \\
        $\Eleak$ & -75 & 0.15 & -75 & mV \\
        $\ENa$ & 50 & 50 & 50 & mV \\
        $\ECa$ & 150 & 150 & 150 & mV \\
        $\EK$ & -80 & -80 & -80 & mV \\
        $\ECaNMDA$ & 20 & 20 & 20 & mV \\
        $\gleak$ & 0.74 & 0.44 & 0.44 & $\mu$F/mm$^2$ \\
        $\Cm$ & 0.01 & 0.01 & 0.01 & $\mu$F/mm$^2$ \\
        Soma diameter $\pm$ stdev & 21 $\pm$ 2.1 & 7 $\pm$ 0.7 & 7 $\pm$ 0.7 & $\mu$m \\
        $\gNa$ initial segment & 2500 & 2500 & 2500 & $\mu$S/mm$^2$ \\
        $\gK$ initial segment & 83 & 5010 & 5010 & $\mu$S/mm$^2$ \\
        $\gNa$ soma & 150 & 150 & 150 & $\mu$S/mm$^2$ \\
        $\gK$ soma & 250 & 1000 & 1000 & $\mu$S/mm$^2$ \\
        $\gNMDA$ & 75.0 & 75.0 & - & $\mu$S/mm$^2$ \\
        $\CaV$ influx rate & 1.00 & 1.00 & 1.00 & mV$^{-1}$ms$^{-1}$mm$^{-2}$ \\
        $\CaNMDA$ influx rate & 2.96 & 0.0106 & - & s$^{-1}$mV$^{-1}\mu$S$^{-1}$ \\
        $\CaV$ decay rate & 6.3 & 4 & - & s$^{-1}$ \\
        $\CaNMDA$ decay rate & 1 & 1 & - & s$^{-1}$ \\
        $\gK$ ($\CaV$) & 29.4 & 105 & 0.368 & nS \\
        $\gK$ ($\CaNMDA$) & 40 & 40 & - & nS \\
        \# compartments & 6 & 3 & 3 & \\
        Dendritic area (relative soma) & 4 & 4 & 4 & \\
        Initial segment area (relative soma) & 0.1 & 0.1 & 0.1 & \\
        \bottomrule
    \end{tabular}
    \label{table:l23neuronparams}
\end{table*}

\begin{table*}
	\caption{{\bf Original synapse parameters}}
    \begin{tabular}{l l l l c c c c c}
        \toprule
        Pre $\rightarrow$ Post & Type & Duration [s] & $\tauraise$ [s] & $\taudecay$ [s] & $\Erev$ [mV] & $\tsoU$ & $\taurec$ [s] & $E_\mathrm{slow}$ [mV] \\
        \midrule
        PYR $\rightarrow$ PYR & Kainate/AMPA & 0.0 & 0.0 & 0.006 & 0 & 0.25 & 0.575 & - \\
        PYR $\rightarrow$ PYR & NMDA & 0.02 & 0.005 & 0.150 & 0 & 0.25 & 0.575 & 0.020 \\
        PYR $\rightarrow$ BAS & Kainate/AMPA & 0.0 & 0.0 & 0.006 & 0 & - & - & - \\
        PYR $\rightarrow$ RSNP & Kainate/AMPA & 0.0 & 0.0 & 0.006 & 0 & - & - & - \\
        PYR $\rightarrow$ RSNP & NMDA & 0.02 & 0.005 & 0.150 & 0 & - & - & 0.020 \\
        BAS $\rightarrow$ PYR & GABA & 0.0 & 0.0 & 0.006 & -85 & - & - & - \\
        RSNP $\rightarrow$ PYR & GABA & 0.0 & 0.0 & 0.006 & -85 & - & - & - \\
        \bottomrule
    \end{tabular}
    \label{tab:l23synapseparams}
\end{table*}

\begin{table*}
	\caption{{\bf Original network structure: number of neurons per functional unit}}
    \begin{tabular}{c c c c c c c}
        \toprule
         & HCs & MCs & PYR & BAS & RSNP & total neurons\\
        \midrule
        per MC & - & - & 30 & 1 & 2 & 33 \\
        per HC & - & 8 & 240 & 8 & 16 & 264 \\
        network total & 9 & 72 & 2160 & 72 & 144 & 2376 \\
        \bottomrule
    \end{tabular}
    \label{tab:l23structure}
\end{table*}

\begin{table}
	\caption{{\bf Original network structure: connection probabilities}}
    \begin{tabular}{c c}
        \toprule
        \multicolumn{2}{c}{within an MC} \\
        \midrule
        PYR $\rightarrow$ PYR & 0.25 \\
        RSNP $\rightarrow$ PYR & 0.70 \\
        \toprule
        \multicolumn{2}{c}{between MCs inside the same HC} \\
        \midrule
        PYR $\rightarrow$ BAS & 0.70 \\
        BAS $\rightarrow$ PYR & 0.70 \\
        \toprule
        \multicolumn{2}{c}{between MCs in different HCs} \\
        \midrule
        PYR $\rightarrow$ PYR & 0.30 \\
        PYR $\rightarrow$ RSNP & 0.17 \\
        \bottomrule
    \end{tabular}
    \label{tab:l23connectivity}
\end{table} 

\subsection{Fitted Hardware-Compatible Parameters}
\label{sec:l23_fit_appendix}

\begin{figure*}
  \centering
  \includegraphics[width=\textwidth]{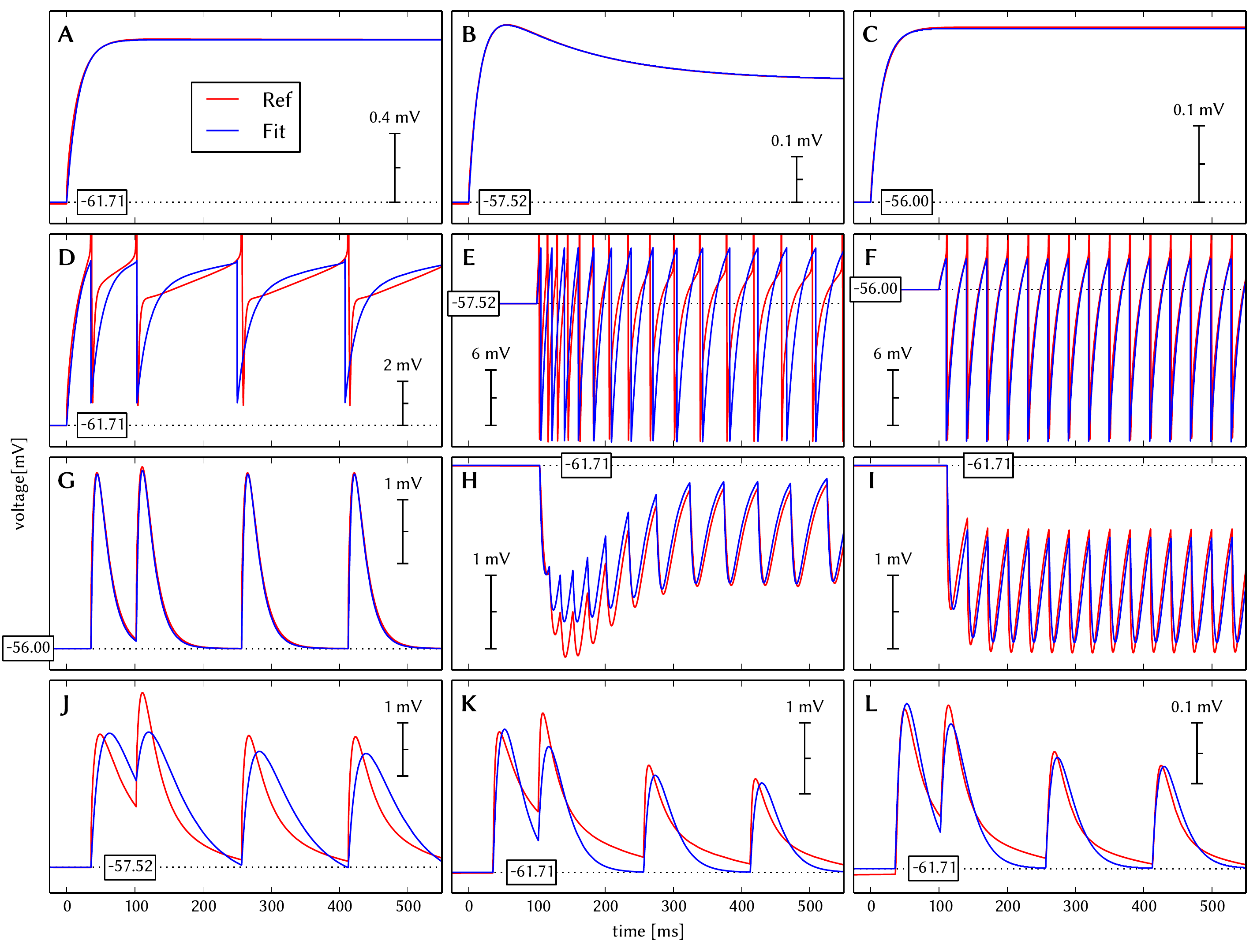}%
  \caption{
    {\bf Comparison of original neuron and synapse dynamics to the fitted dynamics of hardware-compatible models.}
    (\textbf{A - C}) Membrane potential of the three different cell types (PYR, RSNP, and BAS, respectively) under subthreshold current stimulation.
    These were used to determine the rest voltage $\Vrest$, total equivalent membrane capacitance $\Cm$, membrane time constant $\taum$ and the adaptation coupling parameter $a$.
    (\textbf{D - F}) Membrane potential of the three different cell types (PYR, RSNP, and BAS, respectively) under spike-inducing current stimulation.
    While the precise membrane potential time course of the original neuron model can not be reproduced by a single-compartment AdEx neuron, it was possible to reproduce the spike timing and especially average firing rates quite accurately.
    A small deviation of spiking frequency can be observed for RSNP cells during the first 50 ms - in the original model, they adapt slower than their AdEx counterparts.
    From these fits, the values for the absolute refractory period $\tauref$, reset voltage $\Vreset$, threshold voltage $\VT$, slope factor $\deltaT$, spike-triggered adaptation $b$ and adaptation time constant $\tau_w$ were extracted.
    (\textbf{G - L}) PSP fit results for all synapse types of the L2/3 model (PYR$\rightarrow$BAS, RSNP$\rightarrow$PYR, BAS$\rightarrow$PYR, PYR$\rightarrow$RSNP, PYR$\rightarrow$PYR within a MC, and PYR$\rightarrow$PYR between MCs, respectively). The output spikes from \textbf{D - F} have been used as input. These fits were used to determine synaptic weights $\wsyn$, time constants $\tausyn$ and the TSO parameters $U$ and $\tauref$.
    Because the hardware synapses only support a single conductance decay time constant, as opposed to the two different time constants in the original model for AMPA/kainate and NMDA, we have chosen an intermediate value for $\tausyn$, which constitutes the main reason for the difference in PSP shapes. A second reason lies in the saturating nature of synaptic conductances in the original model, which can not be emulated on the hardware without affecting the required TSO parameters (see \Cref{sec:l23_fit}).
  \label{fig:l23fit_appendix}
}
\end{figure*}

\Cref{tab:l23neuronparamsfitted}, \Cref{tab:l23synapseparamsfitted} and \Cref{tab:l23stimulusparamsfitted} contain all parameters required for the fits described in \Cref{sec:l23_fit}.
All fits were performed by minimizing the $L^2$-norm of the distance between the simulated traces (\Cref{fig:l23fit_appendix} \textbf{A - C, G - L}) or between spike timings (\Cref{fig:l23fit_appendix} \textbf{D - F}).

The diffuse background stimulus was generated by Poisson spike sources at a total rate of \SI{300}{\hertz} per PYR cell.

Apart from random noise, the PYR cells further receive input from other PYR cells in cortical layer 4.
The input intensity was calculated from the number of cells in layer 4 likely to project onto layer 2/3, which was estimated to be around 30 with a rate of approximately \SI{10}{\hertz} and a connection density of $25\,\%$ \cite{lundqvist2006attractor}.

Therefore, a Poisson process with $75\,\text{Hz}$ was used for each PYR cell input.
Since we used static synapses for the Poisson input, the synaptic weights for source-PYR connections were chosen as 30\% of PYR-PYR connections within the MCs.
This was verified for compliance the original model from \cite{lundqvist2006attractor}, which uses 7 to 8 sources per stimulated PYR cell with a rate of  $10\,\text{Hz}$ each and depressing synapses.
For each stimulus event in the pattern completion and rivalry experiments (described below), layer 4 cells were set to fire for $60\,\text{ms}$.
In each stimulated MC, 6 PYR cells were targeted from layer 4.

\Cref{tab:fittedfrequencies} shows the average firing rates for the different cell types in the network when only certain synapses are active.

\begin{table*}
	\caption{{\bf Fitted neuron parameters for the L2/3 model}}
  \begin{tabular}{l c c c l l}
    \toprule
    Parameter & PYR & RSNP & BAS & Unit & Comment \\
    \midrule
    $\Cm$ & 0.179 & 0.0072 & 0.00688 & nF & from the fits in \Cref{fig:l23fit_appendix} \tb{A}-\tb{C} \\
    $\Ereve$ & 0.0 & 0.0 & 0.0 & mV & difference to original model compensated by synaptic weights \\
    $\Erevi$ & -80.0 & - & - & mV & difference to original model compensated by synaptic weights \\
    $\taum$ & 16.89 & 15.32 & 15.64 & ms & from the fits in \Cref{fig:l23fit_appendix} \tb{A}-\tb{C} \\
    $\tauref$ & 0.16 & 0.16 & 0.16 & ms & minimum available in hardware at the used speedup \\
    $\tausyne$ & 17.5 & 66.6 & 6.0 & ms & see paragraph "Synapses" \\
    $\tausyni$ & 6.0 & - & - & ms & see paragraph "Synapses" \\
    $\Vreset$ & -60.7 & -72.5 & -72.5 & mV & from the fits in \Cref{fig:l23fit_appendix} \tb{D}-\tb{F} \\
    $\Vrest$ & -61.71 & -57.52 & -56.0 & mV & from the fits in \Cref{fig:l23fit_appendix} \tb{D}-\tb{F} \\
    a & 0.0 & 0.28 & 0.0 & nS & see fig from the fit in \Cref{fig:l23fit_appendix} \tb{B} \\
    b & 0.0132 & 0.00103 & 0.0 & nA & from the fits in \Cref{fig:l23fit_appendix} \tb{D}, \tb{E} \\
    $\deltaT$ & 0.0 & 0.0 & 0.0 & mV & from the fits in \Cref{fig:l23fit_appendix} \tb{D}-\tb{F} \\
    $\tauw$ & 196.0 & 250.0 & 0.0 & ms & from the fits in \Cref{fig:l23fit_appendix} \tb{D}, \tb{E} \\
    $\Vspike$ & -53.0 & -51.0 & -52.5 & mV & from the fits in \Cref{fig:l23fit_appendix} \tb{D}-\tb{F} \\
    $\VT$ & - & - & - & mV & not used since $\deltaT=0$\\
    \bottomrule
  \end{tabular}
  \label{tab:l23neuronparamsfitted}
\end{table*}

\begin{table*}
  \caption{{\bf Fitted synapse parameters for the L2/3 model}}
  \begin{tabular}{c c c c c c c}
    \toprule
    Pre-Post & type & weight [$\mu$S] & $\tausyn$ [ms] & U & $\taurec$ [ms] & $\taufacil$ [ms] \\
    \midrule
    PYR-PYR (local) & exc & 0.004125 & 17.5 & 0.27 & 575. & 0. \\
    PYR-PYR (global) & exc & 0.000615 & 17.5 & 0.27 & 575. & 0. \\
    PYR-BAS & exc & 0.000092 & 6.0 & - & - & - \\
    PYR-RSNP & exc & 0.000024 & 66.6 & - & - & - \\
    BAS-PYR & inh & 0.0061 & 6.0 & - & - & - \\
    RSNP-PYR & inh & 0.0032 & 6.0 & - & - & - \\
    background-PYR & exc & 0.000224 & 17.5 & - & - & - \\
    \bottomrule
  \end{tabular}
  \label{tab:l23synapseparamsfitted}
\end{table*} 

\begin{table}
	\caption{{\bf Stimulus parameters for the L2/3 model}
    }
  \begin{tabular}{c c}
    \toprule
    \multicolumn{2}{l}{\textbf{Background}}\\
    \midrule
    \# of sources per PYR & 1\\
    rate & $\SI{300}{\hertz}$\\
    weight & $\SI{0.000224}{\micro\siemens}$\\
    \midrule
    \multicolumn{2}{l}{\textbf{Shared background pool}}\\
    \midrule
    \# of sources per PYR & $100$ out of $5000$ total\\
    rate & $\SI{3}{\hertz}$\\
    weight & $\SI{0.000224}{\micro\siemens}$\\
    \midrule
    \multicolumn{2}{l}{\textbf{L4}}\\
    \midrule
    \# of sources per MC & 5\\
    $p_{\text{L4} \rightarrow \text{PYR}}$ & $0.75$ \\
    weight & $\SI{0.0012375}{\micro\siemens}$\\
    & ($30\%$ local PYR$\rightarrow$PYR)\\
    \bottomrule
  \end{tabular}
  \label{tab:l23stimulusparamsfitted}
\end{table} 

\begin{table*}
  \caption{{\bf Average firing rates (in Hz) of the different cell types of the L2/3 model with only certain synapses active}}
  \begin{tabular}{c c c c c}
    \toprule
    setup no. & active synapses & $\nu_\mathrm{PYR}$ & $\nu_\mathrm{RSNP}$ & $\nu_\mathrm{BAS}$ \\
    \midrule
    1 & background-PYR, PYR-BAS, PYR-RSNP & 0.738 $\pm$ 0.096 & 57.946 $\pm$ 6.993 & 4.655 $\pm$ 1.081 \\
    2 & same as 1 + BAS-PYR & 0.174 $\pm$ 0.021 & 13.430 $\pm$ 1.910 & 1.119 $\pm$ 0.441 \\
    3 & same as 1 + RSNP-PYR & 0.257 $\pm$ 0.037 & 20.375 $\pm$ 2.536 & 1.783 $\pm$ 0.954 \\
    4 & same as 2 + 3 + PYR-PYR (local) & 0.200 $\pm$ 0.030 & 14.679 $\pm$ 2.261 & 1.258 $\pm$ 0.544 \\
    5 & same as 2 + 3 + PYR-PYR (global) & 0.204 $\pm$ 0.078 & 14.954 $\pm$ 5.680 & 1.337 $\pm$ 0.625 \\
    \bottomrule
  \end{tabular}
  \label{tab:fittedfrequencies}
\end{table*}

\subsection{Delays}
\label{sec:l23_delays}
Each connection within the same MC was set to have constant synaptic delay of $\SI{0.5}{\milli\second}$. Additionally, axonal delays for connections between different MCs were realized by taking into account their spatial distance at an average axonal propagation speed of $\SI{200}{\micro\meter/\milli\second}$.
Both the HCs in the whole network as well as the MCs within a single HC are laid out on a hexagonal grid with a edge length of $\SI{500}{\micro\meter}$ (HC$\leftrightarrow$HC) / $\SI{60}{\micro\meter}$ (MC$\leftrightarrow$MC). In the default network (9HC$\times$9MC) this leads to delays between $\SI{0.5}{\milli\second}$ and $\SI{8}{\milli\second}$.

\subsection{Scaling}
\label{sec:methods-l23-scaling}

Due to the modularity of this network model, several straightforward possibilities exist for increasing or decreasing its size without affecting its basic functionality.
One can vary the total number of neurons simply by modifying the number of cells per MC.
One can also vary the number of MCs per attractor by varying the total number of HCs.
And finally, one can change the number of attractors by changing the number of MCs per HC accordingly.

All such changes need to be accompanied by corresponding modifications in connectivity in order to preserve the network dynamics.
This has been done by keeping the average input current per neuron within an active attractor constant, which is equivalent to conserving the fan-in for every neuron from every one of its afferent populations and leads to the scaling rules shown in \Cref{tab:scalingrules}.
In order to facilitate a comparison with the original results from \cite{lundqvist2006attractor} and \cite{lundqvist2010bistable}, we have only considered homogeneous changes, meaning that all modules (MCs, HCs) were equal in size and symmetrically connected.

The connections to the BAS cells required special treatment for two reasons.
Firstly, during an active state, they receive input from a single MC, but are excited by all MCs in a HC during the competition period between active attractors.
Only one aspect can be preserved when scaling and we have considered the dynamics during UP states as most important, leading to a "PYR $\rightarrow$ BAS" scaling rule independent of $\Nmc$.
Secondly, because PYR cells in MCs only project to the nearest 8 BAS cells, there are always precisely 8 active BAS cells per HC within an active attractor, which yields a simple "BAS $\rightarrow$ PYR" scaling rule.
When decreasing the number of attractors however, the number of existing BAS cells per HC also decreases, making an appropriate connection density scaling necessary.
This is the reason for the two different "BAS $\rightarrow$ PYR" scaling rules found in \Cref{tab:scalingrules}.

\begin{table}
    \caption{
		{\bf Scaling rules for the connection densities of the L2/3 model}
	}
    \begin{tabular}{ll}
        \toprule
        \text{Connection} & \text{ Scaled conn. prob.} $\tilde p$ \\
        \midrule
        PYR $\rightarrow$ PYR (same MC) & $29 / (\Npyr-1) \cdot p$ \\
        PYR $\rightarrow$ PYR (different MC) & $30 / \Npyr \cdot 8 / (\Nhc-1) \cdot p$ \\
        PYR $\rightarrow$ RSNP & $30 / \Npyr \cdot 8 / (\Nhc-1) \cdot p $ \\
        PYR $\rightarrow$ BAS & $30 / \Npyr \cdot p$ \\
        RSNP $\rightarrow$ PYR & $2 / \Nrsnp \cdot p$ \\
        BAS $\rightarrow$ PYR (Enlarging) & $1 / \Nbas \cdot p$ \\
        BAS $\rightarrow$ PYR (Shrinking) & $1 / \Nbas \cdot 8 / \Nmc \cdot p$ \\
        \bottomrule
    \end{tabular}
	\flushleft{
		$N_x$ represents the number of units of type $x$ (the original values are found in \Cref{tab:l23structure}).
		$p$ represents the original connection probability as found in table \Cref{tab:l23connectivity}.
		Whenever a scaled probability $\tilde p$ exceeded $1$, it was clipped to $1$, but the weights of the corresponding synapses were also increased by $\tilde \wsyn = \wsyn \cdot \tilde p$.
		}
    \label{tab:scalingrules}
\end{table}

\Cref{tab:l23_scaling_mapping} shows the combinations of $\Nhc$ and $\Nmc$ used for the quantification of synapse loss after mapping the L2/3 model onto the hardware in \cref{figure:sweepL23}.
In these mapping sweeps the diffusive background noise was modeled, as for the large-scale network ported to the ESS (\cref{sec:l23alldistortions}), with a background pool of \num{5000} Poisson sources and every PYR cell receiving input from 100 of the sources. 
\begin{table}
  \caption{{\bf Scaling table for the L2/3 model used for the synapse loss estimation in \cref{figure:sweepL23}}}
  \centering
  \begin{tabular}[h]{r r r}
    \toprule
	$\Nhc$ &  $\Nmc$ &  total neurons\\
    \midrule
	18&2&\num{1188}\\
	9&6&\num{1782}\\
	27&3&\num{2673}\\
	18&6&\num{3564}\\
	36&4&\num{4752}\\
	9&18&\num{5346}\\
	18&12&\num{7128}\\
	27&9&\num{8019}\\
	18&18&\num{10692}\\
	18&36&\num{21384}\\
	36&24&\num{28512}\\
	36&36&\num{42768}\\
	27&54&\num{48114}\\
	45&45&\num{66825}\\
    \bottomrule
  \end{tabular}
  \label{tab:l23_scaling_mapping}
\end{table}

\subsection{UP-state detection}
\label{sec:l23-upstate_detection}

One crucial element of the analysis is the detection of UP-states from which various other properties such as dwell times, competition times as well as average spike-rates in UP- and DOWN-states are determined.
The method of choice for detecting UP-states is based on the fact that the mean spike rate of an attractor during an UP-state is much higher than the spike rate in all remaining patterns in their corresponding DOWN-states, whereas -- in times of competition -- two or more attractors have elevated but rather similar spike rates.
A measure which quantifies this relationship is the standard deviation $\sigma$ of all mean spike rates per attractor \emph{at a given \mbox{time $t$}}.
The attractor with index $i$ is then said to be in an UP-state at time $t$ if the following relation holds:

\begin{align}
	r_i(t) > c\cdot \sigma (t) > \max_{r \in \{ 1, \dots, N_\text{MC}\} \textbackslash i} r_k(t) \quad ,
\end{align}
where $r_i(t)$ is the rate of attractor $i$ at time $t$ and $c$ is a numerical constant which is set to 1.

This method of detection has several advantages: it is based exclusively on spike trains (and not voltages or conductances, which are more difficult to read out and require much more storage space), it has a clear notion of there being at most one UP-state at any given time and it is completely local (in time), meaning that a very large value somewhere on the time axis cannot bias the detection at other times.

In small networks with randomly spiking neurons, it might happen by chance that all but one of the spike rates lie below the (approximately) constant standard deviation.
These falsely detected UP-states are very short and can thus easily be filtered out by requiring a minimal duration for UP states, which we chose at $100\,ms$.
This value was chosen after investigating dwell time histograms, as it distinguishes reliably between random fluctuations and actual active attractors.

\subsection{Pattern Completion}
\label{sec:l23-pattern_completion}

Pattern completion is a basic property of associative-memory networks.
By only stimulating a subset of PYR cells pwithin a pattern, the complete pattern is recalled.
The activity first spreads within the stimulated MCs, turning them dominant in their corresponding HCs.
After that, the activity spreads further to other HCs -- while the already dominating MCs stabilize each other through mutual stimulation -- activating the whole pattern while suppressing all others.
All PYR cells in the corresponding attractor hence enter an UP-state.

To verify the pattern completion ability of the network, a series of simulations was performed.
In order to reduce the occurrence of spontaneously activating attractors -- which would interfere with the activation of the stimulated attractor -- competition was investigated in larger networks of size 25HC$\times$25MC, as they exhibit almost no spontaneous attractors (the competition time fractions are much higher, see \Cref{fig:comparison} \textbf{H}).

For each network, all of the 25 patterns were stimulated one by one in random order.
The time between consecutive stimuli was chosen to be $1000\,\text{ms}$ to ensure minimal influence between patterns.
The number of stimulated MCs (one per HC) was varied over the course of multiple simulations.

After simulation, each network was analyzed for successfully activated patterns.
An activation attempt was said to be successful if the stimulated pattern was measured as active within $200\,\text{ms}$ after the stimulus onset.
If another pattern was active up to $\SI{75}{\milli\second}$ or if the stimulated pattern had already been active between $20-500\,\text{ms}$ prior to the stimulus onset, the attempt was deemed invalid and ignored during the calculation of success ratios.
This was done to take into account the fact that a pattern is more difficult to activate when another one is already active or while it is still recovering from a prior activation.
From all valid attempts the success probability (assuming a binomial distribution of successful trials) was estimated using the Wilson interval
\begin{equation}
  \label{eq:wilson}
  \tilde p = \frac{1}{1+\frac{z^2}{n}} \left[\hat{p}+\frac{z^2}{2n} \pm z\sqrt{\frac{\hat{p}(1-\hat{p})}{n}+\frac{z^2}{4n^2}} \right]
\end{equation}
where $\hat{p}$ represents the success ratio, $n$ the number of valid attempts and $z=1$ the desired quantile.

For most experiments (regular, synaptic weight noise and homogeneous synaptic loss) the number of invalid activations was always below 5 (out of 25).
The only exception was the PYR population size scaling: starting at 15 PYR cells, the validity rate roughly halves for every reduction in size (by 5 PYR cells per step) due to the increased occurrence of spontaneous attractors.
For simulations carried out on the ESS, only 10 patterns out of 25 were stimulated.
Out of these 10 attempts, only 5 were valid, on average.

\subsection{Pattern Rivalry / Attentional Blink}
\label{sec:l23-pattern_rivalry}

Another important feature of the L2/3 model is its ability to reproduce the attentional blink phenomenon, i.e., the inability of one pattern, stimulated by layer 4 input, to terminate another already active pattern and become active itself.
This phenomenon was investigated through a series of different networks of same size as in \Cref{sec:l23-pattern_completion} (25HC$\times$25MC).
For each network, 24 out of 25 patterns were randomly assigned to 12 pairs.
Afterwards, pattern rivalry was tested on all of these pairs in intervals of $1000\,\text{ms}$. 

Let the two patterns in each pair be denoted $A$ and $B$.
In order to guarantee an immediate activation of pattern $A$, 6 out of 25 HCs were stimulated (as then all completion attempts are successful, see \Cref{fig:comparison} \textbf{N}).
Then, after a certain delay $\Delta T$, pattern $B$ was stimulated with a varying amount of HCs.
Both the number of stimulated HCs as well as the delay $\Delta T$ were varied for each network.

The same way as in \Cref{sec:l23-pattern_completion}, each network was then analyzed as to whether pattern $B$ was successfully activated or not.
If the competing pattern $B$ was activated within $200\,\text{ms}$ after the stimulus onset and stayed active for at least $100\,\text{ms}$, the attempt was counted as successful, otherwise it was deemed unsuccessful.
As before, attempts during which spontaneously activated patterns intervened were ignored.
From all successful and unsuccessful attempts, the success probability was then estimated the same way as in pattern completion, using \Cref{eq:wilson}.

The validity ratios for pattern rivalry are not significantly different from those discussed in \Cref{sec:l23-pattern_completion}.
Most experiments (regular, synaptic weight noise and homogeneous synaptic loss) have 10 to 12 valid attempts (out of 12).
As before, for the PYR population size scaling experiments, the number of valid attempts dropped progressively ($8.2\pm1.7$, $4.8\pm2.1$ and $2.2\pm1.5$ valid attempts for 15, 10 and 5 PYR per MCs respectively).
Simulations carried out on the ESS had an average of 4 (distorted case) and 6 (compensated case) valid attempts (out of 10).

Different network configurations have been compared in terms of \emph{attentional blink} by estimating the 0.5 iso-probability contour in the following way.
For every delay $\deltaT$, the transition point from below to above $0.5$ probability for successful activation of the second pattern was estimated by linearly interpolating between the two nearest data points with a success ratio of above and below $0.5$, respectively.
In case there were several such transition points only the one with the highest stimulus was considered.
If no transition point could be identified, the transition was fixed at at either $25$ or $0$ stimulated MCs, depending on whether all success ratios were above or below $0.5$.
When there were no valid attempts for a certain delay/stimulus pair, its success probability estimate was replaced by the median of all valid activation attempts for that particular time delay $\Delta T$ (this only occurred sporadically in ESS and PYR population size scaling with less than 15 PYR cells per MC).
After identifying the transition point for every time delay $\deltaT$, intermediate values were interpolated linearly.
Finally, the interpolated values were Gauss-filtered ($\mu = 0.25\;\times$ step size for $\deltaT$ in the dataset) to better approximate the true 0.5 iso-probability contour.

\subsection{Star plots}
\label{sec:l23starplots}

While the spiking activity of many cells can be visualized quite well in raster plots, illustrating the temporal evolution of their membrane potentials is less straightforward.
Here, we have chosen to use so-called star plots for visualizing both average voltages and average firing rates of entire cell populations.

\begin{figure}
    \centering          
    \includegraphics[width=\columnwidth]{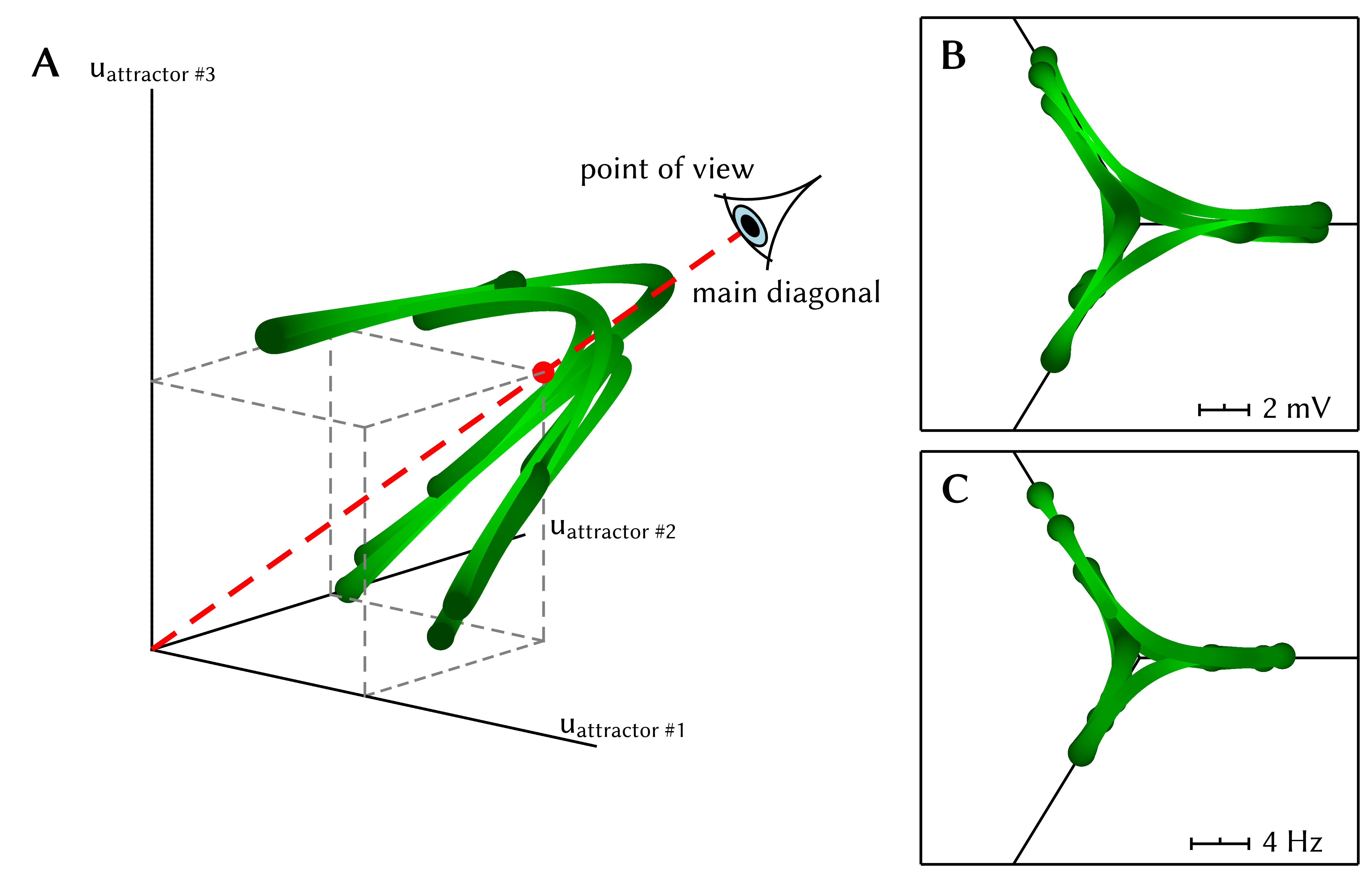}
    \caption{
		{\bf Visualization of the star plot as a projection in the case of a three-dimensional state space.}
        (\textbf{A}) Illustration of the view point with average membrane voltage data plotted in three-dimensional Cartesian coordinates. The data was taken from a (9HC$\times$3MC)-network and covers a \SI{2.5}{\second} period of network activity.
        (\textbf{B}) Resulting star plot from regular view point.
        (\textbf{C}) Star plot of the corresponding average attractor rate data.
    \label{fig:starplotvisualization}
        }
\end{figure}

In a system evolving in an abstract space with 3 dimensions, a star plot represents the orthogonal projection of the state space trajectory along the main diagonal of the corresponding Cartesian coordinate system onto a plane perpendicular to it.
For $n$ dimensions, points $\mathbf{x}$ in the star plot are no longer projections of the states $\mathbf{z}$, but are rather calculated as
\begin{equation}
    \mathbf{x} = \sum_{i=1}^n z_i \left( \cos \frac{2 \pi i}{n}, \sin \frac{2 \pi i}{n} \right)
\end{equation}
A visualization for $n=3$ is illustrated in \Cref{fig:starplotvisualization}.

In case of the L2/3 network, the number of dimensions is given by the number of attractors, with each axis describing some particular feature of the corresponding attractor (such as the average voltage or spike rate of the constituent PYR cells).

In addition to the position in state space, the state space velocity is also encoded in a star plot by both the thickness and the color of the trajectory. Especially in the case of the L2/3 network, this can be very useful in visualizing e.g. attractor stability or competition times.
Here, both line thickness and lightness were chosen proportional to $(\mathrm{const}+e^{-|d\mathbf{x}|/dt})$, with $\mathbf{x}$ being the position in state space.

\Cref{fig:synloss} \tb{B} and \tb{C} show two characteristic examples of star plots used for visualizing the dynamics of the L2/3 network.

\subsection{Average synaptic conductance due to Poisson stimulation}

For a single Poisson source with rate $\nu_i$ connected to the neuron by a synapse with weight $w_i$ and time constant $\tausyn$, the conductance course can be viewed as a sum of independent random variables, each of them representing the conductance change caused by a single spike.
In the limit of large $\nu_i$, the central limit theorem guarantees the convergence of the conductance distribution to a Gaussian, with moments given by

\begin{align}
    \expect{\gsyn_i} &= \sum_{\mathrm{spk} \, s} \expect{ w_i \Theta(t-t_s) \exp \left(-\frac{t-t_s}{\tausyn} \right) } \nonumber \\
                     &= \lim_{T \to \infty} \frac{\expect{N}}{T} w_i \int_0^T \exp \left( -\frac{t}{\tausyn} \right) \, dt \nonumber \\
                     &= w_i \nu_i \tausyn \quad . \label{eqn:cond_mean_single} \\
    \var{\gsyn_i}    &= \sum_{\mathrm{spk} \, s} \var{ w_i \Theta(t-t_s) \exp \left(-\frac{t-t_s}{\tausyn} \right) } \nonumber\\
                     &= \lim_{T \to \infty} \expect{N} \left\{ \expect{ \left[ w_i \Theta(t) \exp \left(-\frac{t}{\tausyn} \right) \right]^2 } \right. \nonumber \\
                     &+ \left. \expect{ \left[ w_i \Theta(t) \exp \left(-\frac{t}{\tausyn} \right) \right]}^2  \right\} \nonumber \\
                     &= \lim_{T \to \infty} \nu_i T \left\{ \frac{1}{T} w_i^2 \int_0^T \exp \left( -2\frac{t}{\tausyn} \right) \, dt \right. \nonumber \\
                     &- \left. \frac{1}{T^2} \left[ \int_0^T \exp \left( -\frac{t}{\tausyn} \right) \, dt \right] \right\} \nonumber \\
                     &= \frac{w_i^2 \nu_i \tausyn}{2} \quad . \label{eqn:cond_var_single}
\end{align}
Since conductances sum up linearly, N Poisson sources lead to an average conductance of

\begin{align}
    \expect{\gsyn}  &= \expect{\sum_{i=1}^N \gsyn_i} \nonumber \\
                    &= N \expect{w} \expect{\nu} \tausyn \quad .
    \label{eqn:cond_mean_multi}
\end{align}

\subsection{Detailed simulations of synapse loss and PYR population reduction}

\begin{figure*}
        \centering
        \includegraphics[width=\textwidth]{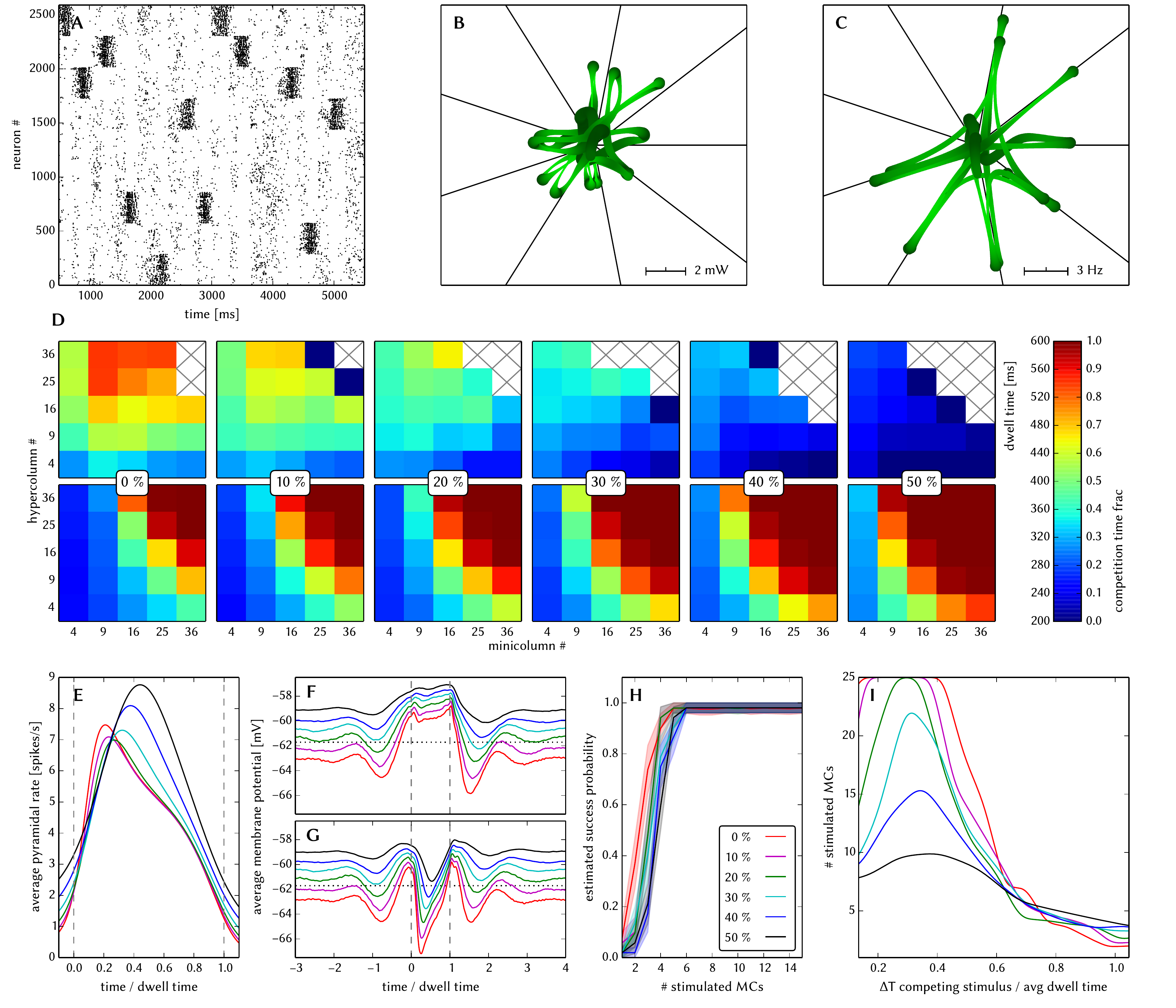}
        \caption{
				{\bf Effects of homogeneous synapse loss on the L2/3 model.}
                Unless explicitly stated otherwise, the default network model (9HC$\times$9MC) was used.
                The topmost 3 figures exemplify the dynamics of the network at 50\% synapse loss, all other figures show the effects of various degrees of synapse loss (0-50\%).
                (\textbf{A}) Raster plot of spiking activity.
                Only PYR cells are shown.
                The MCs are ordered such that those belonging to the same attractor (and \textit{not} those within the same HC) are grouped together.
                (\textbf{B}) Star plot of average PYR cell voltages from a sample of 5 PYR cells per MC.
                (\textbf{C}) Star plot of average PYR cell firing rates.
                (\textbf{D}) Average dwell times and relative competition times for various network sizes.
                (\textbf{E}) Average firing rate of PYR cells during an UP state.
                (\textbf{F}) Average voltage of PYR cells before, during and after their parent attractor is active (UP state).
                (\textbf{G}) Average voltage of PYR cells before, during and after an attractor they do not belong to is active.
                For the previous three plots, the abscissa has been subdivided into multiples of the attractor dwell time.
                In subplots \tb{F} and \tb{G} the dotted line indicates the leak potential $\Vrest$ of the PYR cells.
                (\textbf{H}) Pattern completion in a 25HC$\times$25MC network.
                (\textbf{I}) Attentional blink in a 25HC$\times$25MC network: $p=0.5$ iso-probability contours.
        \label{fig:synloss}
        }
\end{figure*}

\begin{figure*}
        \centering
        \includegraphics[width=\textwidth]{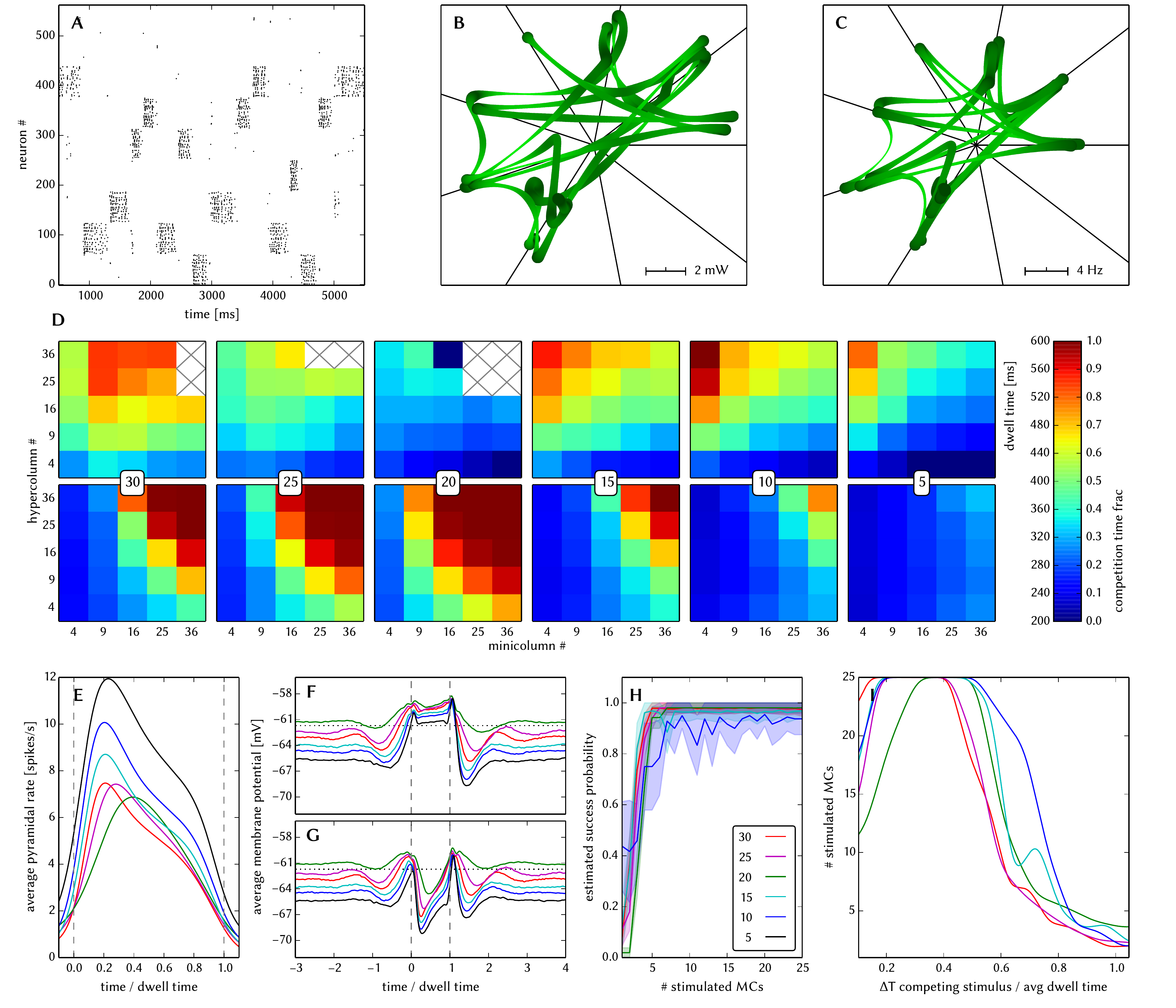}
        \caption{
				{\bf Effects of PYR population size scaling on the L2/3 model.}
                Unless explicitly stated otherwise, the default network model (9HC$\times$9MC) was used.
                The topmost 3 figures exemplify the dynamics of the network at 50\% of its original PYR population size, all other figures show the effects of various degrees of PYR population reduction (0-50\%).
                (\textbf{A}) Raster plot of spiking activity.
                Only PYR cells are shown.
                The MCs are ordered such that those belonging to the same attractor (and \textit{not} those within the same HC) are grouped together.
                (\textbf{B}) Star plot of average PYR cell voltages from a sample of 5 PYR cells per MC.
                (\textbf{C}) Star plot of average PYR cell firing rates.
                (\textbf{D}) Average dwell times and relative competition times for various network sizes.
                (\textbf{E}) Average firing rate of PYR cells during an UP state.
                (\textbf{F}) Average voltage of PYR cells before, during and after their parent attractor is active (UP state).
                (\textbf{G}) Average voltage of PYR cells before, during and after an attractor they do not belong to is active.
                For the previous three plots, the abscissa has been subdivided into multiples of the attractor dwell time.
                In subplots \tb{F} and \tb{G} the dotted line indicates the leak potential $\Vrest$ of the PYR cells.
                (\textbf{H}) Pattern completion in a 25HC$\times$25MC network.
                (\textbf{I}) Attentional blink in a 25HC$\times$25MC network: $p=0.5$ iso-probability contours.
                In \tb{H} and \tb{I}, the dataset for 5 PYR cells per MC was omitted because of its extremely low validity rate.
        \label{fig:pyrscaling}
        }
\end{figure*}

\Cref{fig:synloss} and \ref{fig:pyrscaling} show the effects of various levels of synapse loss and PYR population reduction, respectively.

\label{sec:l23_synloss_appendix}

\subsection{Synaptic weight noise}
\label{sec:l23_fixed_pattern_noise}
\begin{figure*}
  \centering
  \includegraphics{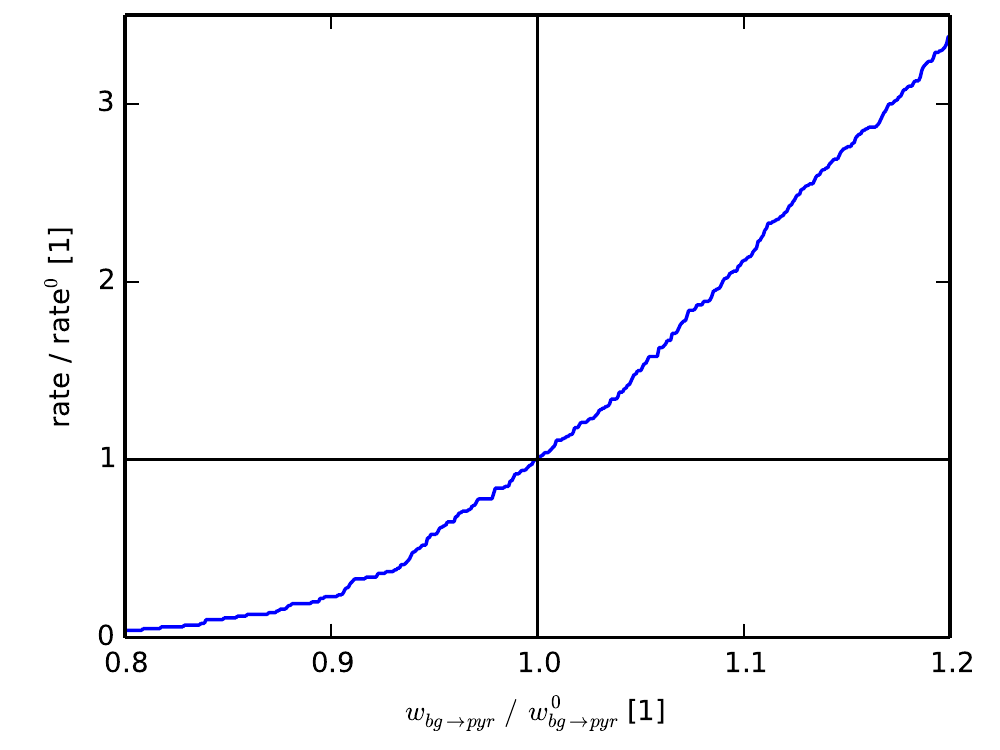}
  \caption{%
    Single PYR cell firing rate for different synaptic input weights. Each weight configuration was simulated for $\SI{100}{\second}$.
    \label{fig:pyr_rate_versus_weight}
  }
\end{figure*}

As can be seen in \Cref{fig:pyr_rate_versus_weight}, the firing rate of single PYR cells is highly dependent on the synaptic input weight that connects them to their respective Poisson source.
For example, a variation of $20\%$ in the input weight can cause the firing rate to either effectively vanish or more than triple.
This heavily distorts network dynamics as PYR cells within MCs will exhibit highly disparate firing rates, thereby disrupting the network's ability to maintain stable UP states (in which all participating PYR cells should fire roughly with the same rate).

\cleardoublepage

\clearpage
\section{Synfire chain with feed-forward inhibition}
\setcounter{page}{1}
\setcounter{table}{0}
\setcounter{equation}{0}
\setcounter{figure}{0}
\label{sec:methods-ffi}
\subsection{Model parameters}
\label{sec:methods-ffi-parameters}
\begin{table}
	\caption{{\bf Neuron parameters used in the synfire chain benchmark model}}
    \centering 
    \begin{tabular}{l c c}
        \toprule
        Parameter & Value& Unit             \\
        \midrule
        $\Cm$     & 0.29 & \nano\farad   \\
        $\tauref$ & 2    & \milli\second   \\
        $\Vspike$ & -57  & \milli\volt   \\
        $\Er$     & -70  & \milli\volt   \\
        $\Vrest$  & -70  & \milli\volt   \\
        $\taum$   & 10   & \milli\second  \\
        $\Ereve$  & 0    & \milli\volt     \\
        $\Erevi$  & -75  & \milli\volt   \\
        $\tausyne$& 1.5  & \milli\second \\
        $\tausyni$& 10   & \milli\second  \\
        \bottomrule
    \end{tabular}
    \label{tab:synfire-neuron-params}
\end{table}
\begin{table}
	\caption{{\bf Projection properties for the feed-forward synfire chain}
    }
    \centering
    \begin{tabular}[h]{l l l l} 
        \toprule 
        Projection & weight         & incoming & delay \\ 
                   & \micro\siemens & synapses  & \milli\second \\
        \midrule 
        RS$_n$ $\to$ RS$_{ n + 1 } $ & 0.001  & 60 & 20 \\
        RS$_n$ $\to$ FS$_{ n + 1 } $ & 0.0035 & 60 & 20 \\ 
        FS$_n$ $\to$ RS$_{ n } $     & 0.002  & 25 & 4  \\ 
        \bottomrule
    \end{tabular}
    \label{tab:synfire_ffi_connectivity}
\end{table}

The neuron and connectivity parameters are given in
\cref{tab:synfire-neuron-params} and
\cref{tab:synfire_ffi_connectivity}.

\subsection{Network scaling}
\label{sec:method_synfire_scaling}

In the default setup studied in this article, the synfire chain consists of 6 groups of 125 neurons (100 excitatory and 25 inhibitory).
In order to quantify the amount of synapse loss after mapping the network to the BrainScaleS wafer-scale hardware for different network sizes, we define the following network scaling rules.
When increasing the network size, we vary both the number of synfire groups and the number of neurons per group while keeping the number of incoming synapses per neuron constant (cf. \Cref{tab:synfire_ffi_connectivity}).
The fraction of inhibitory neurons always amounts to \SI{20}{\%}.
Neuron and synapse parameters are not altered.
\Cref{tab:synfire_scaling} lists the combinations of group size and group count used for the synapse loss estimation in \cref{fig:ffi_ess_all_comp}~\textbf{A}.

The background Poisson stimulus is scaled as follows.
For the hardware implementation of the synfire chain we can not use one individual Poisson source for each neuron due to input bandwidth limitations.
Instead, we assume one pool of 32 Poisson sources for each synfire group, and each neuron receives input from 8 random sources from that pool.
The size of the background pool is then scaled with the number of neurons per synfire group, while always drawing 8 sources from the pool per neuron.
This scaling of the background pool was chosen to make the total number of background sources proportional to the total number of neurons and independent of the group count.
\begin{table}
  \caption{{\bf Scaling table for the synfire chain used for the synapse loss estimation in \cref{fig:ffi_ess_all_comp}~\textbf{A}}}
  \centering
  \begin{tabular}[h]{r r r}
    \toprule
	groups &  group size &  total neurons\\
    \midrule
	8 & 125 & \num{1000} \\
	16 & 125 & \num{2000} \\
	24 & 125 & \num{3000} \\
	20 & 200 & \num{4000} \\
	25 & 200 & \num{5000} \\
	15 & 400 & \num{6000} \\
	20 & 350 & \num{7000} \\
	20 & 400 & \num{8000} \\
	30 & 300 & \num{9000} \\
	25 & 400 & \num{10000} \\
    \midrule
	20 & \num{500} & \num{10000} \\
	40 & \num{500} & \num{20000} \\
	60 & \num{500} & \num{30000} \\
	40 & \num{1000} & \num{40000} \\
	50 & \num{1000} & \num{50000} \\
	30 & \num{2000} & \num{60000} \\
	20 & \num{3500} & \num{70000} \\
	20 & \num{4000} & \num{80000} \\
	30 & \num{3000} & \num{90000} \\
	25 & \num{4000} & \num{100000} \\
    \bottomrule
  \end{tabular}
  \label{tab:synfire_scaling}
\end{table}

\subsection{Additional simulation}
\subsubsection{All distortion mechanisms}
\label{sec:synfire_appendix_all} 
\begin{figure}[htpb!]
    \centering
    \includegraphics{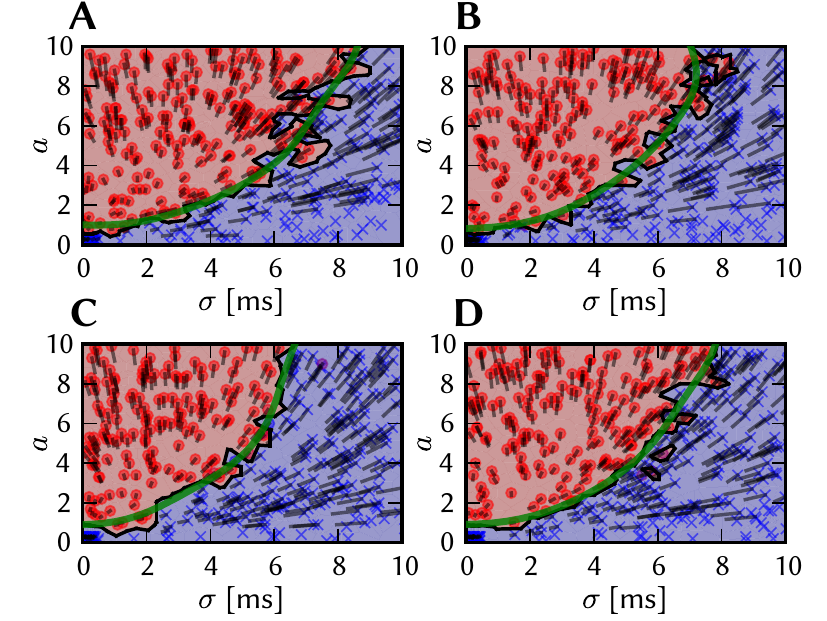}
    \caption{
		{\bf $(\sigma, a)$ state space of the synfire chain model with all compensation methods applied for four different levels of distortion.}
        %
        \textbf{(A)}
        \SI{30}{\%} synapse loss, \SI{20}{\%} weight noise 
        \textbf{(B)}                    
        \SI{30}{\%} synapse loss, \SI{50}{\%} weight noise 
        \textbf{(C)}                    
        \SI{50}{\%} synapse loss, \SI{20}{\%} weight noise 
        \textbf{(D)}                    
        \SI{50}{\%} synapse loss, \SI{50}{\%} weight noise 
    \label{fig:ffi_all_comp}
    }
\end{figure}

To check that the compensation methods do not interfere with each
other, all distortion mechanisms were applied simultaneously with weight noise values of
\SI{20}{\percent} and \SI{50}{\percent} and synapse loss values of
\SI{30}{\percent} and \SI{50}{\percent}, with an axonal delay of
\SI{1.0}{\ms}.
Without compensation no stable region exists in all four cases.
\cref{fig:ffi_all_comp} shows the result with all compensation methods applied.
When several methods required modification of a network parameter, all
modifications were applied.
For instance, in the case of the synaptic weight which was scaled by
both synapse loss and delay compensation methods, both scaling
factors were multiplied.
\Cref{fig:ffi_all_comp} shows the restoration of input selectivity
in all four cases.

\begin{figure}[htpb!]
    \centering
    \includegraphics{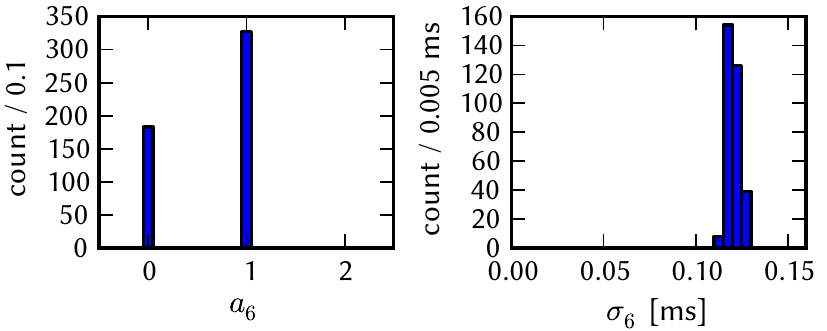}
    \caption{
		{\bf Distribution of $a_6$ and $\sigma_6$ in the reference experiment for the synfire chain model.}
    	\label{fig:synfire_end_a}
	}
\end{figure}

\begin{figure}[ht!]
    \centering
    \includegraphics{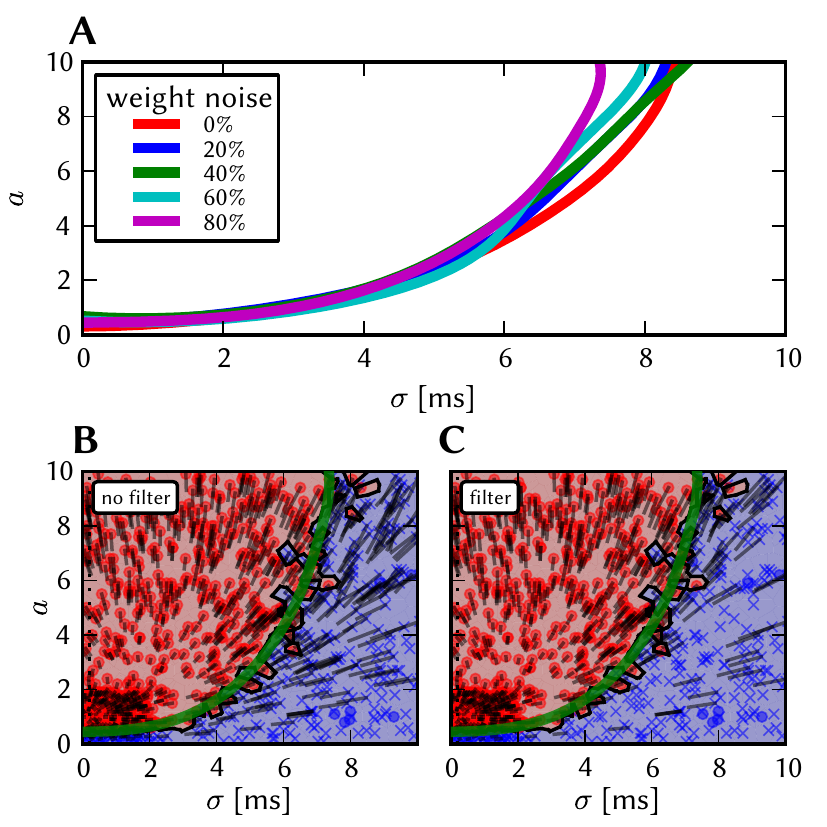}
    \caption{
		{\bf Demonstration of spontaneous event filter in the weight noise compensation (\cref{sec:synfire_spike_filter}).}
        \textbf{(A)} The same experiment as in \cref{fig:ffi_weight_noise} \textbf{C} 
        (weight noise with active compensation) but without the 
        filter for background spikes.
        The separatrix locations are comparable as the filter does not influence the result
        significantly in the compensated case.
        (\textbf{B}, \textbf{C}) Complete state space response for weight noise of 80\%, once 
        with, once without filter.
        This demonstrates that the applied filter does not affect the
        result in the compensated case.
    \label{fig:ffi_weight_noise_comp}
    }
\end{figure}

\subsubsection{Separatrix fit}
\label{sec:synfire_separatrix_fit}

To compare different separatrices, the $a$-values of the last group
are characterized as successful ($+1$) or extinguished ($-1$) and the resulting values
interpolated and smoothed by a gaussian kernel with a standard
deviation $(\SI{1.5}{ms}, 1.5)$ in the $(\sigma, a)$ space.
The iso-contour line of the resulting surface at a value of $0$ is
used as an approximation of the separatrix location, as shown
in \cref{fig:synfire_ffi_schema} \textbf{C} together with the individual simulation results.
Data points with $\sigma \le \SI{0.2}{\ms}$ were not included in the
fit to avoid distortions induced by bandwidth limitations in ESS simulations (\cref{sec:synfire-ess}) 
from affecting the fit quality.
The data points are still shown individually as blue dots and regions,
e.g., in \cref{fig:ffi_ess_all_comp}.
This modification was also included in the software simulations for consistency.
Cases in which the separatrix does not capture the relevant behavior,
e.g., if the separation is not reliable in a large region of the
state space, are shown separately.

\subsubsection{Weight noise compensation}

\Cref{fig:ffi_weight_noise_comp} \textbf{A} shows the separatrix in the case of compensated weight noise.

\subsubsection{Filtering of spontaneous activity}
\label{sec:synfire_spike_filter}

To prevent spontaneous background events from impeding the analysis,
spikes are discarded as part of spontaneous activity if less then $N$
spikes in the same excitatory group occur in a time window of $\pm T$.
The utilized values for $N$ and $T$ are given at each point where the
filter is applied;
They are chosen such that authentic synchronous volleys with $a \ge 0.5$
(which would be counted as successful propagation, as defined above)
are not removed.
\Cref{fig:ffi_weight_noise_comp} \textbf{B} and \textbf{C} show that
the influence of the filter for spontaneous activity is minimal in the
compensated case.

\subsubsection{Further ESS simulations}

\begin{figure}[ht!]
    \centering
\includegraphics{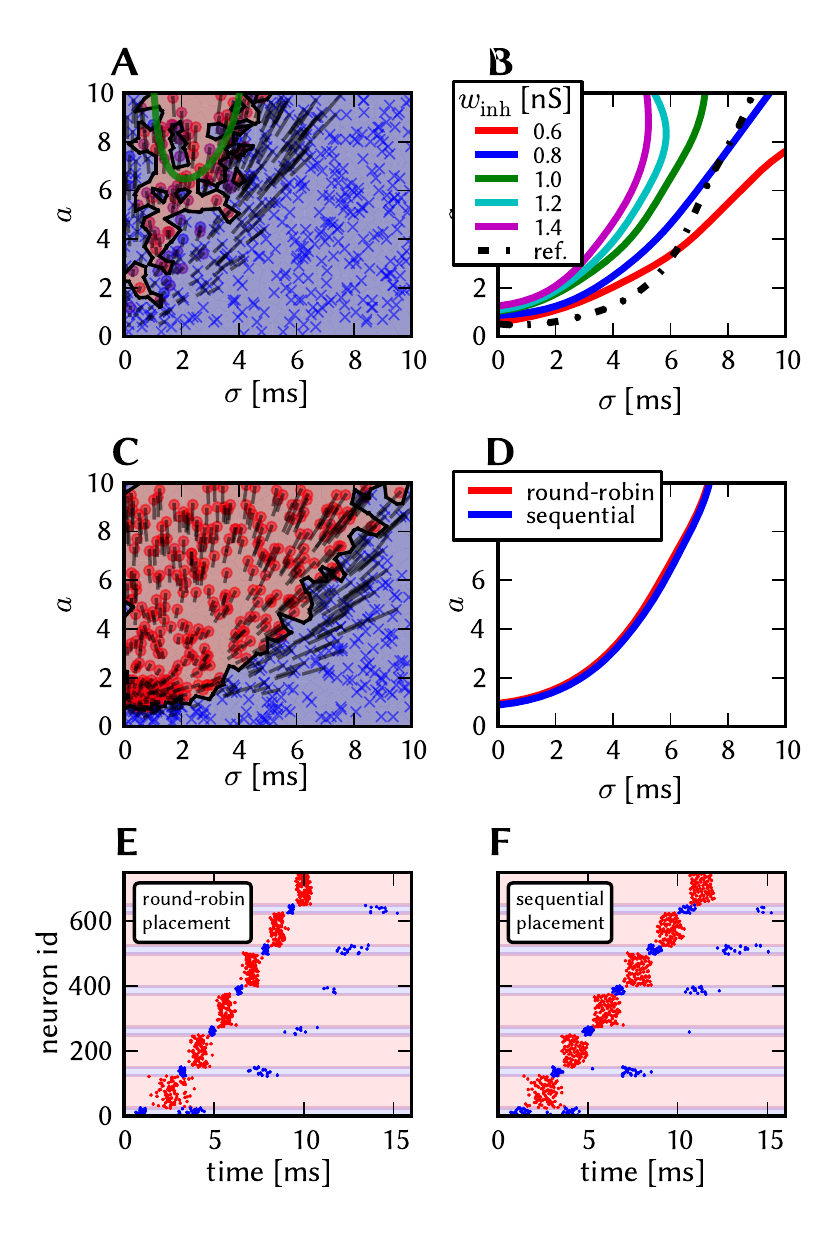}
    \caption{
		{\bf Additional simulations of the feed-forward synfire chain on the ESS without synapse loss:}
		(\textbf{A}) ($\sigma$,$a$) state space on the ESS with default parameters and 20\% weight noise.
		(\textbf{B}) After compensation of for all distortion mechanisms, different separatrices are possible by setting different values of the inhibitory weight.
		(\textbf{C}) Compensated state space belonging to the blue separatrix in \textbf{B}.
        $w$ refers to the synaptic weight of local inhibition.\
		(\textbf{D}-\textbf{F}) Investigation of effects of spike loss and jitter by using two different approaches for neuron placement.
		(\textbf{D}) Separatrices for round-robin and sequential neuron placement with parameters as for the green curve in \textbf{B}.
		Raster plots for round-robin (\textbf{E}) and sequential (\textbf{F}) neuron placement. Stimulus parameters: $a_0=1$ and $\sigma_0=\SI{1}{\milli\second}$.
    \label{fig:ffi_ess_noise_and_delay_comp}
    }
\end{figure}

\paragraph{Distortion and compensation without synapse loss}
For the ESS simulation in \cref{sec:synfire-ess} we enforced a certain amount of synapse loss by restricting the synfire chain network to very limited hardware resources.
However, due to its feed-forward structure, the network can be easily mapped onto the BrainScaleS hardware without any synapse loss (\Cref{fig:ffi_ess_all_comp} \textbf{A}).
Thus, we also investigated the network without synapse loss, such that the active distortion mechanisms in the ESS simulations were synaptic weight noise, non-configurable axonal delays as well as spike loss and jitter.
The state space of the distorted network (\Cref{fig:ffi_ess_noise_and_delay_comp} \textbf{A}) contains only a small and loosely connected region of sustained activity which indicates unreliable separation.
Applying the compensation mechanism for synaptic weight noise and axonal delays fully restores the filter property of the synfire chain, as can be seen in \cref{fig:ffi_ess_noise_and_delay_comp} \textbf{B}, where different separatrices mimic different delay-dependent realizations.
Compared to the compensation for all distortion mechanisms, the compensated state space without synapse loss does not show any flaws (\textbf{C}).

\paragraph{Effect of spike loss and jitter}
\label{hardware_synfire:jitter}

We investigated the effect of spike loss and jitter in the HICANN, where the spikes of the neurons connected to the same on-wafer routing bus are processed subsequently (\cref{sec:communication_infrastructure}), which can lead to spike time jitter and in rare cases to spike loss when firing is highly synchronized.

Which 64 neurons inject their spikes into a routing bus is determined by the placement of the neurons on the HICANN.
Hence, in order to study the effect of spike loss and jitter, we simulated the synfire chain network in two different placement setups:
First, neurons of the same synfire group were placed sequentially onto the same routing bus, and second, neurons were distributed in a round-robin manner over different routing buses, such that neurons of different groups injected their spikes into one routing bus.
Hence, we expect the spiking activity on each routing bus to be more synchronous in the first case than in the second.
In both setups, the utilized hardware and the number of neurons per routing bus was equal, allowing a fair competition between both.
The separatrices for the two different placement strategies with otherwise identical parameters are virtually indistinguishable (\Cref{fig:ffi_ess_noise_and_delay_comp} \textbf{D}).
Nevertheless, the raster plots (\Cref{fig:ffi_ess_noise_and_delay_comp} \textbf{E} and \textbf{F}) reveal the effect of the introduced jitter:
For sequential placement, the spread of spike times within a group is roughly double than for round-robin placement and also the onset of the volley in the last group comes \SI{1.5}{\ms} later.
In contrast to the reference simulation (cf. \cref{fig:synfire_end_a}), the fixed point of succesful propagation is not (\SI{0.12}{\ms},1)  but (\SI{0.21}{\ms},1) for round-robin and (\SI{0.36}{\ms},1) for sequential placement.

We conclude that, especially for dense pulses, the subsequent processing of spikes in the hardware leads to a temporal spread of the pulse volley, which however has virtually no influence on the filter properties of the synfire chain.

\cleardoublepage

\clearpage
\section{Self-sustained asynchronous irregular activity}
\setcounter{page}{1}
\setcounter{table}{0}
\setcounter{equation}{0}
\setcounter{figure}{0}

\label{sec:methods-ai}

\subsection{Network simulation setup}
The default model consists of 3920 neurons (\SI{80}{\%} pyramidal and \SI{20}{\%} inhibitory) equally distributed on a two-dimensional lattice of $1\times1$ \milli\meter$^2$ folded to a torus.
The connection probability is distance-dependent and is normalized such that each neuron receives synaptic input from 200 excitatory and 50 inhibitory neurons.
All simulations run for \SI{10}{\second}.
\SI{2}{\%} of all neurons in the network are initially stimulated by one individual Poisson source for \SI{100}{\milli\second} in order to induce initial network activity.
The default size was chosen such that the model can be fully realized on the BrainScaleS hardware without losing any synaptic connections in the mapping step (\Cref{sec:software}), thereby allowing us to compare topologically equivalent software simulations, with the only remaining difference lying in the non-configurable delays and dynamic constraints on the ESS.

\subsubsection{Model parameters}
\label{sec:methods-ai-parameters}
\begin{figure}[ht]
    \centering
	\includegraphics{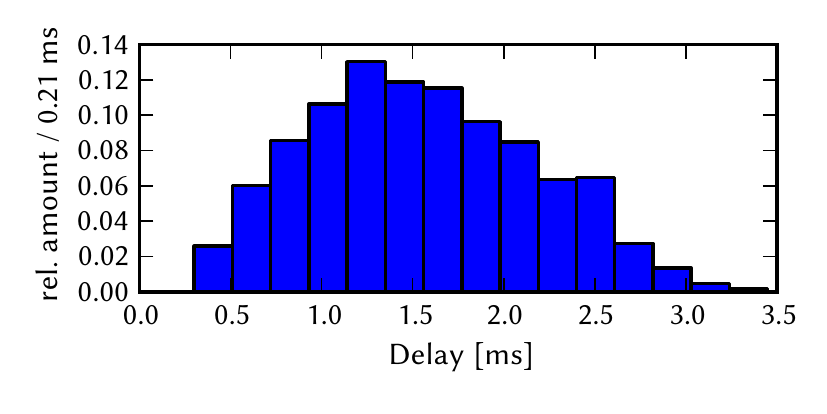}
    \caption{
	{\bf Histogram of delays in the AI network.}
    The mean delay is \SI{1.55}{\ms}.
    \label{fig:ai-delay-histogram}
        }
\end{figure}

The neuron parameters of the AdEx model used in this benchmark are listed in \cref{table:ai-neuron-params} and are equal to those in \cite{muller2012} with the only difference being that excitatory pyramidal cells have neuronal spike-triggered adaptation while inhibitory cells do not.
\begin{table}
    \caption{ {\bf AdEx Neuron parameters used in the AI network}}
    \centering
    \begin{tabular}{l c c l}
        \toprule
        Parameter     & Pyramidal & Inhibitory & Unit          \\
        \midrule                                     
        $\Cm$         & 0.25      & 0.25       & \nano\farad   \\
        $\tauref$     & 5         & 5          & \milli\second \\
        $\Vspike$     & -40       & -40        & \milli\volt   \\
        $\Er$         & -70       & -70        & \milli\volt   \\
        $\Vrest$      & -70       & -70        & \milli\volt   \\
        $\taum$       & 15        & 15         & \milli\second \\
	$a$           & 1         & 1          & \nano\siemens \\
	$b$           & 0.005     & 0          & \nano\ampere  \\
        $\adexdeltaT$ & 2.5       & 2.5        & \milli\volt   \\
        $\tauw$       & 600       & 600        & \milli\second \\
        $\Vthresh$    & -50       & -50        & \milli\volt   \\
        $\Ereve$      & 0         & 0          & \milli\volt   \\
        $\Erevi$      & -80       & -80        & \milli\volt   \\
        $\tausyne$    & 5         & 5          & \milli\second \\
        $\tausyni$    & 5         & 5          & \milli\second \\
        \bottomrule
    \end{tabular}
    \label{table:ai-neuron-params}
\end{table}
Sweeps are performed over the two-dimensional ($\Ge$, $\Gi$) parameter space, with the ranges being \SIrange{3}{11}{\nano\siemens} for $\Ge$ and \SIrange{50}{130}{\nano\siemens} for $\Gi$.
The Poisson sources for the initial network stimulation have a mean rate of $\SI{100}{\hertz}$ and project onto the network's neurons with a synaptic weight of \SI{100}{\nano\siemens}.
The distance-dependent connection probability has a Gaussian profile with a spatial width of $\sigma = \SI{0.2}{\milli\meter}$.
Synaptic delays depend on the distance according to the following equation: $t_\mathrm{delay} = \SI{0.3}{\milli\second} + \frac{d}{v_{\mathrm{prop}}}$, with $d$ being the distance between two cells and $v_\mathrm{prop} = \SI{0.2}{\milli\meter\per\milli\second}$ the spike propagation velocity.
The distribution of delays is shown in \cref{fig:ai-delay-histogram}, the average delay in the network amounts to \SI{1.55}{\ms}.

\subsubsection{Network scaling}
\label{sec:methods-ai-scaling}
When the network is scaled up in size, we only increase the number of neurons while keeping the number of afferent synapses per neuron constant.
All other parameters concerning the connectivity do not change, including the size of the cortical sheet, the distance-dependent delays and connection probability, as well as the ratio of excitatory to inhibitory cells.
Neuron and synapse parameters remain unaltered.


\subsection{Functionality criteria}
\label{sec:methods-ai-criteria}
The survival time is defined as the last spike time in the network.
If the network survives until the end of the simulation, we consider it as self-sustaining.
Additionally, several criteria are employed to characterize the network's activity, regularity and synchrony.

The mean firing rate of all pyramidal neurons is used to classify the overall activity of the network.
The variance of the firing rates across the pyramidal neurons measures the homogeneity of their response.
For a better comparison, we look at the relative variance, i.e., the coefficient of variation of the firing rates $\cvrate = \frac{\sigma(\nu)}{\overline{\nu}}$, where $\overline{\nu}$ and $\sigma(\nu)$ are the mean and standard deviation of the average firing rates $\nu$ of the inidividual neurons.

The coefficient of variation of interspike intervals ($\cvisi$) serves as an indicator of spiking regularity.
It is calculated via

\begin{align}
    \cvisi = \frac{1}{N} \sum_{i = 1}^N{ \frac{\sigma_i(\textrm{ISI})}{\overline{\textrm{ISI}_i} } }
\label{eq:cv-isi}
\end{align}
${\sigma_i(\textrm{ISI})}$ is the standard deviation of interspike intervals in the $i$-th spike train, while ${{\overline{\textrm{ISI}}}_i}$ is the mean interspike interval in the same spike train.
$N$ is the number of averaged spike trains which is set to the number of pyramidal cells for each simulation.
$\cvisi$ is 0 for a regular spike train and approaches 1 for a sufficiently long Poisson spike train.

The correlation coefficient $\text{CC}$ is defined via

\begin{align}
    \text{CC} = \frac{1}{P} \sum_{j,k}^P{ \frac{ \text{Cov}(S_j, S_k)} {\sigma(S_j) \sigma(S_k)} }
\label{eq:cc}
\end{align}
The sum runs over $P = 5000$ randomly chosen pairs of spike trains $(j, k)$ from the excitatory population.
$S_i$ is the time-binned spike count in the $i$-th spike train with a bin width of $\Delta = \SI{5}{\ms}$. $\sigma(S_i)$ denotes the standard deviation of $S_i$, and $\text{Cov}(S_j, S_k)$ the covariance of $S_j$ and $S_k$.
$\text{CC}$ approaches 0 for sufficiently long independent spike trains and is 1 for linearly dependent $(S_j, S_k)$.
The simulation results were cross-checked with a bin width of $\Delta = \SI{2}{\ms}$.

The power spectrum $S(\omega)$ of a spike train is calculated via

\begin{align}
   A_k &=  \sum_{m=0}^{N-1} r_m \exp\left(-2\pi i{mk \over N}\right) \qquad k = 0,\ldots,N-1
   \\
   \omega_k &:= \frac{2 \pi k}{N \Delta}
   \\
   S(\omega_k) &:= |A_k|^2 N \Delta
   \label{eq:power_spectrum}
\end{align}
using the time-binned population firing rate 
$r_i$ with $i \in \left\{0, ..., N - 1\right\}$ 
with a bin width of $\Delta$ for a spike train
of length $N \Delta$ (see, e.g. 3.1.4 in \cite{rieke1997spikes}).
For the AI network we used a bin width of $\Delta=\SI{1}{\ms}$ for calculating the raw power spectra, and a $\sigma=\SI{5}{\Hz}$ for the Gauss-filtered versions which where then used to determine the peak frequency (i.e. the first non-zero peak in the power spectrum).

In case of the L2/3 model, the power spectra were calculated from Gauss-filtered ($\sigma=\SI{5}{\milli\second}$) spike data with a bin width of $\Delta=\SI{0.1}{\milli\second}$ and (unless otherwise stated) smoothed with a $\sigma=\SI{0.3}{\milli\second}$ Gauss-filter.

For all statistics, the first second of the simulation is left out, i.e. only the 9 seconds from \SIrange{1}{10}{\second} are considered.
If the network did not survive until the end of the simulation, the firing rate was calculated between \SI{1}{\second} and the survival time, or between \SI{0.1}{\second} and the survival time for the case when the latter was smaller than \SI{1}{\second}.
%
%

\subsection{Iterative compensation}
\label{sec:methods-ai-iterative-comp}
\begin{figure}[ht]
    \centering
	\includegraphics{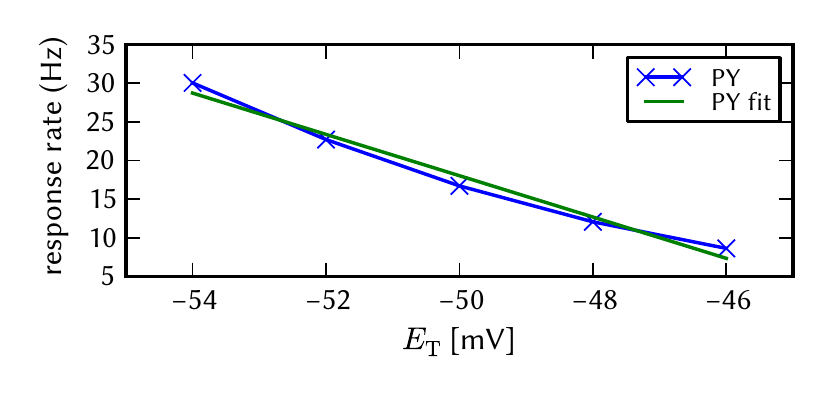}
    \caption{
		{\bf Example compensation factor assertion for the state ($\Ge=\SI{9}{\nano\siemens}$, $\Gi=\SI{90}{\nano\siemens}$) of the AI network:}
		The Figure shows the response rate of an excitatory neuron stimulated by 200 excitatory and 50 inhibitory Poisson sources with rate \SI{12.38}{\Hz} depending on its spike initiation threshold $\Vthresh$.
		The slope $m=-2.6745\frac{\Hz}{\mV}$ of the linear fit is then used to calculate the compensation factor $c_\mathrm{comp}=\frac{0.5}{m}=-0.18695\frac{\mV}{\Hz}$.
    \label{fig:ai-compensation-v-thresh}
        }
\end{figure}

In the so-called iterative compensation, we sequentially modify individual parameters such that the response of each neuron is modified to match its target response.
In our case, we iteratively change the spike detection voltage $\Vthresh$ such that the firing rate of each neuron is shifted towards the target rate.
At each step, the threshold voltage  is adapted as follows:

\begin{align}
\Vthresh^\mathrm{n+1,i} = \Vthresh^\mathrm{n,i} + (\nu^\mathrm{tgt} - \nu^\mathrm{n,i})c_\mathrm{comp}
\label{eq:iterative-comp}
\end{align}
where $\Vthresh^\mathrm{n,i}$ and $\nu^\mathrm{n,i}$ are the threshold voltage and firing rate of neuron i of the n-th step,
 $\nu^\mathrm{tgt}$ is the target rate for all neurons of a population
 and $c_\mathrm{comp}$ is a compensation factor that links the firing response and the threshold voltage.
The target rate $\nu^\mathrm{tgt}$ is computed separately for the excitatory and inhibitory population from the reference simulations (\Cref{sec:ai-criteria}).
We choose the compensation factor for each ($\Ge, \Gi$) state in the following manner:
Similar to the mean-field approach in \cref{sec:ai_comp_strat}, we consider the response rate of an excitatory neuron given a network firing rate of $\nu^\mathrm{tgt}$, that is, the neuron is stimulated by 200 excitatory and 50 inhibitory Poisson sources with rate $\nu^\mathrm{tgt}$.
We then vary the threshold voltage of said neuron between \SI{-54}{\mV} and \SI{-46}{\mV} and thereby determine the dependency of the response rate on the threshold voltage.
From a linear fit of this dependency, we extract the slope $m$, and set the compensation factor to $c_\mathrm{comp}=\frac{0.5}{m}$ (\cref{fig:ai-compensation-v-thresh}).
The factor of $0.5$ was chosen to limit the change of the mean rate in each step in order to avoid oscillations in the compensation procedure.
Whenever we changed the spike initiation voltage $\Vthresh$, we shifted the spike detection voltage $\Vspike$ equally.

We remark that this compensation method requires the parameters for every individual neuron to be fine-tunable.
This is the case for the BrainScaleS wafer-scale hardware, where the AdEx parameters of every hardware neuron are independently configurable with sufficient precision by means of analog floating gate memories (\Cref{sec:neuromorphic_components}), in contrast to the synaptic weights which are restricted to a 4-bit precision in typical operation mode.

\subsection{Further simulations}
\subsubsection{Network size scaling behavior}
\label{sec:methods_ai_scaling_sims}
\begin{figure}[ht]
    \centering
	\includegraphics{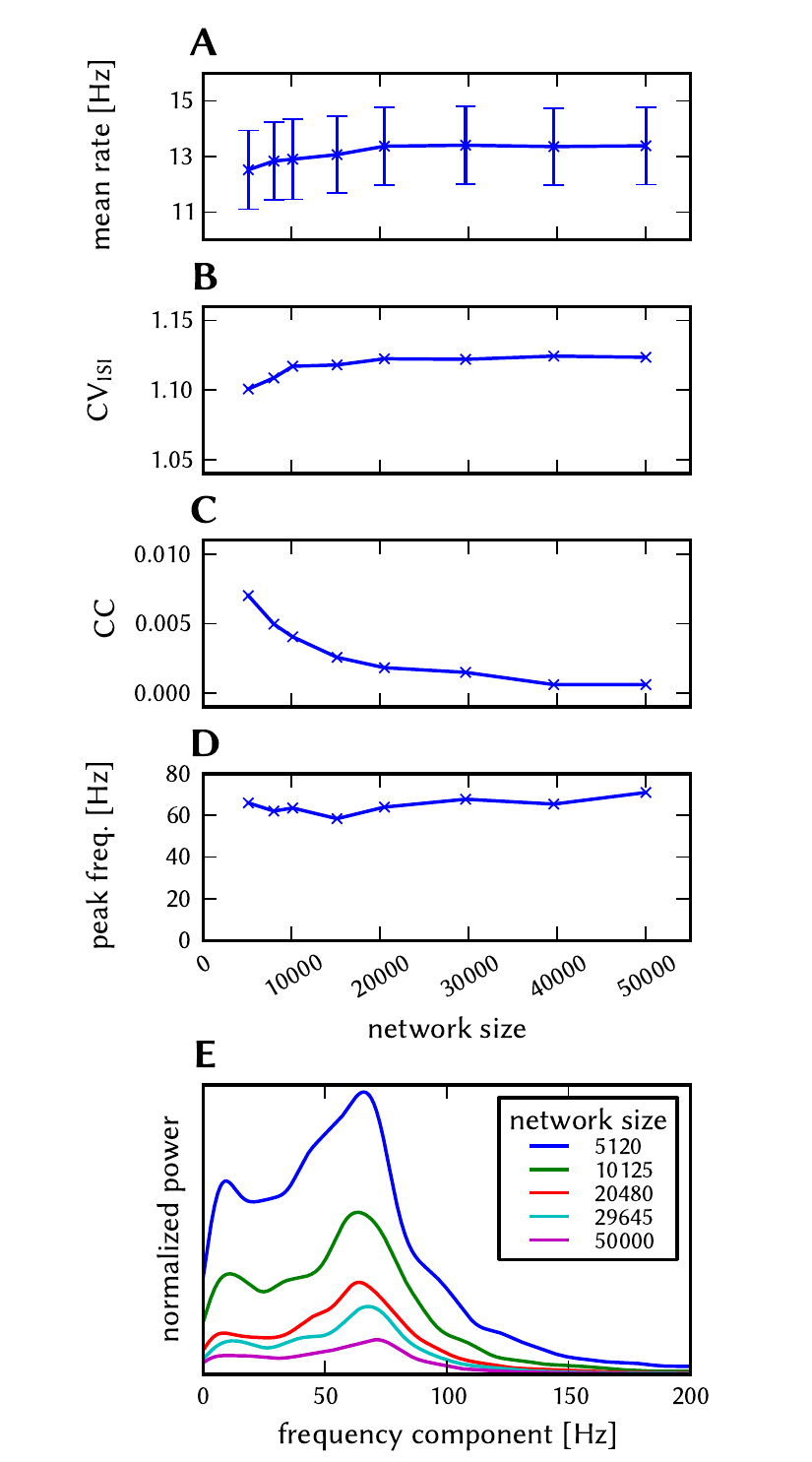}
	\caption{
		{\bf Network size scaling behavior of the AI network for the (\SI{9}{\nano\siemens}, \SI{90}{\nano\siemens}) state.}
		Mean and variance of firing rate across the PY neurons (\textbf{A}),
		coefficient of variance of inter-spike intervals (\textbf{B}), 
		coefficient of pairwise cross-correlation (\textbf{C}), and power spectrum of global activity: full spectra (\textbf{D}) and peak frequencies (\textbf{E}).
	\label{fig:ai-network_size_scaling}
	}
\end{figure}
To investigate what happens when the network is scaled according to the rules given in \cref{sec:methods-ai-scaling}, we pick one state of the ($\Ge$, $\Gi$) space and vary the network size between \num{5000} and \num{50000} neurons.
 The results for the (\SI{9}{\nano\siemens}, \SI{90}{\nano\siemens}) state can be seen in \cref{fig:ai-network_size_scaling}:
 The mean firing rate slightly increases with size until its saturates, while the variance of the firing rate across neurons remains approximately constant (\textbf{A}).
 Like the firing rate, the irregularity ($\cvisi$) increases and saturates with size (\textbf{B}).
 The synchronicity (CC) decreases with size, as one would expect (\textbf{C}).
 The power spectrum of global activity exhibits the same profile for all sizes, however the power is scaled inversely to the network size (\textbf{D} and \textbf{E}).

\subsubsection{Non-configurable axonal delays}
\label{sec:methods_ai_delay_sims}
\begin{figure}[ht]
    \centering
	\includegraphics{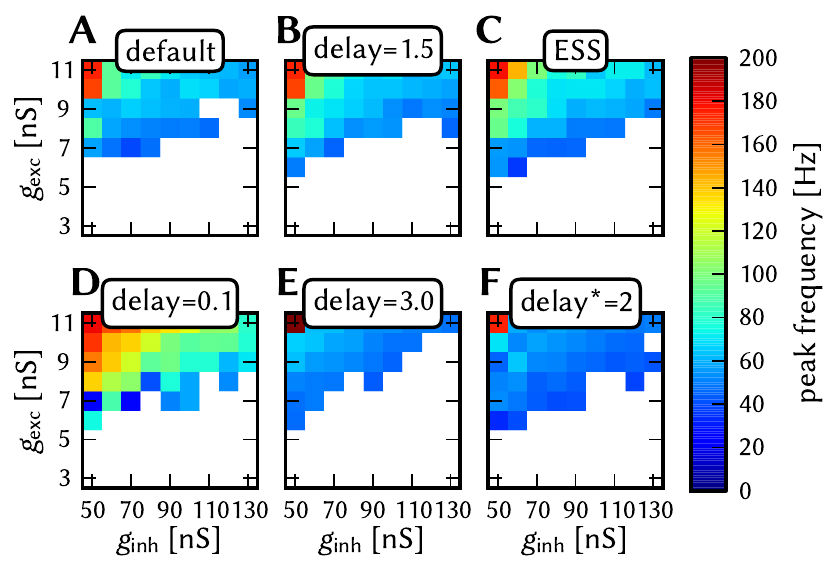}
    \caption{
		{\bf Effects of axonal delays on the AI network.}
		($\Ge$, $\Gi$) spaces with the peak frequency of the global pyramidal activity for different axonal delay setups:
		default with distance-dependent delays (\textbf{A}),
		constant delay of \SI{1.5}{\ms} (\textbf{B}),
		simulation on the ESS where delay is not configurable (\textbf{C}),
		constant delay of \SI{0.1}{\ms} (\textbf{D}),
		constant delay of \SI{3.0}{\ms} (\textbf{E}),
		distance-dependent delays scaled by factor of 2 with respect to default setup (\textbf{F}).
    \label{fig:ai-delay_distortion}
	}
\end{figure}
In \cref{sec:ai-delay-distortion} we argue that non-configurable delays on the BrainScaleS hardware only have a minimal effect on the AI network because the average delay in the model matches the estimated average delay on the hardware.
Here, we provide the simulation results and further investigations on the influence of the delay on the network dynamics.
For the analysis of the effects of non-configurable delay we repeated the ($\Ge$, $\Gi$) sweep with all synaptic delays set to \SI{1.5}{\ms}, cf. \cref{sec:investigated_distortions}.
This distortion mechanism only affected the power spectrum of global actitivity but not the other criteria such that we show only the peak frequency parameter spaces in \cref{fig:ai-delay_distortion}.
The distorted network (\textbf{B}) with a constant delay of \SI{1.5}{\ms} is not significantly different from the default network with distance-dependent delays (\textbf{A}), both the region of sustained activity and the position of the peak in the power spectrum are in good match.
The same holds for ESS simulations (\textbf{C}), where non-configurable delays were the only active distortion mechanism.

To further investigate the influence of the delays, we ran additional simulations where all delays in the network were set to \SI{0.1}{\ms} (\textbf{D}), and \SI{3}{\ms} (\textbf{E}), respectively.
Lowering delays increases the speed of activity propagation such that the position of the peak in the power spectrum is shifted towards higher frequencies.
For higher delays the peak frequency decreases analogously, but also the region of sustained activity diminishes significantly.
(\textbf{F}) shows simulations with distance-dependent delays scaled by a factor of 2 with respect to the baseline model, thus having an average delay of \SI{3.1}{\ms} (cf. \cref{fig:ai-delay-histogram}).
While the peak frequency is in good agreement with the \SI{3}{\ms} simulations, the region of sustained states is extended and even larger than in the baseline setup.
Herewith our simulations affirm that distance-dependent delays in fact do expand the region of self-sustained states in the ($\Ge$, $\Gi$) space (cf. \cref{sec:ai-model-description}).

\subsubsection{Combining distortion mechanisms: synapse loss and synaptic weight noise}
\label{sec:methods_ai_further_sims}

We also investigate what happens when both synapse loss and synaptic weight noise are active at the same time.
Additionally, we test up to which extent we can compensate for both sources of distortions.
To do so we scaled both mechanisms up to \SI{90}{\%} and tried to restore the original behavior for two states: (\SI{9}{\nano\siemens}, \SI{90}{\nano\siemens}) and (\SI{10}{\nano\siemens}, \SI{70}{\nano\siemens}).

\begin{figure}[ht]
    \centering
		\includegraphics{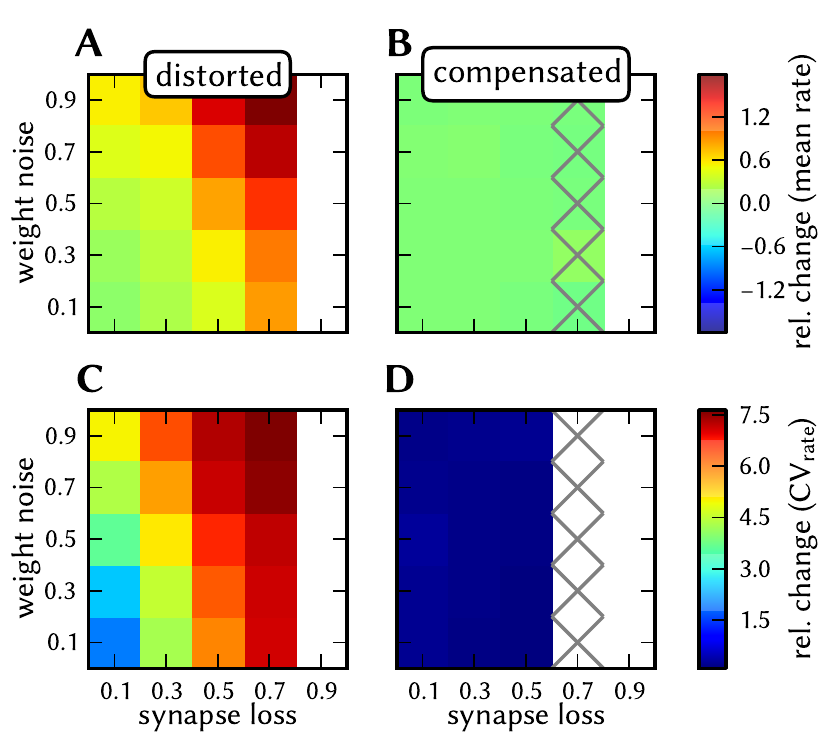}
    \caption{
		{\bf Compensation for combined distortion mechanisms in the AI network with the iterative method.}
		Sweep over synapse loss and synaptic weight noise for the ($\Ge=\SI{9}{\nano\siemens}$, $\Gi=\SI{90}{\nano\siemens}$) state.
		Relative change of the firing rate with respect to the undistorted network for distorted (\textbf{A}) and compensated (\textbf{B}) simulations.
		Relative change of $\cvrate$ with respect to the undistorted network for distorted (\textbf{C}) and compensated (\textbf{D}) simulations.
		The compensated simulations refer to the 10th step of iterative compensation.
		White data points stand for networks where the distorted network did not survive. Data points marked with a cross denote cases where the compensated network did not survive.
        \label{fig:ai-mixed_comp_9_90}
    }
\end{figure}

The relative change 
 of the mean rate and $\cvrate$ are shown in \cref{fig:ai-mixed_comp_9_90} for the (\SI{9}{\nano\siemens}, \SI{90}{\nano\siemens}) state.
 For this state, synapse loss compensation works fine up to a level of \SI{50}{\%}: the relative change of the mean rate and $\cvrate$ are close to 0.
The compensation fails for synapse losses of \SI{70}{\%} and above: when the original firing rate is recovered, the network is unstable, i.e. it does not survive until the end of the experiment.
The amount of synaptic weight noise has no effect on this behavior.
We remark that, during the iterative compensation, there are stable networks with a slightly higher firing rate than the target rate: the network becomes unstable when approaching its target rate.
This is in accordance with observations from the \SI{50}{\%} loss parameter space compensation in \cref{fig:ai-loss_distortion2},
 where the region of sustained activity is smaller than before, i.e. requiring a higher frequency for fewer synapses.
We also note that our iterative compensation algorithm does not recover distorted networks that die out shortly after initial stimulation
(cf. the \SI{90}{\%} synapse loss column in \cref{fig:ai-mixed_comp_9_90} \textbf{A}).
Synaptic weight noise does not pose a problem to the iterative compensation: In all cases the mean rate could be fully recovered and the variance of firing rates close to the original level, with the relative difference of $\cvrate$ being smaller than 1.5.

\begin{figure}[ht]
    \centering
		\includegraphics{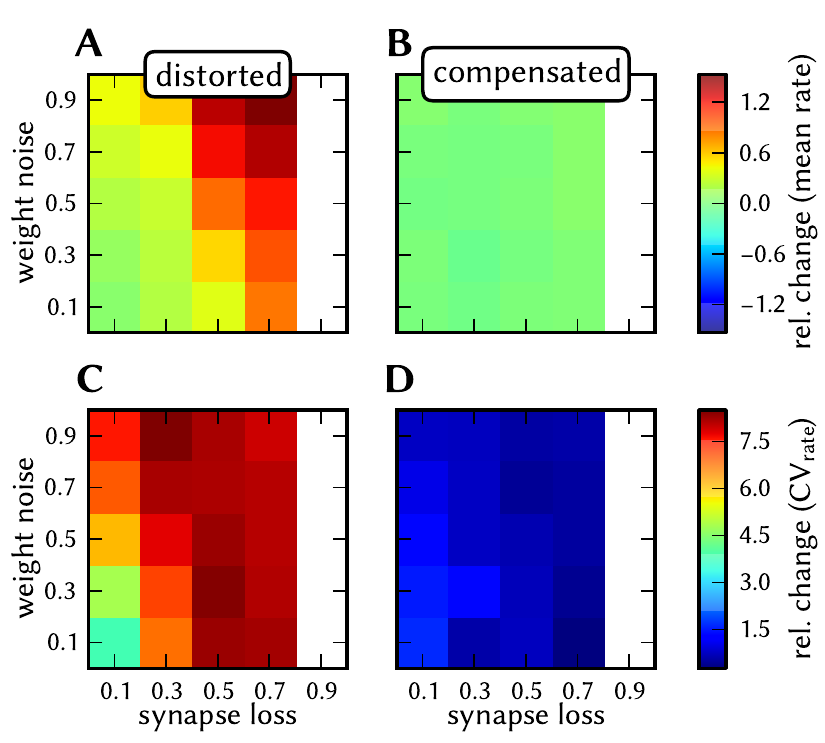}
    \caption{
		{\bf Compensation for combined distortion mechanisms in the AI network with the iterative method.}
		Sweep over synapse loss and synaptic weight noise for the ($\Ge=\SI{10}{\nano\siemens}$, $\Gi=\SI{70}{\nano\siemens}$) state.
		Relative change of the firing rate with respect to the undistorted network for distorted (\textbf{A}) and compensated (\textbf{B}) simulations.
		Relative change of $\cvrate$ with respect to the undistorted network for distorted (\textbf{C}) and compensated (\textbf{D}) simulations.
		The compensated simulations refer to the 10th step of iterative compensation.
		White data points stand for cases where the distorted network did not survive.
        \label{fig:ai-mixed_comp_10_70}
    }
\end{figure}

For the (\SI{10}{\nano\siemens}, \SI{70}{\nano\siemens}) state, compensation was capable of restoring a synapse loss including \SI{70}{\%}, cf. \cref{fig:ai-mixed_comp_10_70}.
Interestingly, the reduction of the variation of firing rates after 10 compensation steps performed slightly better when starting with a higher synapse loss.

We summarize that the iterative method effectively compensates the distortions induced by synapse loss combined with synaptic weight noise, at least when the synapse loss does not exceed \SI{50}{\%}.
Furthermore, we expect these results to hold also for a large area in the ($\Ge$, $\Gi$) space where the network is in the asynchronous regime.

\end{appendices}

\end{document}